\documentclass[
    aps,physrev,
    superscriptaddress,
    floatfix,
]{revtex4-2}
\usepackage{tikz}
\usetikzlibrary{quantikz2}

\usepackage{graphicx} % Required for inserting images
\usepackage{amsmath}
\usepackage{amssymb}
\usepackage{amsfonts}
\usepackage{xcolor}
\usepackage{url}
\usepackage{listings}
\usepackage{siunitx}
\usepackage[normalem]{ulem}
\usepackage{xcolor}  % for marking added text in blue
\usepackage{hyperref}  % by importing hyperref all cross-referenced elements become hyperlinked. load last.

\newtheorem{lemma}{Lemma}
\newtheorem{assumption}{Assumption} 



\begin{document}

\newtheorem{theorem}{Theorem}

\title{Operator Learning with Variational Quantum Circuits}

\author{Jorgen Wu}
\email{jow243@pitt.edu}
\affiliation{Mechanical Engineering, University of Pittsburgh, Pittsburgh, PA, USA}

\author{Masoud Barati}
\email{masoud.barati@pitt.edu}
\affiliation{Electrical and Computer Engineering, University of Pittsburgh, Pittsburgh, PA, USA}

\author{Peyman Givi}
\email{peg10@pitt.edu}
\affiliation{Mechanical Engineering and Petroleum Engineering, University of Pittsburgh, Pittsburgh, PA, USA}

\date{September 2026}

\begin{abstract}
A new methodology  is developed  for quantum machine learning which enables variational quantum circuits to learn linear and non-linear solution operators to differential equations.
This is devised by extending the framework of the DeepONet, a technique for modeling differential equations.
The branch and trunk neural networks are replaced by variational quantum circuits, leveraging the quantum universal approximation theorem.
The methodology is tested on several differential equations and yields low approximation \& generalization errors even with shallow circuit depths. In addition, it does not encounter barren plateaus during training.
The model displays improved error and parameter scaling relative to the classical DeepONet.  This is due to the intrinsic property that the approximation error depends on the dimension of the input function space, rather than the number of sensor locations.
A theoretical framework is established in the form of explicit error bounds, which measure the contributions of the truncation, branch, and trunk circuit errors to the total operator approximation error.
Full derivations for the error bounds are developed in the norms $L^2$, $L^p$, $C^0$, and Sobolev $H^k$, along with a classical-quantum comparative analysis (as Supplemental Materials).
\end{abstract}

\maketitle  % for revtex this must be called after abstract. for normal blank template, put before?
\newpage

\parskip0.1in
\parindent 0pt

\section{Introduction} \label{sec:introduction}

In machine learning, the universal approximation theorem states that a sufficiently wide and deep neural network can approximate any continuous function with arbitrary precision \cite{chenchen_continuousfunctions,chenchen_universalapproximation,hornik_ffnn_uat, Cybenko1989ApproximationBS}. Several variants of this theorem have been proposed for networks of arbitrarily bounded widths and depths \cite{pmlr-v125-kidger20a, GRIPENBERG2003260,barron1992neural, barron1994approximation}; see \cite{gühring2020expressivitydeepneuralnetworks} for a review.  ``Operator learning'', a subfield of machine learning, approximates the solution of a family of differential equations by learning the solution operator $\mathcal{G}$  \cite{BOULLE202483}.
These operators provide a map between the spaces of input and output functions. The universal approximation theorem provides a mathematical basis for representing any nonlinear operator by means of proper basis functions and coefficients.
The ``Deep Operator Network'' (DeepONet) \cite{lu2021deeponet} is an architecture that provides a  means of learning highly nonlinear and complex operators for entire families of partial differential equations (PDEs).  These networks yield lower generalization errors than those of simple feed-forward neural networks, motivating their application to a wide range of problems.  Examples include modeling of chemical kinetics and combustion \cite{chemical_kinetics_deeponet, combustion_acceleration_deeponet}, gas dynamics in diesel engines \cite{Kumar_diesel_deeponet_2023}, stresses and temperatures in manufactured parts \cite{elastoplastic_deeponet, thermal_mechanical_deeponet}, and electromagnetics \cite{electromagnetic_deeponet}. In these applications, the DeepONet has proven useful in producing the solution faster than direct  solvers with comparable accuracies. 

There has also been significant progress in the development of quantum machine learning (QML) algorithms \cite{zaman2025qmlsurvey}. Variational quantum circuits (VQCs), consisting of quantum circuits controlled by classically optimized parameters (typically rotation angles of quantum gates), have experienced widespread use due to their ability to  explore the exponentially large Hilbert space offered by quantum devices. With the use of VQCs, variational quantum algorithms (VQAs) have proven effective for quantum computing on  NISQ devices  \cite{Cerezo_2021_VQA}.
The hardware-efficient ansatz (HEA) is a general form of VQC, intended to reduce the effect of hardware noise by selecting native gates and operations. This ansatz (and VQCs in general) can exhibit the undesirable behavior of barren plateaus \cite{mcclean_barren_2018, you_exponentially_2021}.
This is the quantum equivalent of the vanishing gradient in classical neural networks.
As such, the success of the methodology is somewhat dependent on the initial guessed values of the trainable parameters \cite{anschuetz_quantum_traps_2022}.
Nevertheless, VQCs can be very effective with careful selection of the feature map \cite{Abbas_2021}  when the data being modeled satisfy an area law of entanglement \cite{Leone2024practicalusefulness}.
The VQCs have been used in a variety of QML solutions of classical problems, often demonstrated in the MNIST handwritten digit data set or in a smaller subset \cite{farhi2018classificationquantumneuralnetworks, dilip_data_2022, marshall_high_2023, mordacci_multi-class_2024}.  In addition to approximation of arbitrary functions \cite{datareuploadinguniversal, onequbituniversal}, VQCs have also been used in deep reinforcement learning settings \cite{vqc_deep_reinforcement_learning_2020}, as well as physics-informed neural networks \cite{wang2023expertsguidetrainingphysicsinformed} to learn solutions to complex functions \cite{Setty_2025, Berger2025}. They have also been applied to differential machine learning \cite{Savine2020DifferentialML}, where models are trained not only on target outputs, but also on the derivatives of those outputs with respect to inputs \cite{sakuma_quantum_differential}.

Despite significant progress in QML, the extent of work on operator learning methods with quantum circuits is not very significant.
Examples of previous work include the quantum Fourier network \cite{jain_quantum_fno_2023} and the quantum DeepONet \cite{xiao_quantum_deeponet_2025}.
These works employ a unary (one-hot) encoding scheme, in which the elements of an input vector are loaded as the amplitudes of basis states with a single qubit set to $|1\rangle$ using $n$ qubits for a vector of dimensions $n$.
The quantum orthogonal network of \cite{Landman2022quantummethodsorthogonal}, which acts within this unary subspace via reconfigurable beam-splitter (RBS) gates, is used to implement the matrix multiplications underlying neural network layers on a quantum device.
However, this method requires mid-circuit measurements, as nonlinear activation functions and the addition of bias terms must be calculated classically.

In this work, a new methodology, termed the  ``quantum operator network'' (QONet), is developed, which provides an alternative approach to quantum operator learning.
In this method, the branch and trunk circuits in the DeepONet are replaced with VQCs, resulting in a hybrid classical-quantum architecture.
The capability of the QONet to learn both linear and non-linear operators is demonstrated through numerical experiments, with performance assessment of the model with varying circuit depth and the number of qubits.
Explicit error bounds  are derived in Appendix \ref{apdx:qonet-error-bounds} with proofs across four function space norms ($L^2$, $L^p$, $C^0$, $H^k$).  A comparative analysis with classical DeepONet bounds is furnished in the Supplemental Material \cite{supplement_error_bounds}. 
A consistent distinction is maintained between the approximation-theoretic and learning-theoretic statements. The results based on the former, such as those underlying the error bounds in Section~\ref{sec:error_bounds} and Appendix~A, certify the existence of circuit parameters that realize a target accuracy at a given circuit size.  The analogous results in the Supplemental Material are of the same type. The corresponding learning-theoretic statements, which would additionally guarantee that gradient-based optimization recovers such parameters, are not addressed. The accuracy attained by the trained QONet is considered by numerical experiments in Section~\ref{sec:results}.  % under Adam optimization

\section{Preliminaries}

\subsection{Universal Approximation of Functions \& Operators} \label{sec:UAT}

The general form of the universal approximation theorem for neural networks \cite{chenchen_continuousfunctions} and operators \cite{chenchen_universalapproximation} is presented below.
Theorem \ref{thm:uat_nn} and its variants describe the versatility of neural networks and their ability to learn complex relationships in data, and  Theorem \ref{thm:uat_op} states that any continuous operator can be represented as the sum of products of basis functionals that act on input functions and basis functions acting  on a query location.
% provides a mathematical basis for representing any non-linear operator.

\begin{widetext}
\begin{theorem}[Universal Approximation Theorem for Neural Networks] \label{thm:uat_nn}
Suppose that $\sigma$ is a continuous non-polynomial function, $X$ is a Banach Space, $K \subset X$ is a compact set, $V$ is a compact set in $C(K)$ (endowed with the $C^0$ sup-norm topology), and $f$ is a continuous function defined on $V$. Then, for any $\epsilon > 0$, there are a positive integer $n$, with $m$ points $x_1$ through $x_m \in K$, and real constants $c_i$, $\theta_i$, $\varepsilon_{ij}$, for $i \in [1,n]$, $j \in [1,m]$ such that the following holds for all $u \in V$.
\begin{equation} \nonumber
    \left| f(u) - \sum_{i=1}^n c_i \sigma\left( \sum_{j=1}^m \varepsilon_{ij} u(x_j) + \theta_i \right) \right| < \epsilon
\end{equation}
\end{theorem}

\begin{theorem}[Universal Approximation Theorem for Operators] \label{thm:uat_op}
Suppose $\sigma$ is a continuous non-polynomial function, $X$ is a Banach space, $K_1 \subset X$ and $K_2 \subset \mathbb{R}^d$ are two compact sets in $X$ and $\mathbb{R}^d$ respectively, $V$ is a compact set in $C(K_1)$, $G$ is a nonlinear continuous operator that maps $V$ into $C(K_2)$. For any $\epsilon > 0$, there exist positive integers $n$, $p$, $m$, and constants $c_i^k$, $\varepsilon_{ij}^k$, $\theta_i^k$, $\varsigma_k \in \mathbb{R}$, $w_k \in \mathbb{R}^d$, $x_j \in K_1$, $i \in [1,n]$, $k \in [1,p]$, $j \in [1.m]$, so that:
\begin{equation} \nonumber
    \left| G(u)(y) - \sum_{k=1}^{p} \sum_{i=1}^n c_i^k \sigma \left( \sum_{j=1}^m \varepsilon_{ij}^k u(x_j) + \theta_i^k \right) \sigma(w_k \cdot y + \varsigma_k) \right|<\epsilon
\end{equation}
\end{theorem}
\end{widetext}

% \added{Theorem~\ref{thm:uat_op} establishes the existence of a sum-of-products representation; the bilinear inner-product form adopted in the DeepONet and in the QONet developed here is one architectural realization consistent with this representation, rather than a direct corollary of the theorem.}
The formulation of the DeepONet arises naturally from Theorem \ref{thm:uat_op} and  establishes the existence of a sum-of-products representation.  This formulation is commonly implemented as the inner product of two neural networks, known as the branch and trunk.
The branch is fed the discretized initial condition, and the trunk is fed the query coordinates.  The bilinear inner-product form adopted in the DeepONet and in the QONet developed here is one architectural realization consistent with this representation, rather than a direct corollary of the theorem.  The (unstacked) DeepONet is  expressed as:
\begin{equation}
    G(u)(y) \approx  \langle \mathbf{b}, \mathbf{t} \rangle  + b_0
\end{equation}
where $\mathbf{b} \in \mathbb{R}^k$ is the output of the branch network, $\mathbf{t} \in \mathbb{R}^k$ is the output of the trunk network, $\langle \cdot, \cdot \rangle$ denotes  the standard inner product and $b_0 \in \mathbb{R}$ is an additional scalar bias term. An activation function may also be applied to the output layer of the trunk network.

\subsection{Quantum Circuits as Universal Approximators} \label{sec:QUAT}

% Quantum circuits have been proposed as universal approximators.  \textcolor{blue}{NEED A REFERENCE\cite{XXXX} -> switched sentence ordering -JW }.
Data reuploading schemes have been developed that allow circuits composed of only a single qubit to approximate arbitrary continuous complex functions \cite{datareuploadinguniversal, onequbituniversal}.
These techniques allow quantum circuits to behave as universal approximators.  The universality of the approach is due to the fact that a quantum circuit can be considered as a partial Fourier series, where the expressivity of the circuit is directly related to how many Fourier coefficients (and thus how large a frequency spectrum) the circuit can access \cite{Schuld_2021}.
Under this trigonometric interpretation, the output of a quantum circuit is expressed  as:
\begin{equation}
\begin{split}
    f_{\theta}(x) &= \langle 0 | U^{\dagger}(x; \theta) M U(x; \theta) | 0 \rangle \\
    &= \sum_{w\in\Omega} c_{w}(\theta)e^{iwx}
\end{split}
\end{equation}
where $U(x;\theta)$ is an arbitrary unitary operator parameterized by the input variable $x$ and the trainable parameters $\theta$, and $M$ is a Hermitian operator (observable).
As shown in \cite{Manzano2025approximationquantum, Goto_2021}, quantum circuits can approximate any continuous function.
Precise error bounds on the approximation  are identified in \cite{Gonon_2025}, which states that a VQC with $O(\varepsilon^{-2})$ weights and $O(log_2[\varepsilon^{-1}])$ qubits can approximate functions with integrable Fourier transforms to error $\varepsilon > 0$.
Theorem 3 from \cite{Manzano2025approximationquantum} is expressed (with slight alterations for clarity):    
% \begin{widetext}
\begin{theorem}[$C^0$ Convergence of VQCs] \label{thm:converge_quantum_circuit}
Let $(H_m|m \in N)$ be a universal Hamiltonian family where $H_m$ acts on $m$ subsystems of dimension $d$. The associated family of quantum models $f_m$ is then:
\begin{equation}
    f_m(x) = \langle \Gamma | S_{H_m}^{\dagger}(x) M S_{H_m}(x) | \Gamma \rangle 
\end{equation}
where $|\Gamma\rangle = W(\theta)|0\rangle$ is the state produced by the trainable block in the circuit, and $S(x) = e^{-x_1H} \otimes \cdots \otimes e^{-x_NH}$ is the feature map.
For all functions $f^* \in C^0(U)$, where U is compactly contained in the closed cube $[0, 2\pi]^N$, for all $\varepsilon > 0$ there exists some $m' \in N$, state $|\Gamma\rangle \in \mathbb{C}^{m'}$, and observable $M$ so that $f_{m'}$ converges uniformly to $f^*$:
\begin{equation}
\begin{split}
    \sup_{x \in [0,2\pi]^N} ||f_{m'}(x) - f^*(x)|| < \varepsilon
\end{split}
\end{equation}
\end{theorem}
% \end{widetext}
This theorem states that it is possible to find a finite VQC for which the approximation error is smaller than some arbitrary error $\varepsilon$ at all points of interest.  In addition to $C^0$, the work in \cite{Manzano2025approximationquantum} also contains proofs of convergence in $L^p$ and in the Sobolev space $H^k$, along with generalization bounds for those  spaces.
% \textcolor{blue}{Theorem~\ref{thm:converge_quantum_circuit} establishes the existence of circuits realizing approximate Fourier modes, while Theorem~4 of~\cite{Manzano2025approximationquantum} provides the Sobolev-type polynomial rates used in the branch and trunk approximation bounds.}  Both
These theorems certify the existence of parameter values that realize a prescribed accuracy without addressing their attainability by gradient-based optimization.
% It is important to note that the feature map used is of the form $S(x) = e^{-x_1H} \otimes \cdots \otimes e^{-x_NH}$, which generally refers to quantum states resulting from the tensor product of parameterized unitary matrices.
% This does not include amplitude and unary encoding schemes, suggesting that they do not allow a quantum circuit to behave as a universal approximator.
Another proposed form of the universal approximation theorem for quantum circuits is the following from \cite{onequbituniversal}, which also defines the fundamental UAT gate.

\begin{theorem}[Quantum UAT] \label{thm:quat}
Let $f$, $\phi$, be any pair of functions $f:I_m \rightarrow [0,1]$ and $\phi:I_m \rightarrow [0,2\pi)$, such that $z(\vec{x})=f(\vec{x})e^{i\phi(\vec{x})}$ is a continuous complex function on $I_m$, with $I_m=[0,1]^m$.
Then there is an integer N and a set of parameters $\{ \vec{\theta_1}, \vec{\theta_2}, \cdots, \vec{\theta_N}, \}$ such that, for any $\epsilon>0$,
\begin{equation}
    \left| f(\vec{x})e^{i\phi(\vec{x})} - \langle1|\prod_{i=1}^N U^{\text{UAT}}(\vec{x}_i, \vec{\theta}_i) |0\rangle \right| < \epsilon
\end{equation}
where $U^{\text{UAT}}$ is the fundamental UAT gate, defined as
\begin{equation}
    U^{\text{UAT}}(\vec{x};\vec{\omega},\alpha,\varphi) = R_Y(2\varphi)R_Z(2\vec{\omega} \cdot \vec{x} + 2\alpha) 
\end{equation}
\end{theorem}
This theorem also describes the data reuploading scheme for a single qubit.
Such a circuit consists of repeated groups, where a gate with a learnable parameter is followed by a gate that is parameterized by data.
It is important to note that neither Theorem \ref{thm:converge_quantum_circuit} nor \ref{thm:quat} utilizes amplitude embedding.
Instead, they utilize an angle encoding scheme, where data are loaded onto the circuit as the parameter of a rotation gate.

\subsection{Quantum Approximation of Operators} \label{sec:quantum_approximation_of_operators}

The DeepONet is a high-level framework that does not strictly define the structure of branch and trunk networks \cite{lu2021deeponet}.
This allows the subnetworks to adopt their own inductive biases where appropriate, such as CNNs for image data or RNNs for time series data.
Regardless of the specific subnetwork structures,  the DeepONet is fundamentally the inner product of two universal approximators.
The key characteristics of these approximators are that they are continuous, sufficiently expressive, and can achieve arbitrarily small approximation errors within the domain of applicability.
As described in Section \ref{sec:QUAT}, VQCs also possess these characteristics, as long as they are constructed in a manner that complies with Theorems \ref{thm:converge_quantum_circuit} and \ref{thm:quat}.
It follows that if VQCs are universal approximators, they can be substituted into the DeepONet architecture. The corresponding model is termed  the Quantum Operator Network (QONet), which is formulated as follows: %by the definition of $G(u)(y)$
\begin{equation}
\begin{split}
    G(u)(y) \approx \sum_{k=1}^n \langle B | Z_k | B \rangle \langle T | Z_k | T \rangle + b_0 \\
    \text{where:} \\
    b_0 \in \mathbb{R} \\
    u(x),\; x \in K_1 \subset \mathbb{R}^{d_1} \\
    y \in K_2 \subset \mathbb{R}^{d_2}
\end{split}
\end{equation}
The notation $Z_k$ denotes the single-qubit Pauli-$Z$ observable on the $k$-th qubit, $Z_k = I^{\otimes (k-1)} \otimes Z \otimes I^{\otimes (n-k)}$.
Here, $n$ denotes the number of qubits (and thus the number of outputs in the expansion), $d_1$ is the dimension of the input function domain and $d_2$ is the dimension of the output query domain.
In the numerical experiments conducted here, $d_1 = 1$ and $d_2 = 1$.
The output states of the branch and trunk circuits are represented by the quantum states $|B\rangle$ and $|T\rangle$. 

A subtle but important feature in the QONet is that the branch and trunk produce each a $n$-dimensional vector of Pauli-$Z$ expectation values from a single quantum circuit, rather than from $n$ independent circuits.
Specifically, $b_k(u) = \langle B(u)|Z_k|B(u)\rangle$ for $k=1,\ldots,n$ are simultaneous measurements of the same quantum state $|B(u)\rangle$.
This is consistent with the classical DeepONet and offers benefits for quantum circuits as well, since measurements of the expectation values of individual qubits bypass many of the difficulties associated with taking the expectation value of a single product Hamiltonian \cite{yen_deterministic_2023}.
Evaluation of the expectation values of single qubits is  also faster and less resource-intensive than recovering a full quantum state \cite{williams2024addressingreadoutproblemquantum}.
However, one might ask whether these correlated outputs can independently approximate $n$ arbitrary functions.
This can be established by noting that under the data reuploading structure as employed here, each qubit's expectation value $\langle Z_k \rangle$ is an independent partial Fourier series whose coefficients are controlled by the variational parameters \cite{Schuld_2021}.
The entangling gates in the trainable blocks provide sufficient coupling between qubits to allow independent optimization of each output component.  This is further supported by the results in \cite{Manzano2025approximationquantum, Goto_2021}, which establish that VQCs with angle encoding and sufficient depth can approximate arbitrary continuous vector-valued functions. Therefore, the vector output $n$-dimensional $[b_1(u), \ldots, b_n(u)]$ of a single branch PQC constitutes a universal approximator for vector-valued functions, and the same holds for the trunk circuit outputs $[t_1(y), \ldots, t_n(y)]$.

The $n$ outputs $\{b_k(u)\}_{k=1}^n$ are not independent, as all arise from the same quantum state $|B(u)\rangle$. For example, if $|B(u)\rangle = |0\rangle^{\otimes n}$, then $\langle Z_k\rangle = 1$ for every $k$, and the achievable set of expectation-value tuples is a strict subset of $\{-1,1\}^n$. Accordingly, full vector-valued universality is not claimed from the scalar results of~\cite{Manzano2025approximationquantum, Goto_2021, Schuld_2021} alone. The weaker property used in the error bounds developed below is the following: under data reuploading with entangling layers, each scalar expectation $\langle Z_k\rangle$ is a partial Fourier series in $u$ whose coefficients are controlled by an independent subset of the variational parameters, while the entangling gates provide cross-qubit coupling sufficient for the joint coefficient tuple to span an open neighborhood of any target $(c_{\omega_1}(u), \ldots, c_{\omega_n}(u))$ at sufficient circuit depth. A complete proof of multi-output universality for this particular ansatz remains an open problem. The error bounds below are therefore formulated conditional on the existence of parameter settings realizing the prescribed scalar approximations, consistent with the existential character of Theorems~\ref{thm:converge_quantum_circuit} and~\ref{thm:quat}.

It is also worth noting that the Pauli-$Z$ expectation values are bounded in $[-1, 1]$, whereas classical neural networks with unbounded activations (e.g. ReLU) can produce outputs of arbitrary magnitude.
This bounded output range does not affect universality, as it can simply be considered analogous to bounded activations such as hyperbolic tangent or sigmoid functions.
In addition, any bounded continuous function in a compact domain can be scaled to lie within $[-1,1]$, and the data scaling described in Section \ref{sec:DataGeneration} ensures that both inputs and targets fall within this range. The bias term $b_0$ in the QONet definition provides additional flexibility to shift the output range as needed.

\subsubsection{Operator Approximation Error Bounds} \label{sec:error_bounds}

The substitution of VQCs into the DeepONet framework admits explicit error bounds. Consider a target operator $\mathcal{G}: V \to L^2(K_2)$, where $V \subset L^p(K_1)$ is a compact set of input functions.
Suppose $\mathcal{G}(u)(y)$ admits a Fourier expansion $\mathcal{G}(u)(y) = \sum_{\omega \in \mathbb{Z}^{d_2}} c_\omega(u) \varphi_\omega(y)$ with coefficient decay $|c_\omega(u)| \leq C_0 |\omega|^{-s}$ for $s > d_2/2$.
The bilinear structure of the QONet approximation $\mathcal{G}_\theta(u)(y) = \sum_{k=1}^n b_k(u) t_k(y) + b_0$, adopted here as a specific architectural realization consistent with Theorem~\ref{thm:uat_op}) incurs two sources of errors:

\textbf{1. Truncation Error:} By retaining only $n$ Fourier modes (with $|\omega| \leq K$, $n \propto K^{d_2}$):
\begin{equation} \label{eqn:qonet_truncation_error}
    \|\mathcal{G}(u) - \mathcal{G}_n(u)\|_{L^2} \leq C_4 \, n^{-(s/d_2 - 1/2)},
\end{equation}
where $\mathcal{G}_n(u)$ is the $n$-term truncated Fourier series.

\textbf{2. Circuit Approximation Error:} Let $\delta_b = \max_k |c_{\omega_k}(u) - b_k(u)|$ be the maximum branch approximation error and $\delta_t = \max_k \|\varphi_{\omega_k} - t_k\|_{L^2}$ be the maximum trunk approximation error. 
Theorems~\ref{thm:converge_quantum_circuit} and~\ref{thm:quat} guaranty, in an existential sense, that $\delta_b$ and $\delta_t$ decrease as the depth and width of the circuit increase.
% Theorems \ref{thm:converge_quantum_circuit} and \ref{thm:quat} guarantee that both $\delta_b \to 0$ and $\delta_t \to 0$ increase as the depth and width of the circuit increase. 
The circuit approximation error satisfies:
\begin{equation} \label{eqn:qonet_circuit_approximation_error}
    \|\mathcal{G}_n(u) - \mathcal{G}_\theta(u)\|_{L^2} \leq \sqrt{n}\,\delta_b + S(n)\,\delta_t \;+\; \sqrt{n}\, \delta_b\, \delta_t,
\end{equation}
where the final term originates from the two-term bilinear decomposition $c_k\phi_k - b_k t_k = (c_k - b_k)\phi_k + b_k(\phi_k - t_k)$ combined with $|b_k| \leq |c_k| + |c_k - b_k|$. Although retained explicitly, under the standing assumption $\delta_b, \delta_t \ll 1$, this term is of the second order and is asymptotically small.

Here, $S(n) = \sum_{k=1}^n |c_{\omega_k}(u)|$ depends on the smoothness of the target operator. Specifically, $S(n) = O(1)$ when $s > d_2$, $S(n) = O(\log n)$ when $s = d_2$, and $S(n) = O(n^{1-s/d_2})$ when $d_2/2 < s < d_2$. The elementary derivation of these three regimes is given in Appendix~A. The total $L^2$ error bound is therefore:
\begin{equation} \label{eqn:total_error_bound}
    \varepsilon_{L^2}(n, \delta_b, \delta_t) \leq C_4 \, n^{-(s/d_2 - 1/2)} + \sqrt{n}\,\delta_b + S(n)\,\delta_t \;+\; \sqrt{n}\,\delta_b\,\delta_t.
\end{equation}
For  $d_2 = 1$, and when the operators are sufficiently smooth ($s > 1$), $S(n) = O(1)$, the trunk error contributes only a constant factor.

\noindent\emph{Assumption on the trained model.} The above bounds identify the trunk targets with a specific orthonormal Fourier basis $\{\varphi_{\omega_k}\}_{k=1}^n$ of the leading modes and the branch targets with the corresponding coefficient functionals $c_{\omega_k}(u)$. They therefore quantify the error of an idealized QONet whose trunks are  driven toward the chosen $\varphi_{\omega_k}$ and whose branches are driven toward the $c_{\omega_k}$. The QONet as actually trained converges to any minimizer of the empirical loss, which need not coincide with this Fourier truncation. The bounds therefore characterize one route by which the QONet can in principle realize the target accuracy; a basis-free SVD-based analysis in the spirit of~\cite{LanthalerMishra2022} is left for future work.

Appendix \ref{apdx:qonet-error-bounds}, along with the Supplemental Material \cite{supplement_error_bounds}, contains complete derivations of this result in addition to analogous bounds under the following three norms:
\begin{itemize}
    \item \textbf{($L^p$ bound):} $\varepsilon_{L^p} \leq C_5\,n^{-(s/d_2 - 1/p)} + C_\Phi\,n\,\delta_b + \widetilde{S}_p(n)\,\delta_t$;
    \item \textbf{($C^0$ bound):} $\varepsilon_{C^0} \leq C_7\,n^{-r} + n\,\delta_b + S_{C^0}(n)\,\delta_t$, which extends Theorem~\ref{thm:converge_quantum_circuit} from the function approximation to the operator approximation;
    \item \textbf{($H^k$ bound):} $\varepsilon_{H^k} \leq C_{11}\,n^{-\alpha/d_2} + C\,n^{1/2+k/d_2}\,\delta_b + S_{H^k}(n)\,\delta_t^{(H^k)}$, relevant for problems requiring smoothness of the solution. The factor $n^{1/2 + k/d_2}$ on the branch term arises from a Bessel-inequality estimate $\bigl\|\sum_k (c_k - b_k)\phi_{\omega_k}\bigr\|_{H^k}^2 \leq (1+K^2)^k \sum_k (\delta_b^{(k)})^2$, with $K \sim n^{1/d_2}$, as explicitly derived in Appendix~\ref{apdx:qonet-error-bounds}.  % of Eq.~\eqref{eq:gamma_q_b_poly} of Appendix-A
\end{itemize}

Additionally, in the Supplemental Material derivations are provided for  parallel error bounds for the classical ANN-based DeepONet with identification of precise conditions under which the quantum architecture achieves superior error scaling, conditional on the explicit structural hypothesis on the target operator introduced below.

\begin{assumption}[Low-Intrinsic-Dimension Operator]\label{ass:low-intrinsic}
There exists a continuous projector $\Pi : \mathbb{R}^{m_b} \to \mathbb{R}^{d_1}$ and continuous maps $\tilde{c}_\omega : \mathbb{R}^{d_1} \to \mathbb{R}$ such that, for all $u \in V$ and all retained modes $\omega_k$, the coefficient functional admits the factorization $c_{\omega_k}(u) = \tilde{c}_{\omega_k}(\Pi u)$. The classical preprocessing $\mathbf{v} = A\mathbf{u} + \mathbf{b}$ of the input functions given to the branch circuit is assumed to admit a linear realization whose first $d_1$ rows approximate $\Pi$.
\end{assumption}

Under Assumption~\ref{ass:low-intrinsic}, the VQC branch effectively approximates a function on $\mathbb{R}^{d_1}$ rather than on $\mathbb{R}^{m_b}$, and consequently the quantum branch approximation rate may be expressed in terms of $d_1$. In the absence of this assumption, the universal-approximation results of~\cite{Manzano2025approximationquantum,Goto_2021} yield rates governed by the actual input dimension of the circuit, namely $m_b$. The condition $m_b^c > 2 d_1$ stated in the Supplemental Material is to be interpreted as conditional on the Assumption~\ref{ass:low-intrinsic}, as it constitutes a structural property of the target operator rather than a property of the architecture alone. Classical architectures equipped with comparable inductive biases (convolutional, equivariant, or neural-operator type) can, in principle, exploit analogous structure.

Under the same structural hypothesis, and provided that the coefficient functionals $u \mapsto c_\omega(u)$ admit an analytic extension to a complex strip of width $\rho > 0$ (in the sense of~\cite{Schwab2019,Wojtaszczyk2020}), the classical Fourier approximation theory yields an existential decay $\delta_b = O(e^{-\alpha P_b})$ for the branch VQC under angle encoding and data reuploading; this complements the polynomial rate above. The constant $\alpha$ is determined by the analyticity radius $\rho$ and is not estimated in the present work.

\section{Experiments}

\subsection{Variational Quantum Circuit Construction} \label{VQCconstruction}
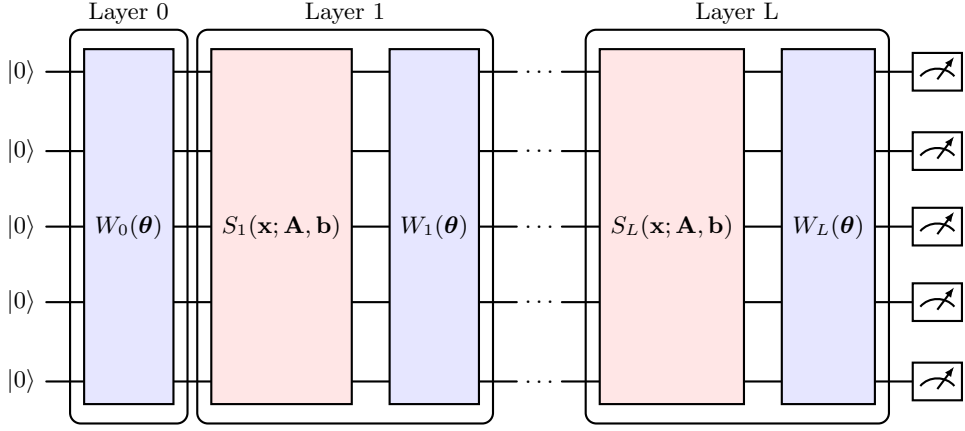
\begin{figure*}
    \centering
    \begin{quantikz}
        \lstick{$|0\rangle$} & \gate[5, style={fill=blue!10}]{W_0(\boldsymbol{\theta})}\gategroup[5, style={rounded corners, inner xsep=2pt}]{Layer 0} & \gate[5, style={fill=red!10}]{S_1(\mathbf{x};\mathbf{A},\mathbf{b})}\gategroup[5, steps=2, style={rounded corners, inner xsep=2pt}]{Layer 1} & \gate[5, style={fill=blue!10}]{W_1(\boldsymbol{\theta})} & \ \ldots\ & \gate[5, style={fill=red!10}]{S_L(\mathbf{x};\mathbf{A},\mathbf{b})}\gategroup[5, steps=2, style={rounded corners, inner xsep=2pt}]{Layer L} & \gate[5, style={fill=blue!10}]{W_L(\boldsymbol{\theta})} & \meter{} \\
        \lstick{$|0\rangle$} & & & & \ \ldots\ & & & \meter{} \\
        \lstick{$|0\rangle$} & & & & \ \ldots\ & & & \meter{} \\
        \lstick{$|0\rangle$} & & & & \ \ldots\ & & & \meter{} \\
        \lstick{$|0\rangle$} & & & & \ \ldots\ & & & \meter{}
    \end{quantikz}
    \caption{The variational quantum circuits  have a repeated structure, consisting of alternating trainable blocks and data encoding blocks. This example is shown for five qubits.}
    \label{fig_vqcstructure}
\end{figure*}

The variational circuits comprising the branch and trunk of the QONet are constructed using the repeating structure described in \cite{datareuploadinguniversal, onequbituniversal, Schuld_2021, Manzano2025approximationquantum}, where the circuit consists of alternating layers of trainable blocks and data encoding blocks.
An example is shown in Figure \ref{fig_vqcstructure}.
The output of this circuit can be written as:
% The same notation will be used consistently throughout this document.
\begin{equation}
\begin{split}
    |\psi\rangle = W_L(\boldsymbol{\theta}) S_L(\mathbf{x};\mathbf{A}, \mathbf{b}) \dots W_1(\boldsymbol{\theta}) S_1(\mathbf{x};\mathbf{A}, \mathbf{b}) W_0(\boldsymbol{\theta}) |0\rangle
\end{split}
\end{equation}
where $|\psi\rangle$ is the quantum state produced by the circuit, $W_L(\boldsymbol{\theta}) \in \mathbb{R}^{2^n \times 2^n}{U(2^n) \subset \mathbb{C}^{2^n \times 2^n}}$ is the unitary operator corresponding to the trainable block parameterized by the vector $\boldsymbol{\theta}$, $S_L(\mathbf{x};\mathbf{A}, \mathbf{b}) \in \mathbb{R}^{2^n \times 2^n}{U(2^n) \subset \mathbb{C}^{2^n \times 2^n}}$ is the unitary operator corresponding to the data encoding block, and $n$ is the number of qubits which corresponds to a (quantum) Hilbert space of dimension $2^n$.
From the ordering above, the trainable block $W_0(\boldsymbol{\theta})$ is applied first to the initial state $|0\rangle^{\otimes n}$, so that the first data-encoding block $S_1$ acts on a non-trivial state. %This ordering is required because $R_Z$ rotations applied directly to $|0\rangle$ contribute only a global phase.

\subsubsection{Trainable Blocks}
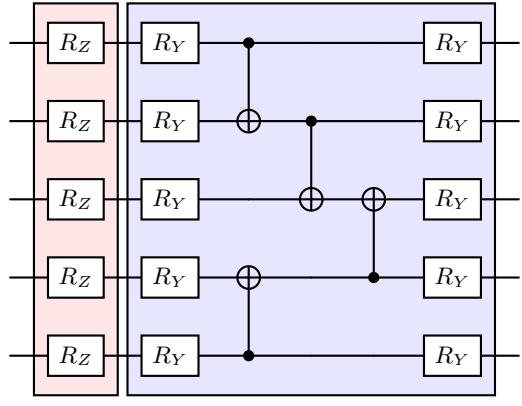
\begin{figure}
    \centering
    \begin{quantikz}
    & \gate{R_Z}\gategroup[5, style={fill=red!10, inner xsep=2pt}, background]{} & \gate{R_Y}\gategroup[5, steps=5, style={fill=blue!10, inner xsep=2pt}, background]{} & \ctrl{1} &         &        & \gate{R_Y} & \\
    & \gate{R_Z} & \gate{R_Y} & \targ{}  & \ctrl{1} &         & \gate{R_Y} & \\
    & \gate{R_Z} & \gate{R_Y} &          & \targ{}  & \targ{} & \gate{R_Y} & \\
    & \gate{R_Z} & \gate{R_Y} & \targ{}  &          & \ctrl{-1} & \gate{R_Y} & \\
    & \gate{R_Z} & \gate{R_Y} & \ctrl{-1} &         &           & \gate{R_Y} & 
\end{quantikz}
    \caption{A close-up view of one layer within the circuit. The data encoding blocks (highlighted in red) consist of only $RZ$ gates. This example is shown for five qubits. Additional classical preprocessing (not shown) is performed to the inputs beforehand. The trainable blocks (highlighted in blue) consist of $RY$ and $CNOT$ gates. This example is illustrated using five qubits with a triangular entanglement pattern. This entanglement pattern is  employed to reduce circuit depth and increase connectivity compared to a linear entanglement pattern.}
    \label{fig:data_and_train_blocks}
\end{figure}

The trainable blocks in both the branch and trunk circuits consist of single-qubit rotations and CNOT gates for entanglement.
This is shown in Figure \ref{fig:data_and_train_blocks} and can be expressed using Equation \ref{eqn_trainblock}, where $W(\boldsymbol{\theta})$ is the product of three distinct layers.
Although this equation describes a linear entanglement pattern, other entanglement patterns can be employed, such as the triangular pattern shown in Figure \ref{fig:data_and_train_blocks} to increase entanglement while minimizing connectivity between non-adjacent qubits.

% \begin{equation} \label{eqn_trainblock}
%     W(\boldsymbol{\theta}) = \left[\bigotimes_{i=1}^n RY_i(\theta_{n+i})\right] \left[\bigotimes_{i=1}^{n-1} CNOT_{i, i+1}\right] \left[\bigotimes_{i=1}^n RY_i(\theta_i)\right]
% \end{equation}
\begin{equation} \label{eqn_trainblock}
    \begin{split}
        W(\boldsymbol{\theta}) = &\left[\bigotimes_{i=1}^n RY_i(\theta_{n+i})\right] %\\
        \left[\bigotimes_{i=1}^{n-1} CNOT_{i, i+1}\right]
        \left[\bigotimes_{i=1}^n RY_i(\theta_i)\right]
        % &\quad \cdot \left[\bigotimes_{i=1}^{n-1} CNOT_{i, i+1}\right] %\\  % use these when in two column format
        % &\quad \cdot \left[\bigotimes_{i=1}^n RY_i(\theta_i)\right]
    \end{split}
\end{equation}

\subsubsection{Data Encoding Blocks} \label{sec:data_encoding}
As discussed in \cite{lu2021deeponet}, the training data for the DeepONet are presented as a tuple consisting of an input function, a query location, and a target output at the query location.
This is expressed as $\left(u(x), y, G(u(x))(y) \right)$, where $u(x)$ is the input function, $y$ is the query location, and $G(u(x))(y)$ is the output of the solution operator in the location of the query location.
The input function is discretized on $n$ grid points, which are typically equally spaced.
The circuits are then constructed with $n$ qubits, as this naturally allows one value of the input function to be encoded per qubit using a rotation gate.
% An angle encoding scheme was chosen because, as discussed in Section \ref{sec:QUAT}, Theorem \ref{thm:converge_quantum_circuit} does not cover amplitude encoding schemes.
% The branch circuit performs classical preprocessing of the input function before loading it onto the circuit via single-qubit rotation gates.
% This is achieved through a weight matrix and a bias vector, similar to a classical neural network.
% The utilization of preprocessing parameters in the data encoding block, in combination with the trainable blocks, can be thought of as an extension of the fundamental UAT gate shown in Theorem \ref{thm:quat} to multiple qubits. 
The data encoding block used in the branch circuit is defined by Equation \ref{eqn_branch_fmap}.
Classical preprocessing of the input function occurs before the result is loaded onto the circuit via single-qubit rotation gates. This is achieved through a weight matrix and a bias vector, similar to those in a classical neural network. Note that the matrix $\mathbf{A}$ is initialized as $-\mathbf{I}$, and all elements of $\mathbf{b}$ are initialized as $\pi/2$. It is not always necessary to optimize the classical weight matrix and bias vector; they can either be ``frozen'' or optimized along with the rest of the variational parameters in the trainable blocks. The use of preprocessing parameters in the data encoding block, in combination with the trainable blocks, can be thought of as an extension of the fundamental UAT gate given by  Theorem \ref{thm:quat} to multiple qubits.
The affine preprocessing $\mathbf{v} = \mathbf{A}\mathbf{u} + \mathbf{b}$ embedded in each encoding block is structurally a small classical feedforward stage attached to the branch circuit. In the parameter-count comparisons of the following section, this property is taken into account: a comparison like-for-like freezes $\mathbf{A}$ and $\mathbf{b}$ or augments the classical baseline with an analogous preprocessing block. Both cases are discussed  where relevant.
\begin{equation} \label{eqn_branch_fmap}
\begin{split}
    S_B(\mathbf{u}(\mathbf{x});\mathbf{A}, \mathbf{b}) = \bigotimes_{i=1}^n RZ_i(v_i) \\
    \text{where} \\
    \mathbf{u}(\mathbf{x}) \in \mathbb{R}^{n} \\
    \mathbf{A} \in \mathbb{R}^{n \times n} \\
    \mathbf{b} \in \mathbb{R}^{n} \\
    \mathbf{v} = (\mathbf{A} \cdot \mathbf{u}(\mathbf{x})+\boldsymbol{b}) \in \mathbb{R}^n \\
\end{split}
\end{equation}

The following feature map in Equation \ref{eqn_trunk_fmap} is  used in the trunk circuit data encoding blocks. This feature map is of the form $S(x)$ specified in Theorem \ref{thm:converge_quantum_circuit}, with $H = \sum_{i=0}^n \sigma_Z^{(i)}$ and $N=1$. Alternatively, it can be considered to be the same feature map as the branch network, except that the scalar input for query location is encoded onto every qubit.

\begin{equation} \label{eqn_trunk_fmap}
\begin{split}
    S_T(y) = \bigotimes_{i=1}^n RZ_i\left(\frac{\pi}{2} - y \right)\\
\end{split}
\end{equation}
It should be noted that this trunk encoding loads the same scalar $y$ onto every qubit, which restricts the trunk to learning functions of a single variable ($d_2 = 1$).
For multi-dimensional output domains ($d_2 > 1$), one could encode different components of $y$ on different qubit subsets, or employ multi-dimensional angle encoding schemes; this extension is left for future work.
Because the same scalar $y$ is loaded onto every qubit, the joint state $|T(y)\rangle$ traces a one-parameter curve in Hilbert space and the vector $(t_1(y),\ldots,t_n(y))$ traces a one-parameter curve in $[-1,1]^n$. However, each component $t_k(y) = \langle T(y)|Z_k|T(y)\rangle$ is a partial Fourier series in $y$ whose accessible frequencies are set by the number of qubits and the depth of the circuit and whose coefficients are controlled by independent components of $\boldsymbol{\theta}$. The $n$ outputs $\{t_k\}$ therefore constitute $n$ smooth scalar functions of $y$ that, under generic settings of the trainable parameters, are linearly independent on $K_2$; their span provides an approximate $n$-dimensional basis on $K_2$. This is the sense in which the trunk circuit plays the role of the trunk neural network in the DeepONet; no claim is made that arbitrary $n$-tuples of values can be realized at a single $y$.
This form of feature map also possesses a geometric interpretation when considering the expectation value of the Pauli-Z operator.  As an example, a single qubit that starts in the $|0\rangle$ state and is rotated by an angle of $\pi/2$ around the y-axis will have an expectation value of $0$ for the Pauli-Z operator.
To obtain Pauli-Z expectation values in the range $0$ and $1$, the qubit must be rotated by an angle less than $\pi/2$, or by an angle $\pi/2-\omega$ where $\omega \in [0, \pi/2]$.
Similarly, for expectation values in the range  $0$ and $-1$, $\omega \in [-\pi/2, 0]$.
Note that while this geometric interpretation suggests that the $RY$ gate should be used in the feature map, it was found that the $RZ$ gate resulted in lower training losses and better overall performance.
The use of a different rotation gate in the trainable blocks or a different expectation value during measurement could change this finding.

\subsubsection{Parameter Count}

The parameter count of the QONet is  determined by considering a circuit with $N$ qubits, $L$ data encoding blocks, and $R=L+1$ trainable blocks.
Each data-encoding block contains $N$ $R_Z$ gates, and each trainable block has $2N$ $R_Y$ gates.
There are a total of $LN+2N(L+1)$ single-qubit rotation gates.
However, $LN$ gates are used for data encoding, so the total number of parameterized single-qubit gates that can be optimized per circuit is only $2N(L+1)$ or $2NR$.
Since there is a branch and trunk circuit, the total number of parameterized single-qubit gates is $4NR$.
% The number of $CNOT$ gates is not relevant to this discussion; regardless, for a linear entanglement layer, a circuit will contain $(L+1)(N-1)$; for a triangular entanglement layer, $2(L+1)(\lfloor N/2\rfloor)$.
The number of classical preprocessing parameters per trainable block in the branch circuit is $W(N^2 + N)$, where $W$ is the number of unique sets of preprocessing parameters.
If these  parameters are not used, then $W$ is simply set to zero
The total number of trainable parameters (degrees of freedom) in the QONet is then calculated as:
\begin{equation} \label{eqn:qonet_dof}
    \text{DOF}_{\text{QONet}} = 4NR + W(N^2+N)
\end{equation}
As such, the QONet has a gate and parameter count that increases approximately linearly with the number of qubits when the circuit depth is held constant. 
Similarly, the gate count increases approximately linearly with the circuit depth when the number of qubits is kept constant.  The number of CNOT gates per entanglement layer depends on the entanglement structure.
Each trainable block has an entanglement section in the middle, resulting in $R=L+1$ entanglement layers.
If a linear entanglement layer is used, there are $N-1$ $CNOT$ gates per layer.
The triangular entanglement layer has $2(\lfloor N/2 \rfloor)$ $CNOT$ gates per layer.
The total number of $CNOT$ gates in the circuit is then $R(N-1)$ for a linear entanglement layer, and $2R(\lfloor N/2 \rfloor)$ for a triangular entanglement layer.

\subsection{Data Generation} \label{sec:DataGeneration}

The samples generated for the input function $u(x)$ must span the chosen function space sufficiently. Unevenly distributed training data often lead to models that generalize poorly.
Here, two function spaces are considered: Chebyshev polynomials of the first kind, and a cosine function.
The set of Chebyshev initial conditions can be generated by randomly sampling coefficients from uniform distributions in the range $[-1, 1]$:
\begin{equation} \label{cheb_init}
    V_{Chebyshev} = \left\{ \sum_{i=0}^{N-1} a_i T_i(x) : |a_i| \leq 1 \right\}
\end{equation}
The set of cosine initial conditions can be generated by randomly selecting a phase, amplitude, and offset of a cosine wave from the  uniform distribution. The cosine function is  chosen because it naturally lends itself to periodic initial conditions.
\begin{align} \label{cosine_init}
    V_{\text{Cosine}} = \big\{ &A\cos(2\pi x + P) + B : \nonumber \\
    &A \sim [0,1], B \sim [-0.5, 0.5], P \sim [0, 2\pi] \big\}
\end{align}
    % V_{Cosine} = \left\{ A\cos(2\pi x + P) + B : A \sim [0,1], B \sim [-0.5, 0.5], P \sim [0, 2\pi] \right\}
The way in which the data are prepared implicitly changes the underlying operator that is  being learned.
For initial value ODEs, given an input function (and assuming the same initial value for all input functions), there is only one possible output function for the model to learn. For 
PDEs with spatial and temporal variations, a single input function generates an entire trajectory of output functions for each time step. With this in mind, the PDE datasets are prepared with a constant prediction horizon, so that each input function at time $t$ is paired with a target function at time $t + \Delta t$.
A model trained on such a dataset learns to function as a solution propagator.
Note that the time-step taken when solving the differential equation as part of generating the dataset does not have to be the same as the prediction horizon; that is to say, $dt$ and $\Delta t$ do not have to be necessarily the same.
For example, the differential equation could be discretized with a forward in time using a time-step $dt=0.01$, while the prediction horizon is $\Delta t=2.0$, which corresponds to a 20-step solution propagator.
The resulting operator is markedly smoother (lower rank) than the single-step propagator. The following results, which show that very shallow QONet circuits achieve low error on this operator, should be interpreted in this light: the model is being asked to represent a heavily-smoothed propagator rather than the unsmoothed step-wise dynamics.
More information on the specifics of the generated datasets can be found in the Appendix \ref{apdx:training-information}.

The generated datasets are then scaled using the maximum absolute scaling, where each dataset is divided by its maximum absolute value. This approach is favored over other methods commonly employed  in classical machine learning \cite{ahsan_effect_2021}, such as min-max scaling, normalization, or robust scaling, because it does not shift the center of the data during scaling. As discussed previously, the feature maps in this work rely on geometric interpretations of qubits. For this reason, it is desirable that an input with a value of zero remain zero after scaling.  All features of the resulting scaled data are contained in the range $[-1, 1]$, although the data may not be extended throughout the range. An implication of Theorem \ref{thm:converge_quantum_circuit}  \cite{Manzano2025approximationquantum} is that a range of $[-\pi/2, \pi/2]$ may be the most appropriate, but $[-1, 1]$ was found to work sufficiently well in simulations.
% A systematic ablation between these two ranges has not been performed and is identified as future work.

\section{Results} \label{sec:results}

Numerical experiments are conducted  using the Pennylane library \cite{bergholm2022pennylaneautomaticdifferentiationhybrid}.
Optimization of variational quantum circuits is carried out using JAX \cite{jax2018github} and Optax.
JAX is  chosen for its ability to pre-compile operations and its pseudo-random number generator, which offer runtime reductions and allow for easy reproduction of results, respectively.
Moreover, the automatic differentiation capability within JAX enables the use of gradient-based optimizers, such as those available in the Optax library.
% The QONet was trained to learn the solution operators for four differential equations.
% The initial conditions were sampled as described in Section \ref{sec:DataGeneration}.
% Those are summarized in Table \ref{tab:datasets}.
% For more information on the specifics of each operator, refer to the following subsections.
Hyperparameters are manually set for each experiment and
 the Adam optimizer \cite{kingma2017adammethodstochasticoptimization} is used in all experiments.   The L-BFGS optimizer \cite{Liu1989_LBFGS} was also considered, but was not used due to its higher memory consumption and longer wall-clock time. Optimization algorithms specifically designed for VQCs, such as ``quantum natural gradient descent'' \cite{Stokes_2020_quantum_natural_gradient, borysenko2025_quantum_natural_gradient}, were not evaluated.  The learning rates are  set experimentally, based on other investigations that have shown  VQCs perform well with relatively higher learning rates (as compared to neural networks) \cite{lockwood2022empiricalreviewoptimizationtechniques}.
Unless otherwise specified, a learning rate of \num{5E-3} or \num{1E-2} is used. Models are sometimes trained for additional epochs after the training loss plateaus to improve generalizability.
This is consistent with \cite{NIPS2017_a5e0ff62_trainlonger}; however, these results pertain to neural networks and as such may not be entirely applicable to VQCs.

The variational parameters are randomly initialized according to Theorem 1 of \cite{Wang_2024_trainabilityenhancement}, indicating that the parameters should be uniformly initialized in the range $[-a\pi, a\pi]$, where $a$ is the solution to $\frac{\sin 2a\pi}{2a\pi} = \frac{S(2L-1)-2}{S(2L+1)}$.  Here, $L$ is the number of repeated trainable blocks in the branch and trunk circuits, and $S$ refers to the number of non-identity components of the Hamiltonian of interest.  This theorem is developed with the  focus on the variational quantum Eigensolver (VQE) algorithm, so it may not be strictly applicable to general VQCs, where the quantity of interest is the expectation value of a single qubit rather than a larger product Hamiltonian. 
Here, the parameter $S$ is set to unity $S=1$, since the output of the VQCs is obtained through the expectation values on single qubits.
It was also found that naively initializing all variational parameters in the range $[-\pi/2, \pi/2]$ works well regardless of the size or depth of the circuit. However, the restricted ranges for parameter initialization performed generally better, which is consistent with the findings in \cite{kashif2024parameterinitialization}. There does not appear to be a consensus on the most effective parameter initialization strategy in QML, unlike classical machine learning, where the initialization of Glorot \cite{pmlr-v9-glorot10a} and He \cite{he2015delvingdeeprectifierssurpassing} are widely used.
Other parameter initialization and optimization strategies, including layer-wise learning \cite{skolik_layerwise_2021} and an identity block strategy \cite{Grant2019initialization} are not considered.

An investigation is also conducted into how the number of trainable blocks in the VQC and the number of qubits affect the error during training.
As discussed previously, VQCs  can be thought of as partial Fourier series, where circuits with more depth have access to more ``frequencies'' or Fourier coefficients.  As such, more layers are expected to enable the circuit to train to smaller approximation errors.  Similarly, more grid points (and thus more qubits) are expected to yield lower errors. However, with simulations involving larger numbers of qubits, computational requirements (both RAM and CPU time) become the limiting factor.  The maximum number of qubits considered here is 15. 

\subsection{Anti-Derivative} \label{sec:results_ad}
\begin{figure}
    \centering
    \includegraphics[width=0.5\textwidth]{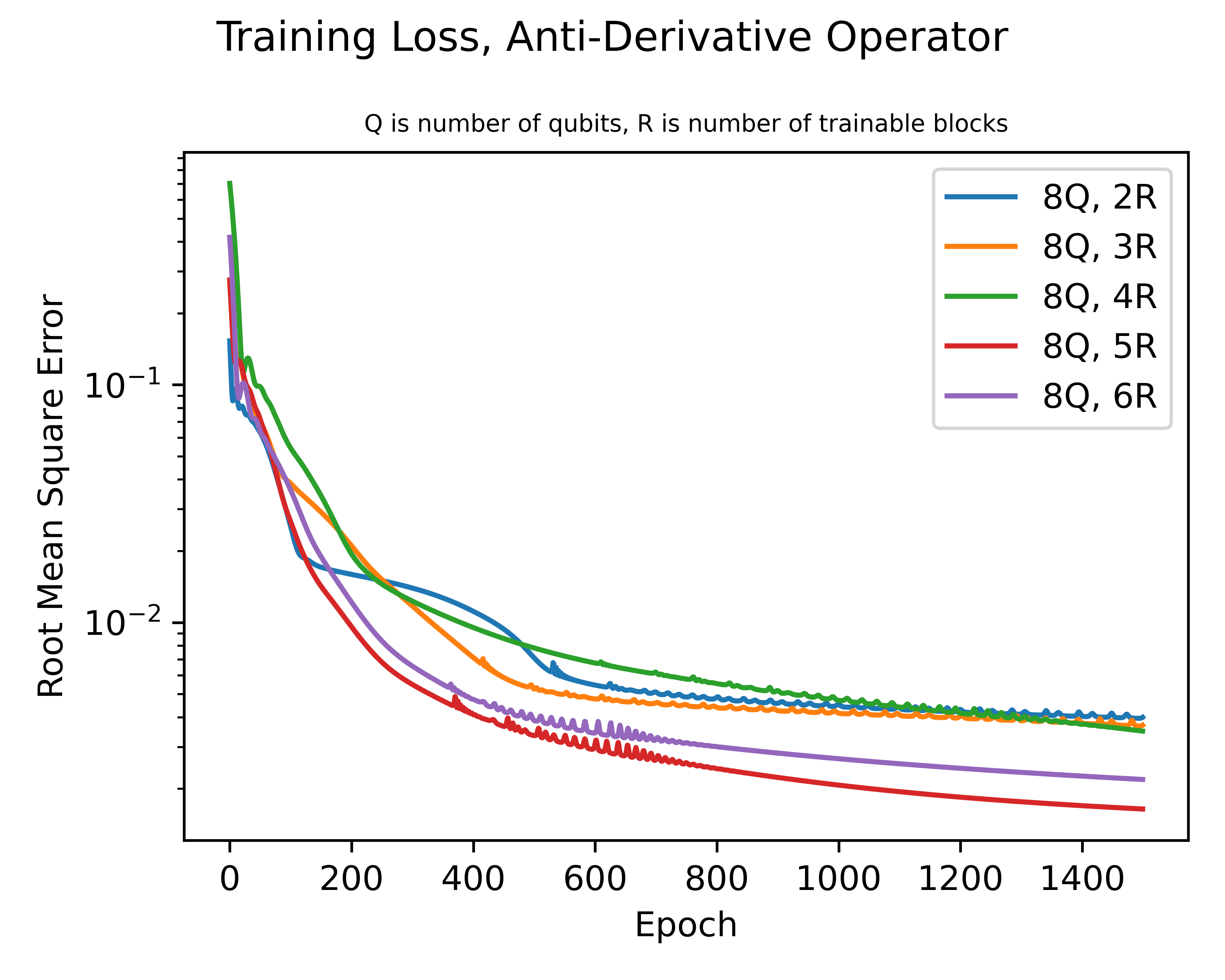}
    \includegraphics[width=0.5\textwidth]{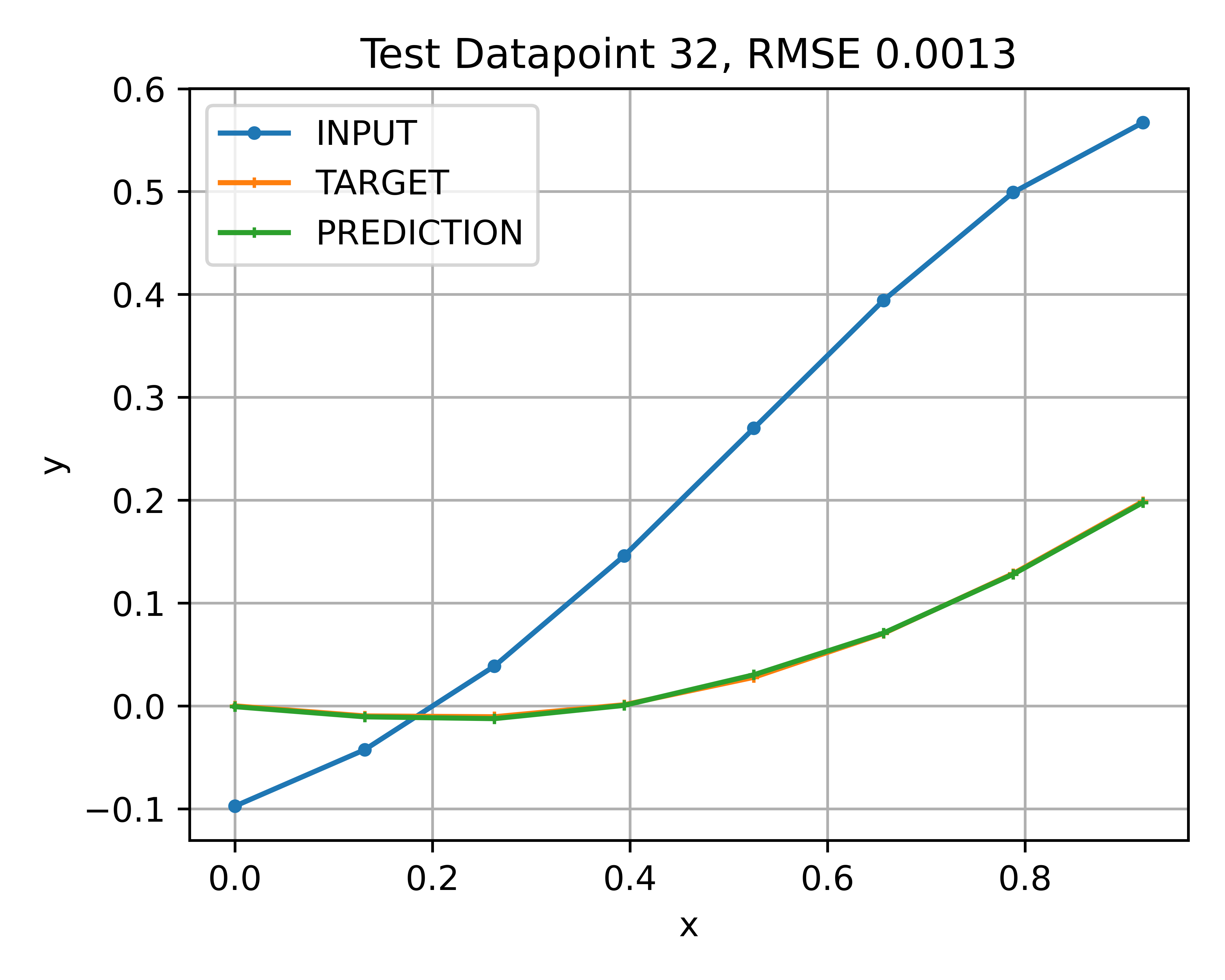}
    \caption{Training losses for the anti-derivative operator with varying circuit depths. A sample test case from the best performing model configuration is shown.}
    \label{fig:loss_vs_repeats_ad}
\end{figure}

The 1D anti-derivative is defined as:

\begin{equation} \label{eqn:ad}
\begin{split}
    \frac{dy}{dx} &= u(x), x\in [0,1]\\
    y(0) &= 0
\end{split}
\end{equation}
The dataset for this experiment is  obtained from a GitHub repository associated with ETH Zurich's Deep Learning in Scientific Computing course \cite{CAMLab2023Dataset}.
%and can be downloaded by running the following commands.
Copies of the data set are also available in the repository of this paper.  The data are down-sampled to eight grid points, and eight qubits are used in each of the branch and trunk circuits. To ensure reproducibility, the original 100-grid trajectories are down-sampled to eight grid points by retaining every 13th point starting from $x = 0$, with the final endpoint preserved; the sampling of training and test pairs is controlled by a fixed JAX-PRNG seed.
% \begin{widetext}
% \begin{verbatim}
% $ wget https://github.com/mroberto166/CAMLab-DLSCTutorials/
%                         raw/main/antiderivative_aligned_train.npz
% $ wget https://github.com/mroberto166/CAMLab-DLSCTutorials/
%                         raw/main/antiderivative_aligned_test.npz
% \end{verbatim}
% \end{widetext}
Training is performed for circuits with two to six trainable blocks. 
For data encoding blocks, $RZ$ gates are used, and the classical pre-processing parameters are not optimized and are left at their initial values. Training losses are shown in Figure \ref{fig:loss_vs_repeats_ad}, and summary statistics on the test data are presented in Table \ref{tab:test_rmse_ad}. 
By comparing training and test losses, it is observed that the model effectively learns the training data and generalizes well to unseen data.
It can also be inferred from the training history that the gradient landscape is well-behaved and smooth, and that barren plateaus are not encountered during training.
In addition, as the depth of the circuit increases, performance improves marginally with respect to training errors, while test errors remain approximately constant.
This suggests that the anti-derivative operator is simple enough that even shallow circuits (with two trainable blocks) achieve a near-optimal performance.
The increased standard deviation at higher depths likely reflects greater sensitivity to parameter initialization in more complex circuits.
% Additionally, as the anti-derivative is a relatively simple operator, increased circuit depth may not be necessary.
% This idea is supported by all circuit depths tested being able to achieve low training and test errors.

\begin{table*}[htbp]
    \caption{RMSE over Test Set, Anti-Derivative Operator}
    \label{tab:test_rmse_ad}
    \begin{ruledtabular}
    \begin{tabular}{cccc} 
     Number of Trainable Blocks & Mean RMSE & Median RMSE & Std. RMSE \\
     \hline
     2 & \num{3.51E-3} & \num{3.04E-3} & \num{2.06E-3} \\ 
     3 & \num{3.41E-3} & \num{2.89E-3} & \num{2.07E-3} \\ 
     4 & \num{3.70E-3} & \num{3.01E-3} & \num{3.15E-3} \\ 
     5 & \num{2.20E-3} & \num{1.50E-3} & \num{3.48E-3} \\ 
     6 & \num{3.43E-3} & \num{2.09E-3} & \num{5.39E-3} \\ 
    \end{tabular}
    \end{ruledtabular}
\end{table*}

\subsection{Diffusion} \label{sec:results_diffusion}
\begin{figure}
    \centering
    \includegraphics[width=0.5\textwidth]{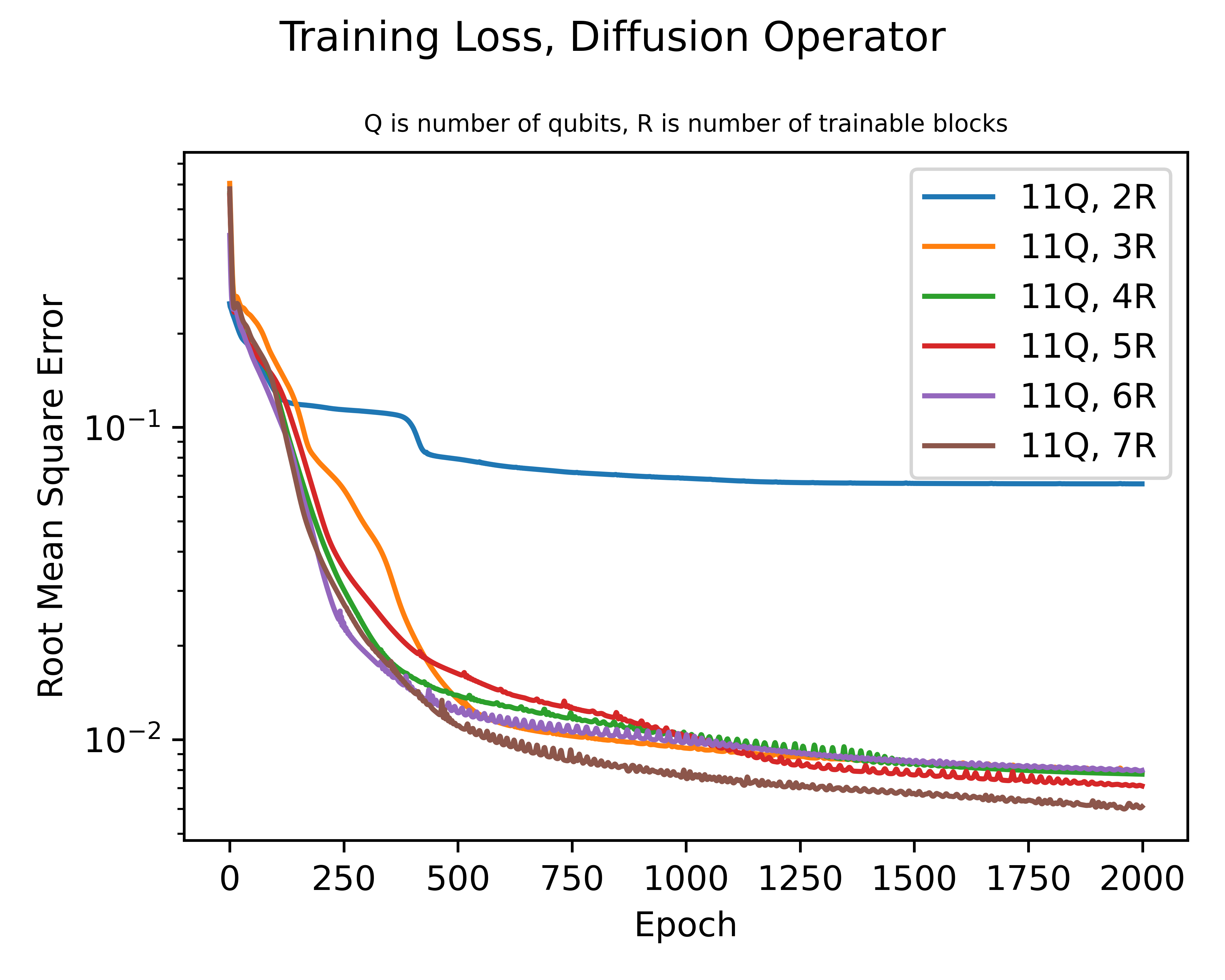}
    \includegraphics[width=0.5\textwidth]{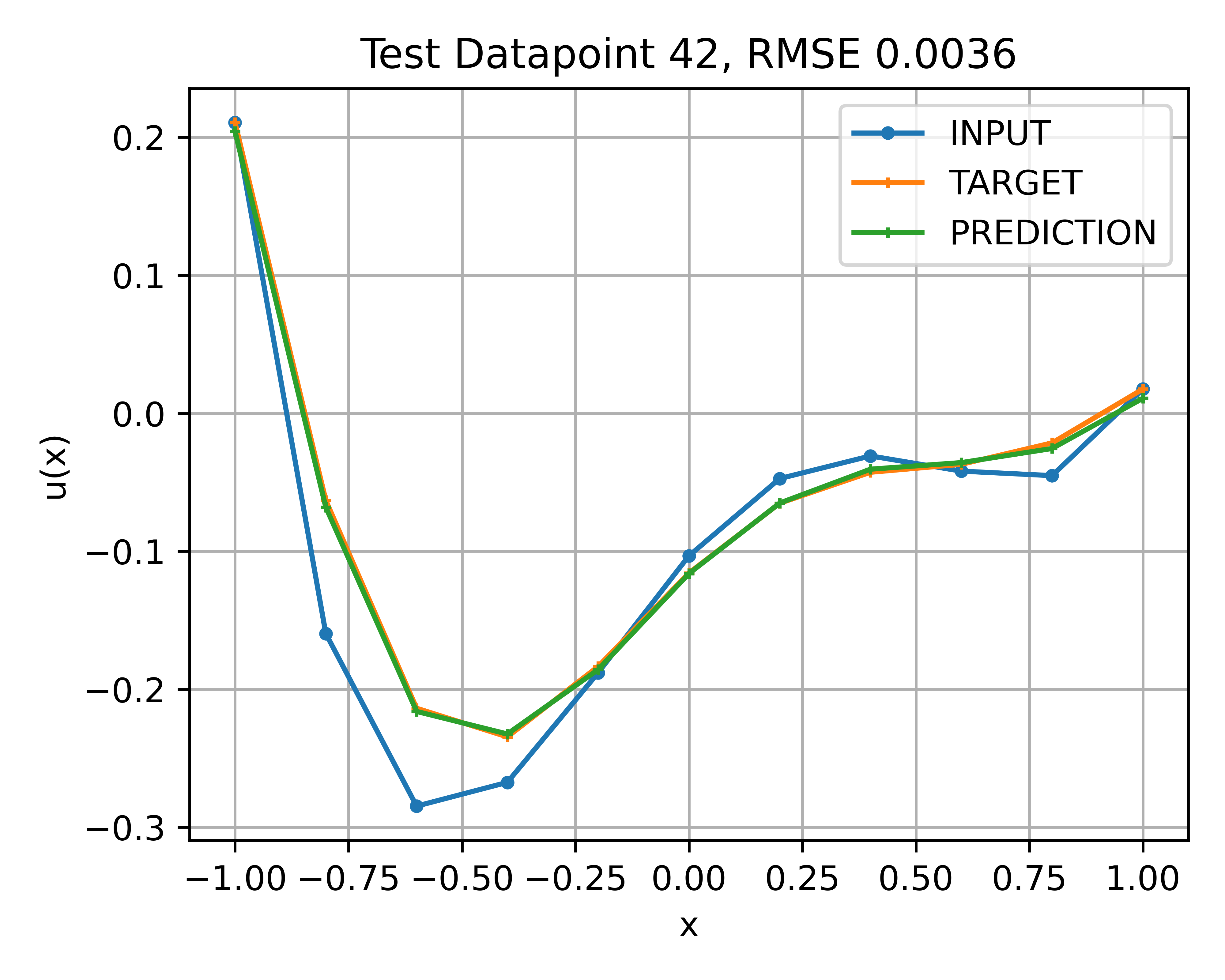}
    \caption{Training losses for the diffusion operator with varying circuit depths. A sample test case from the best performing model configuration is shown.}
    \label{fig:loss_vs_repeats_heat}
\end{figure}

The 1D diffusion equation is given by: 

\begin{equation} \label{eqn:heat}
\begin{split}
    \frac{\partial u(x,t)}{\partial t} &= D\frac{\partial^2 u(x,t)}{\partial x^2}\\
    x \in [-1,1]&,\  t \ge 0 \\
    u(-1, t) &= u(-1, 0)\\
    u(1, t) &= u(1, 0)\\
    u(x,0) &\in V_{Chebyshev}
\end{split}
\end{equation}
where the diffusion coefficient is set to a constant value of $D=0.01$.
The conditions $u(\pm 1, t) = u(\pm 1, 0)$ specify time-invariant Dirichlet boundaries fixed to the initial values at $x = \pm 1$ for all $t \geq 0$. A second-order central finite difference is used to discretize the spatial derivatives and an explicit first order scheme is used for time integration.
Dirichlet boundary conditions are assumed in $x$,  and the initial conditions are sampled from Chebyshev polynomials as described in Section \ref{sec:DataGeneration}. The circuit configuration is identical to the one evaluated for the antiderivative operator, with  11 qubits and two to seven trainable blocks. 
Training losses are shown in Figure \ref{fig:loss_vs_repeats_heat}, and summary statistics on the test data are presented in Table \ref{tab:test_rmse_heat}. 
It is observed that the model effectively learns the training data, generalizes to unseen data well, and behaves well during training. In addition, performance improves as the circuit depth increases.
% \textcolor{blue}{However, in this case, the trial where the branch and trunk circuits each have two trainable blocks stalls during training, as evidenced by the plateau in training loss.}
Note that the training loss for the shallowest circuit configuration appears to stall during training.
This does not appear to be an indication of a barren plateau as observed in \cite{mcclean_barren_2018}, but rather that two trainable blocks may not create sufficiently expressive circuits to learn the diffusion operator. 
Conversely, the trial where the branch and trunk circuits each have four trainable blocks produces similar results to the trial with 7 trainable blocks.
This is likely a consequence of the specific parameter initialization for those trials, demonstrating the importance of parameter initialization and how it can have large effects on results.

\begin{table*}[htbp]
    \caption{RMSE over Test Set, Diffusion Operator}
    \label{tab:test_rmse_heat}
    \begin{ruledtabular}
    \begin{tabular}{cccc} 
     Number of Trainable Blocks & Mean RMSE & Median RMSE & Std. RMSE \\
     \hline
     2 & \num{6.06E-2} & \num{5.43E-2} & \num{3.01E-2} \\ 
     3 & \num{6.73E-3} & \num{5.60E-3} & \num{4.10E-3} \\ 
     4 & \num{7.09E-3} & \num{6.00E-3} & \num{4.43E-3} \\ 
     5 & \num{6.36E-3} & \num{5.03E-3} & \num{4.30E-3} \\ 
     6 & \num{7.40E-3} & \num{5.87E-3} & \num{5.16E-3} \\ 
     7 & \num{6.15E-3} & \num{5.05E-3} & \num{4.16E-3} \\ 
    \end{tabular}
    \end{ruledtabular}
\end{table*}

\subsection{Reaction-Diffusion}
\begin{figure}
    \centering
    \includegraphics[width=0.5\textwidth]{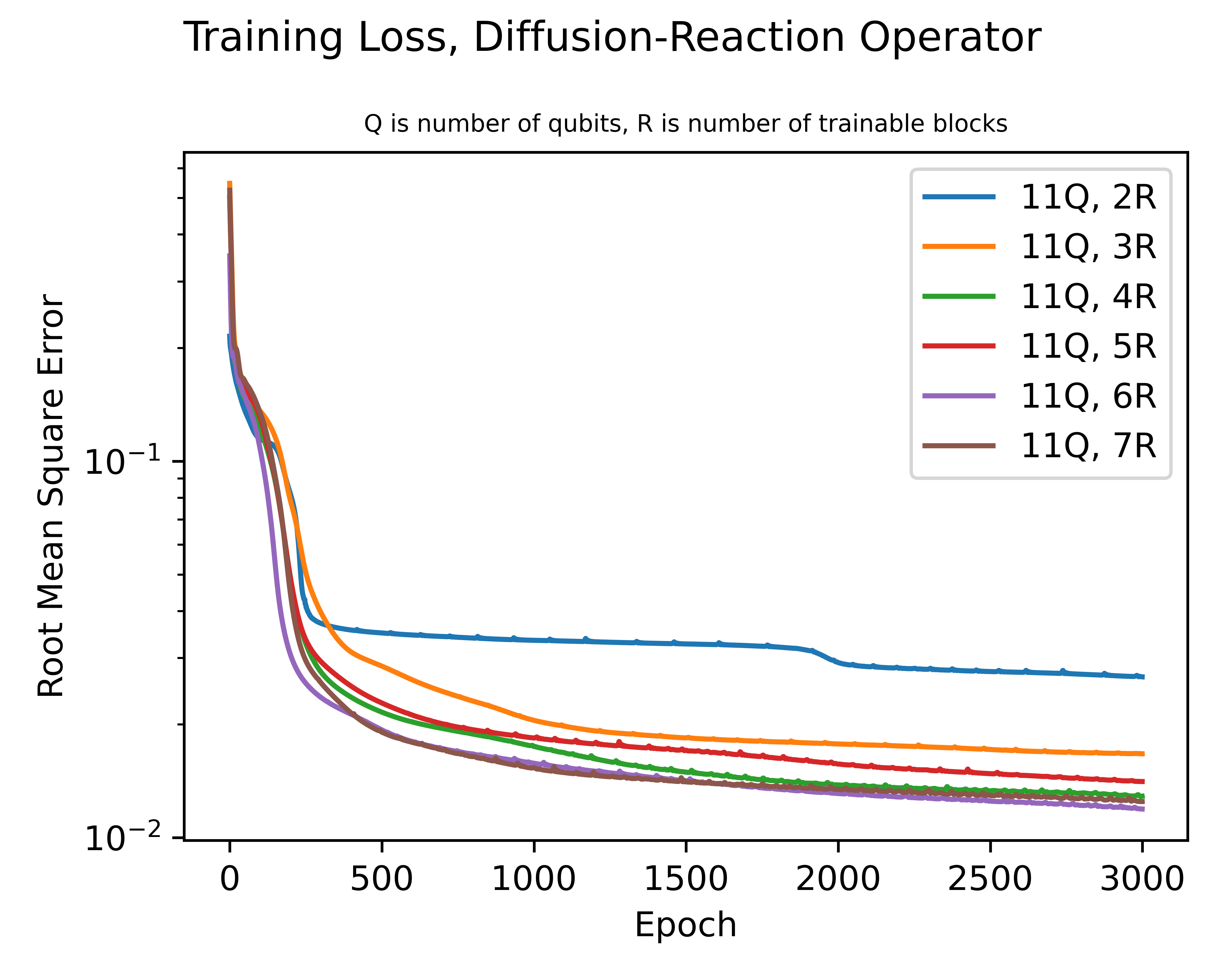}
    \includegraphics[width=0.5\textwidth]{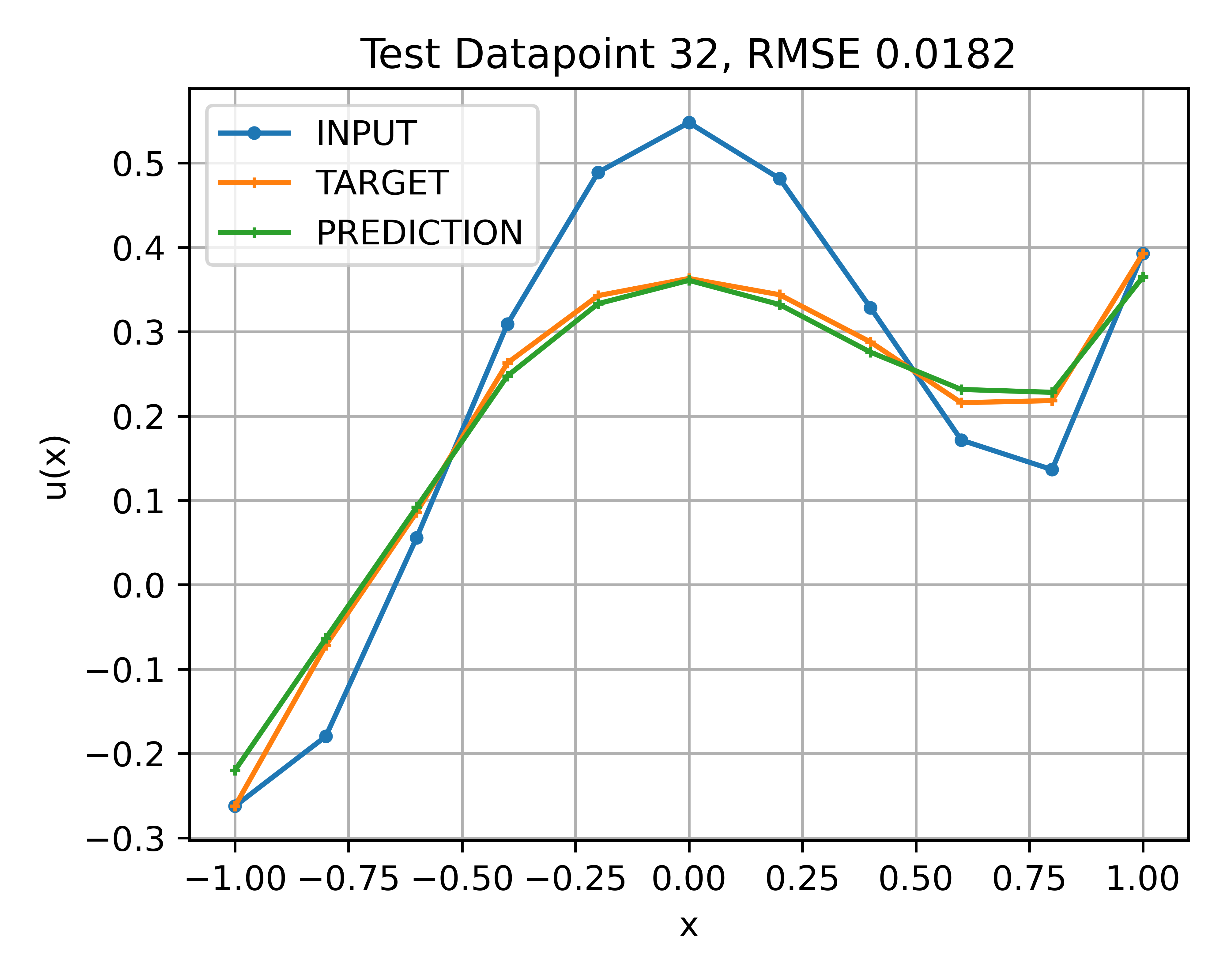}
    \caption{Training losses for the reaction-diffusion operator with varying circuit depths. A sample test case from the best performing model configuration is shown.}
    \label{fig:loss_vs_repeats_dr}
\end{figure}

The 1D reaction-diffusion equation is presented by:
\begin{equation} \label{eqn:dr}
\begin{split}
    \frac{\partial u(x,t)}{\partial t} &= D\frac{\partial^2 u(x,t)}{\partial x^2} + R|u(x,t)|(1-u(x,t))\\
    u(-1, t) &= u(-1, 0)\\
    u(1, t) &= u(1, 0)\\
    u(x,0) &\in V_{Chebyshev}
\end{split}
\end{equation}
with $R=0.5$ and $D=0.01$ chosen arbitrarily.  The discretizations and initializations are the same as those implemented for the diffusion equation.  An absolute value is added to the reaction term to ensure a positive reaction rate.  
%Without this, the reaction term will cause the negative values of $u(x,t)$ to become increasingly more negative.  With this modification, the equation no longer reduces to the Fisher--KPP form, and the learned operator may differ qualitatively from the standard reaction-diffusion equation.  %While this loses some physical meaning, as concentrations cannot be negative, it allows initial conditions to be freely drawn from the Chebyshev polynomial space without restrictions.
% To compare the effect of the non-linear reaction term, the same initial conditions are used as when training the diffusion operator.
Training is done for circuits with 11 qubits and two to seven  trainable blocks. 
For this experiment, the classical preprocessing parameters for the data encoding blocks in the branch circuit are unfrozen and optimized. 
The same set of preprocessing parameters is shared across all data encoding blocks.
 Training losses are shown in Figure \ref{fig:loss_vs_repeats_dr} and summary statistics on the test set are given in Table \ref{tab:test_rmse_dr}.
As the number of trainable blocks increases, the average error over the test set tends to decrease more significantly than  that  for the diffusion operator.
Since the diffusion-reaction operator is more complex than just diffusion, the circuit depth contributes to reducing both the approximation and the generalization error.
Similarly to what was seen in Section \ref{sec:results_diffusion}, the training loss exhibits plateaus for the circuits with only two trainable blocks, indicating that two trainable blocks are not expressive enough to adequately represent the operator.

\begin{table*}[htbp]
    \caption{RMSE over Test Set, Reaction-Diffusion Operator}
    \label{tab:test_rmse_dr}
    \begin{ruledtabular}
    \begin{tabular}{cccc} 
     Number of Trainable Blocks & Mean RMSE & Median RMSE & Std. RMSE \\
     \hline
     2 & \num{2.55E-2} & \num{2.41E-2} & \num{7.97E-3} \\ 
     3 & \num{1.63E-2} & \num{1.53E-2} & \num{5.23E-3} \\ 
     4 & \num{1.27E-2} & \num{1.24E-2} & \num{3.41E-3} \\ 
     5 & \num{1.44E-2} & \num{1.39E-2} & \num{4.29E-3} \\ 
     6 & \num{1.22E-2} & \num{1.18E-2} & \num{3.80E-3} \\ 
     7 & \num{1.29E-2} & \num{1.23E-2} & \num{4.20E-3} \\ 
    \end{tabular}
    \end{ruledtabular}
\end{table*}

\subsection{Burgers' Equation}

The Burgers equation (1-D non-linear convection-diffusion) is given by:
\begin{equation} \label{eqn:burgers}
\begin{split}
    \frac{\partial u(x,t)}{\partial t} + u(x,t) \frac{\partial u(x,t)}{\partial x} &= \frac{1}{\text{Re}} \frac{\partial^2 u(x,t)}{\partial x^2}\\
    u(0,t) &= u(1,t)\\
    u(x,0) &\in V_{Cosine}
\end{split}
\end{equation}

\begin{figure}
    \centering
    \includegraphics[width=0.5\textwidth]{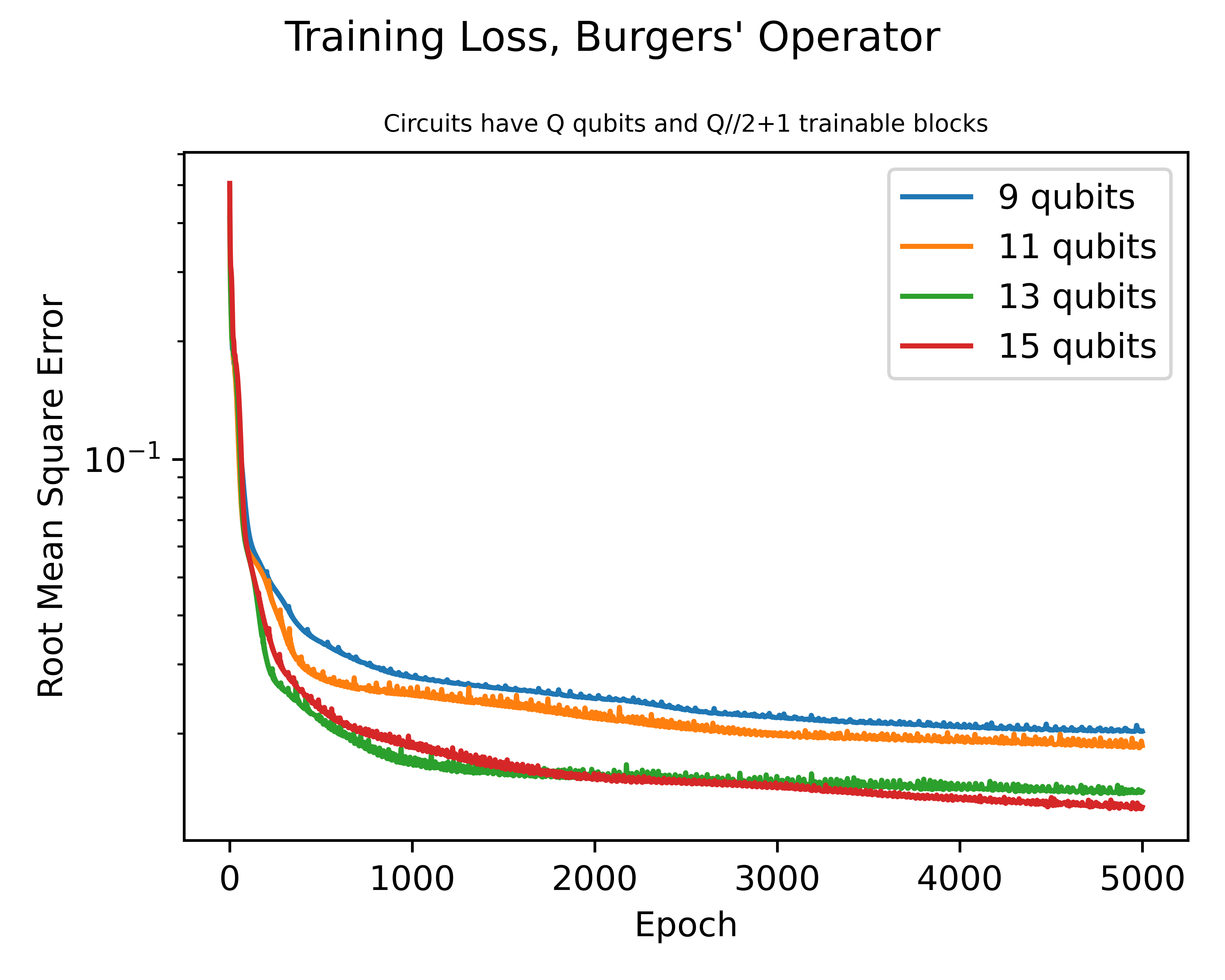}
    \includegraphics[width=0.5\textwidth]{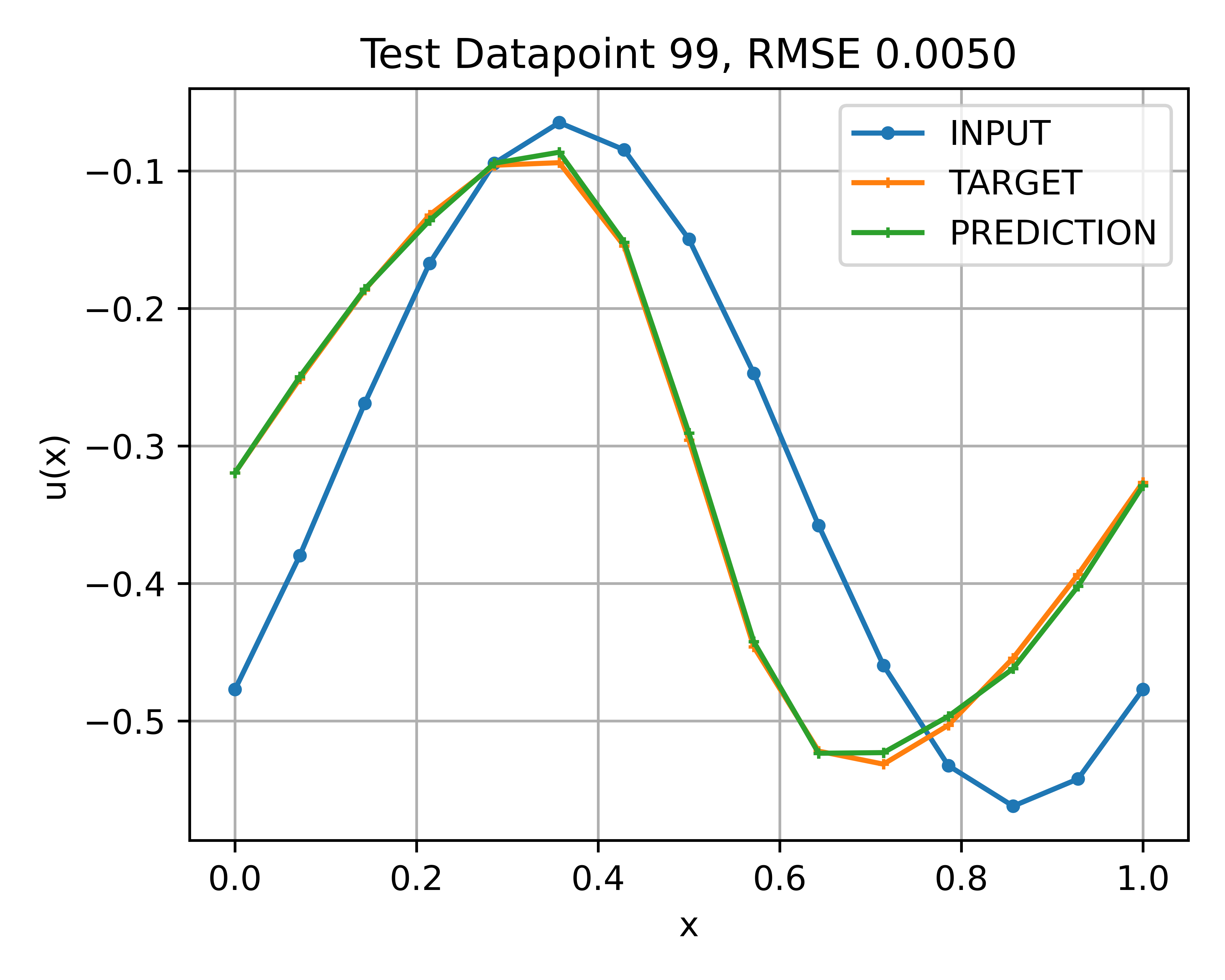}
    \caption{Training losses for the non-linear convection-diffusion operator with varying numbers of qubits and circuit depths. A sample test case from the best performing model configuration is shown.}
    \label{fig:loss_vs_repeats_burgers}
\end{figure}

% \textcolor{blue}{Here, $\nu=0.005$ and $c=0.7$. It should be noted that, while the canonical Burgers' equation corresponds to $c=1$, the scaled convection term used here does not alter the essential physics present in the equation, as both nonlinear convection and linear diffusion terms are still present. This scaling term is functionally equivalent to simulation at a different effective timestep and Reynolds number.}
with $\text{Re}=140$ set arbitrariy.
The discretisation scheme is the same as that for the diffusion reaction equation, except that a first-order upwind scheme is employed for the convection term.
Datasets are generated for 81, 101, 121, and 141 grid points, using the pseudo-random number generator feature in JAX to sample the same initial conditions.
Downsampling by taking every tenth grid point results in datasets with 9, 11, 13, and 15 grid points, which correspond to the number of qubits evaluated.
% \textcolor{blue}{At this point, it should be noted that the slight (essentially negligible)  differences in discretization when generating datasets introduce a different amount of numerical diffusion into each dataset, caused by the first-order upwind scheme.} % removing this sentence seems to be easier
Training is conducted for circuits with 9, 11, 13, and 15 qubits.
The number of trainable blocks is set equal to one more than half the number of qubits, rounded down to the nearest integer.
The classical preprocessing parameters for the data encoding blocks in the branch circuit are unfrozen and optimized.
Each data encoding block in the branch circuit now has its own unique set of preprocessing parameters.
 Training losses are shown in Figure \ref{fig:loss_vs_repeats_burgers} and summary statistics on the test set are shown in Table \ref{tab:test_rmse_burgers}.
As both the number of qubits and the number of trainable blocks increase, the training loss and the average error over the test set decrease.
This confirms that both the model's complexity (quantified by the number of trainable blocks) and resolution (quantified by the number of qubits, or discretization of input and target functions) are key factors in minimizing the approximation and generalization errors.

\begin{table*}[htbp]
    \caption{RMSE over Test Set, Burgers' Operator}  % Non-Linear Convection-Diffusion
    \label{tab:test_rmse_burgers}
    \begin{ruledtabular}
    \begin{tabular}{cccc} 
     Number of Qubits & Mean RMSE & Median RMSE & Std. RMSE \\ %[0.5ex] 
     \hline
     9 & \num{1.59E-2} & \num{9.24E-3} & \num{1.54E-2} \\
     11 & \num{1.51E-2} & \num{9.92E-3} & \num{1.28E-2} \\
     13 & \num{1.21E-2} & \num{7.08E-3} & \num{1.04E-2} \\
     15 & \num{1.19E-2} & \num{7.63E-3} & \num{9.99E-3} \\
    \end{tabular}
    \end{ruledtabular}
\end{table*}

% \begin{table}%[H] add [H] placement to break table across pages
% \caption{\label{}
% \begin{ruledtabular}
% \begin{tabular}{cccc}
% Lines of & table here & ending & with \\
% \end{tabular}
% \end{ruledtabular}
% \end{table}

\subsection{Comparison with DeepONet} \label{sec:compare_deeponet}

An attempt is made to make a comparison between the QONet and the classical DeepONet.  This is done here with the following considerations:
First, the baseline does not attempt to tune hyperparameters used when training the DeepONet (such as the learning rate). Second, the QONet incorporates a learnable classical preprocessing layer ($\mathbf{A}$, $\mathbf{b}$) that is not present in the baseline DeepONet, as noted in Section~\ref{sec:data_encoding}. Third, VQCs and neural networks are not completely analogous, but for simplicity, they are treated as such. Comparisons are made under matched experimental conditions. Both models are trained using the same datasets,  learning rates,  number of epochs,  and number of layers. The number of hidden layers in the branch and trunk neural networks of the DeepONet is varied, similarly to how the depth of the branch and trunk circuits is varied for the QONet. The width of the hidden layers, as well as the output layers, is set to be the same as the input layer.

\begin{figure*}[p]
    \centering
    \includegraphics[width=1.0\textwidth]{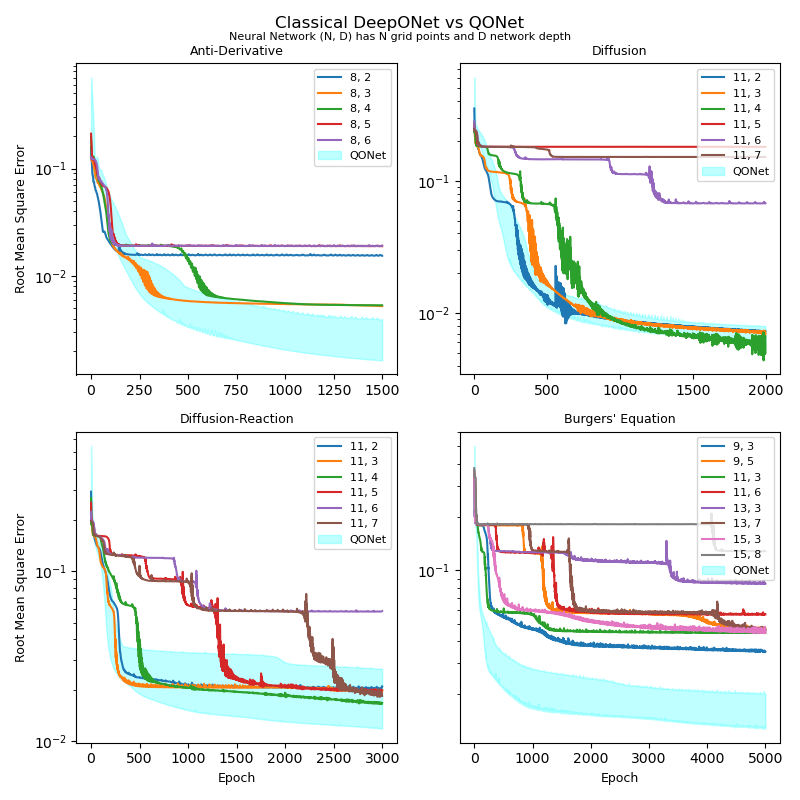}
    \caption{A comparison of the DeepONet with the QONet under matched experimental conditions.
    The numbered pairs $N, D$ on the legend are the configurations of the DeepONet.
    The branch and trunk neural networks each have a width of $N$, which corresponds to the discretization of the input function, and a depth of $D$.
    All QONet training losses shown in Section \ref{sec:results} are overlaid as the shaded blue region, except for the QONet configuration with only two trainable blocks for the diffusion operator.
    While not an exhaustive comparison, it is evident that the DeepONet can exhibit plateaued training losses (implying vanishing gradients or saddle points) and oscillations (indicating convergence difficulties).
    By contrast, the QONet smoothly and consistently achieves a similar or lower loss during training.}
    \label{fig:classical_deeponet_comparison}
\end{figure*}

In Figure \ref{fig:classical_deeponet_comparison},  the training losses for the QONet are identified by a region shaded with light blue.
From the training histories, it is clear that the classical DeepONets are ill-behaved during training, often encountering large plateaus and showing oscillations (which may be indicative of instability).
The wide variety in training behavior also suggests that the DeepONets are extremely sensitive to parameter initialization.
In contrast, the region containing all the training losses from the QONet indicates fast and smooth convergence in training.
The observed contrast in training behavior is consistent with, but does not necessarily constitute a proof of a more favorable optimization landscape for the QONet.   A more rigorous assessment  requires the measurement of $\mathrm{Var}_{\boldsymbol{\theta} \sim p_0}[\partial \mathcal{L}/\partial \boldsymbol{\theta}]$ with  random initializations as the qubit count increases.  

The results generated here highlight the potential of the QONet, as it exhibits fast and stable convergence even on extremely coarse discretizations, with near-monotonic training loss curves.
This suggests an implicit regularization effect inherent to the model architecture, which may act to reduce training data requirements.
Additionally, in all of the experimental trials, the QONet achieved errors similar to (and sometimes lower than) the classical DeepONet while using fewer trainable parameters. The takeaway is that, under matched experimental conditions, the QONet appears to demonstrate more stable training dynamics than a comparably sized classical DeepONet.  In particular, the QONet seems to avoid plateau behavior (Section \ref{sec:absence_of_barran_plateaus}) as observed in its classical counterpart, suggesting potential advantages in its optimization landscape.  This suggests an improved convergence rate as shown in Section \ref{sec:error_bounds}, which warrants future use of the model.  

The parameter count of the DeepONet is simply the sum of the branch and trunk network parameter counts.
As the neural networks are assumed to have a constant width throughout, the branch network has $R(N^2+N)$ parameters, where $N$ is the number of grid points discretizing the input functions.
The trunk network has $(R-1)(N^2+N) + 2N$, where $R$ denotes the number of hidden layers (including the output layer).
The total number of trainable parameters in the DeepONet is then:
\begin{equation} \label{eqn:don_dof}
    \text{DOF}_{\text{DeepONet}} = (N^2+N)R + (N^2+N)(R-1) + 2N
\end{equation}
By comparing Equations \ref{eqn:qonet_dof} and \ref{eqn:don_dof}, it is observed that the QONet has the potential for its parameter count to scale linearly with either depth or the number of qubits (width).
The DeepONet scales quadratically with respect to the number of input points and can only scale linearly with respect to the depth of the network.
This advantage in parameter scaling is most significant for scenarios where either no preprocessing parameters are used, or only one shared set is used.
As the number of unique sets of preprocessing parameters increases, the parameter count of the QONet increases towards, but still remains lower than that of a comparable classical DeepONet.
The linear parameter scaling $P = O(N \cdot L)$ exhibits  a particular  structural advantage of the  QONet.
% The linear parameter scaling ($P = O(N \cdot L)$) as described in Section \ref{sec:scaling} is a key structural advantage inherent to the QONet.
As described in the Supplemental Material \cite{supplement_error_bounds}, with  Assumption~\ref{ass:low-intrinsic}, the quantum branch approximation error scales existentially as $\delta_b = O(P_b^{-r_b/d_1})$ in the Sobolev regime or $\delta_b = O(e^{-\alpha P_b})$ in the analytic regime, where $d_1$ is the \emph{intrinsic} dimension of the input function space.
(The coefficient functionals to be learned by the branch circuit must be Sobolev regular or analytic.) % Sobolev regular for the polynomial regime, and analytic for the analytic regime.
The classical DeepONet branch network with the same parameter budget $P_b$ attains $\delta_b^c = O(P_b^{-2 r_b / m_b^c})$ under Yarotsky's worst-case Sobolev rate~\cite{YAROTSKY2017}, where $m_b^c$ denotes the number of sensor locations. Without Assumption~\ref{ass:low-intrinsic}, the relevant quantum rate is also expressed in terms of $m_b$, and the simple curse-of-dimensionality argument no longer favors the quantum architecture. When Assumption~\ref{ass:low-intrinsic} holds and the classical baseline is taken to be the Yarotsky worst-case rate (so that the classical network does not exploit the same low-intrinsic-dimensionality structure), the architecture comparison favors the QONet. Precise conditions and resource-complexity analysis are given in Theorems~S14--S15 and~S17 of the Supplemental Material.

Figure \ref{fig:error_comparison} illustrates the theoretical error limits of Equation \ref{eqn:total_error_bound} and their classical counterparts (under the Yarotsky worst-case baseline) in six key parameters, using representative values $s=2$, $r_b=2$, $r_t=3$, $m_b^c=8$, and $d_1 = d_2 = 1$ (which are in agreement with the experimental setup employed). \textcolor{black}{The classical worst-case Sobolev rate of $O(P^{-2r/d})$ is used as the reference and is conservative for analytic targets, for which deep tanh and ReLU networks are known to achieve exponential rates~\cite{Mhaskar1996,Schwab2019}.  }Three architectures are compared: the QONet in the analytic regime (exponential convergence), the QONet in the Sobolev regime (polynomial convergence), and the classical DeepONet. The orange markers denote breaking points where the quantum architecture transitions from higher to lower error. The most notable feature is panel (c), which shows the curse of dimensionality: as the number of sensor locations $m_b^c$ increases, the classical error rises monotonically while the quantum error remains constant under Assumption~\ref{ass:low-intrinsic}, since the QONet's approximation rate depends on the intrinsic dimension $d_1$ rather than the discretization resolution.
In the Sobolev polynomial regime, the classical trunk rate $2 r_t / d_2$ is faster than the quantum trunk rate $r_t / d_2$ by a factor of two. However, the total quantum error remains lower in panels (a) and (b) because, under Assumption~\ref{ass:low-intrinsic}, the branch advantage ($r_b/d_1$ vs.\ $2 r_b/m_b^c$, which favors the quantum case when $m_b^c > 2 d_1$) more than compensates for the trunk disadvantage. %The conclusions drawn from these panels are therefore conditional on Assumption~\ref{ass:low-intrinsic}.
For  1D operators ($d_1 = 1$), the quantum advantage condition $m_b^c > 2d_1 = 2$ is satisfied for all experiments whenever Assumption~\ref{ass:low-intrinsic} holds. Detailed individual analysis for each parameter, as well as two-parameter 3D surface plots, are provided in the Supplemental Material \cite{supplement_error_bounds}.

\textcolor{black}{Although the comparisons conducted  in this section favors the QONet, it must be acknowledged that there are some shortcomings in performing direct comparisons in this manner.  VQCs and neural networks are both universal function approximators, but they accomplish this through different means. As such, varying the number of hidden layers, and in turn the parameter count, in the DeepONet may not have the same effect on the expressive power of the branch and trunk as varying the circuit depth does for the QONet.
The same learning rates and number of epochs are used for consistency, but in \cite{lockwood2022empiricalreviewoptimizationtechniques} it is shown that VQCs tolerate higher learning rates than neural networks. It is possible that lower learning rates would produce better results for the DeepONet, and even higher learning rates would further improve the results for the QONet.
Finally, the input functions are downsampled to an extremely coarse discretization of the input functions, which is necessary due to the limitations in simulating quantum circuits. This is discussed further in Appendix \ref{apdx:training-information}.
A side effect of this downsampling is that high-frequency information is lost, which may have a disproportionately adverse effect on the performance of the DeepONet relative to the QONet.
However, if this is the case, it highlights the ability of the QONet to perform well even with only low-resolution data.}

\begin{figure*}[p]
    \centering
    \includegraphics[width=1.0\textwidth]{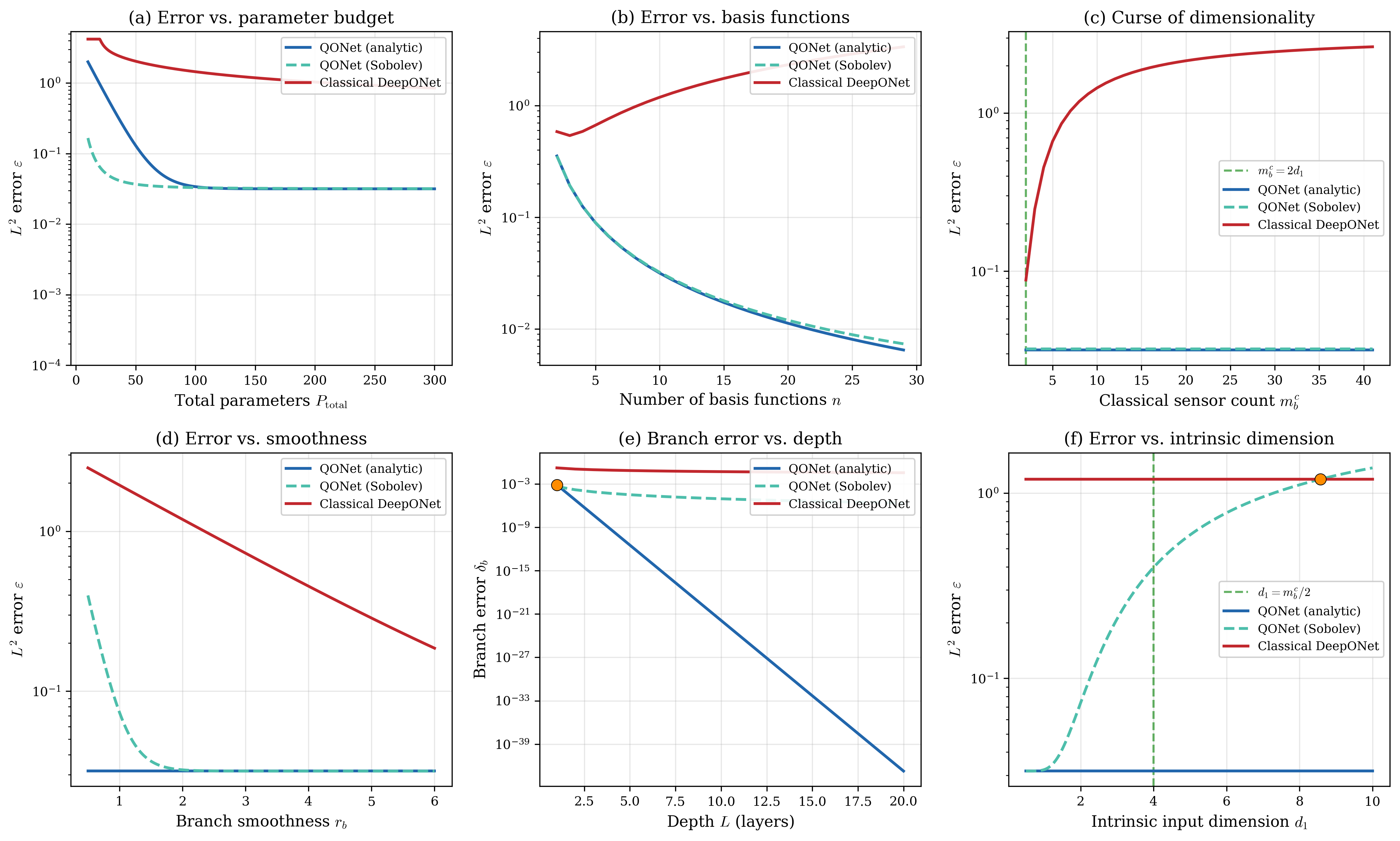}
    \caption{Theoretical $L^2$ error comparison between the QONet and classical DeepONet based on the error bounds derived in Section \ref{sec:error_bounds} and the Supplemental Material. (a) Error vs. total parameter budget. (b) Error vs. number of basis functions. (c) Curse of dimensionality: quantum error is independent of sensor count $m_b^c$ under Assumption~\ref{ass:low-intrinsic}. (d) Error vs. branch smoothness. (e) Branch approximation error vs. circuit/network depth. (f) Error vs. intrinsic input dimension. Breaking points (orange circles) mark where quantum error first becomes lower than classical.
    % Green-shaded regions denote quantum advantage.
    }
    \label{fig:error_comparison}
\end{figure*}

\section{Further Discussions}

\subsection{Role of Classical Preprocessing Parameters} \label{sec:role_of_preprocessing}

The classical preprocessing parameters ($\mathbf{A}$, $\mathbf{b}$) in the branch data encoding blocks are  treated differently in the four experiments, indicating the increasing complexity of the operators being learned.
For the linear anti-derivative and diffusion operators, the classical preprocessing parameters are not optimized and are frozen at their initial values ($\mathbf{A} = -\mathbf{I}$, $\mathbf{b} = \pi/2$).
For the reaction-diffusion operator, the parameters are optimized, with a single shared set of preprocessing parameters across all data encoding blocks. For the  Burgers' equation, each data encoding block is  given its own unique set of preprocessing parameters.
This progression suggests that linear operators can be learned without optimizing the classical preprocessing, whereas nonlinear operators benefit from (or require) adaptive preprocessing. The physical interpretation is that the affine transformation $\mathbf{v} = \mathbf{A}\cdot\mathbf{u} + \mathbf{b}$ allows the circuit to learn an optimal embedding of the input function into rotation angles, which becomes increasingly important as the operator nonlinearity increases. The choice between shared and unique preprocessing parameters per layer represents a bias--variance tradeoff: shared parameters reduce the risk of overfitting but may limit expressivity, while unique parameters increase the model's capacity at the cost of additional trainable parameters (scaling as $O(L \cdot n^2)$ rather than $O(n^2)$). A systematic investigation of this tradeoff is recommended.

\subsection{Absence of Barren Plateaus} \label{sec:absence_of_barran_plateaus}

As noted in Section \ref{sec:introduction}, barren plateaus impose  a significant challenge for variational quantum algorithms, where the variance of the cost function gradient vanishes exponentially with increasing circuit size \cite{mcclean_barren_2018}. In all experiments conducted in this work, the training losses exhibited smooth, monotonically decreasing behavior without extended plateaus (with the exception of deliberately undersized two-block configurations).  Several factors may contribute to this favorable trainability. 
First, the circuits used here are relatively shallow, with at most eight trainable blocks.
This is well below the $O(n)$ depth threshold at which barren plateaus are known to become exponentially likely for random circuits \cite{mcclean_barren_2018, Brand_o_2016}.  % I added Brandao's 2016 paper as citation, but these are both covered by McClean 2018 which is cited earlier -JW
Second, the parameter initialization strategy from \cite{Wang_2024_trainabilityenhancement} restricts the initial parameter range, concentrating the circuit near the identity and thus ensuring non-vanishing initial gradients. Third, and perhaps the most important, the cost function structure uses local observables in the form of Pauli-$Z$ expectation values on individual qubits rather than a global observable involving the entire quantum state. Local cost functions have been shown to mitigate barren plateaus compared to global cost functions \cite{Cerezo_2021_VQA}.  %Collectively, these factors provide a plausible explanation for the smooth training dynamics observed in all experiments.

Within the regime tested, no barren plateaus were observed. Training was smooth and the loss decreased monotonically for all circuits up to $n=15$ qubits and $L=7$ trainable blocks using the initialization of~\cite{Wang_2024_trainabilityenhancement} (with the exception of deliberately undersized, very shallow circuits).
However, this empirical observation does not guarantee the absence of barren plateaus on larger scales~\cite{Cerezo_2021_VQA,Larocca2024}. A direct test would measure the variance of the loss gradient, $\operatorname{Var}_{\theta \sim p_0}\left[\partial \mathcal{L} / \partial \theta_i\right]$, across random initializations as $n$ and $L$ grow. An exponential decay of this variance $\left(\sim 2^{-c n}\right)$ would indicate a barren plateau, while a polynomial decay would indicate a trainable circuit. This measurement is left for future work.
For deeper or wider circuits, standard mitigation strategies can be applied, such as layerwise learning~\cite{skolik_layerwise_2021} or identity-block initialization~\cite{Grant2019initialization}.

\subsection{Scaling \& Practical Considerations} \label{sec:scaling}

\begin{figure*}[p]
    \centering
    \includegraphics[width=1.0\textwidth]{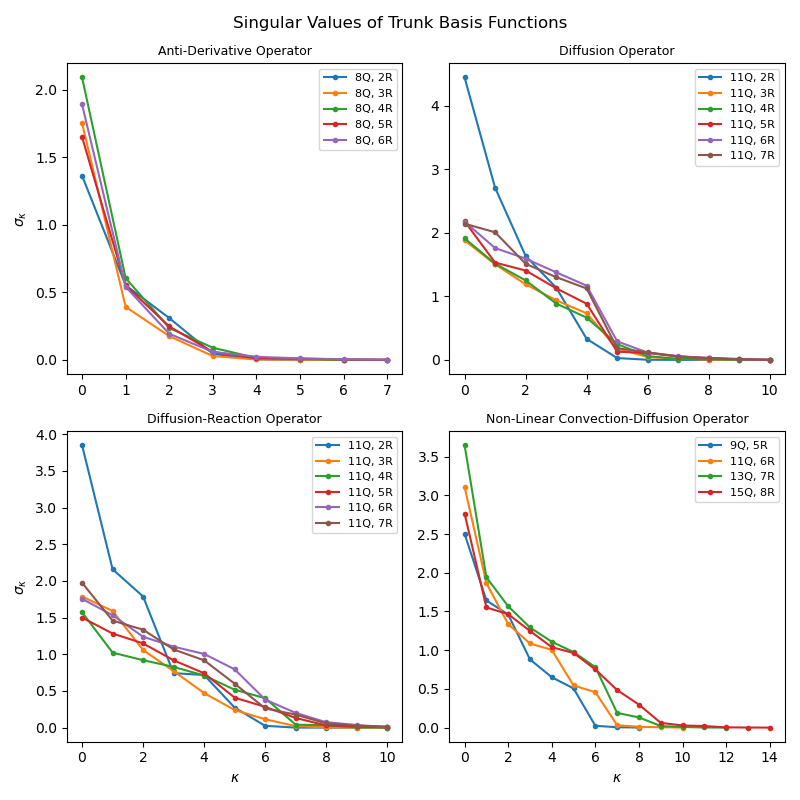}
    \caption{The singular values of the basis functions learned by the trunk circuit of the QONet for each trial configuration and operator. The decay in the singular values suggests that basis functions can be ranked by their contribution to the function space spanned by the output of the trunk circuit. The decay rate can provide insight into how descriptive the basis learned by the trunk is. Generally, increasing circuit depth (and number of qubits) appears to decrease the decay rate, indicating a more descriptive basis.}
    \label{fig:trunkbasissingularvalues}
\end{figure*}

When evaluating the performance of DeepONet-based architectures, it may be of interest to quantify how well the basis functions learned by the trunk circuit represent the solution space.
This can be investigated by considering the singular values of these functions, similar to the work in \cite{machine-learning-based-spectral-methods_2023}.
These singular values are shown in Figure \ref{fig:trunkbasissingularvalues} and are calculated from the $n \times N_y$ trunk-output matrix $[t_k(y_j)]_{k,j}$ evaluated on the test-set query grid. With $n$ qubits, there are at most $\min(n, N_y)$ nonzero singular values, and the curves  terminate at $\min(n, N_y)$.
Slower decaying values generally indicate a more descriptive set of functions with less redundancy and overlap.  It can be observed that as the number of trainable blocks per circuit increases for the anti-derivative, diffusion, and diffusion-reaction operators, the singular values tend to decay more slowly.
In particular, deliberately undersized models (with only two trainable blocks) for the diffusion and diffusion-reaction operators have singular values that decay faster than those for the larger circuits.  This demonstrates the effect of the increasing model complexity, showing that the trunk circuit can learn a more expressive set of basis functions as both the number of trainable blocks and the number of qubits increase. 

 Although Figure \ref{fig:trunkbasissingularvalues} demonstrates how the expressiveness of the QONet changes with increasing model size, the complexity order is also useful to understand how the model scales beyond the range of qubits and circuit depths.  
As discussed in Section \ref{sec:compare_deeponet}, the structure of the QONet allows the parameter count to scale linearly, significantly reducing the parameter count relative to the classical DeepONet when classical preprocessing is not applied to the branch inputs.
This benefit is problem-dependent and is especially notable when the operator being learned is linear.
If the classical preprocessing parameters in each data encoding block are unfrozen and optimized, the total parameter count then scales quadratically.
The preprocessing parameters are applied by a matrix multiplication operation that entails a cubic time complexity.
On the surface, this appears to contradict the parameter scaling advantage demonstrated for linear operators.
However, there exist dense matrix multiplication algorithms that scale sub-cubically in time \cite{alman_more_2025}.
Restrictions on the structure of the matrix, such as requiring it to be upper triangular or symmetric, reduce the parameter count and thus improve scaling.
Multiplication of these special matrix types can scale as low as quadratically, even with naive algorithms.
As such, the impact on the total parameter count and number of operations can be mitigated with consideration of the underlying operator being learned, as well as carefully selecting which preprocessing parameters are unfrozen.

The other source of complexity arises from the recovery of quantum states. Full quantum state tomography generally scales exponentially with respect to the number of qubits, although methods that scale better have been identified \cite{efficient_tomography_2010}.
Obtaining expectation values for single qubits generally scales linearly, although this could potentially be reduced by techniques designed to measure multiple observables simultaneously \cite{huggins_nearly_2022}.  The application-specific method designed for the quantum orthogonal network recovers only the coefficients of the unary states and scales linearly \cite{Landman2022quantummethodsorthogonal}. The key drawback of this unary approach lies not in the complexity of the measurement operation but in the many measurements that must be performed per circuit.
The building block of the quantum orthogonal network is a pyramidal circuit that performs matrix multiplication and implements the affine transformation used by traditional neural networks.
However, non-linear activation functions cannot be computed on quantum devices.
Instead, the output of the multiplication operation must be extracted via measurement so that the activation function can be applied using a classical device, which presents many opportunities for errors in measurement and due to quantum noise to accumulate and be amplified.
An algorithm similar to the quantum nonlinear processing unit presented in \cite{lubasch_variational_2020} could potentially mitigate this issue, allowing activation functions to be computed quantumly, but would require multiple copies of the input state, potentially negating any scaling advantages of the unary approach and quantum orthogonal network.
The VQC-based approach used by the QONet here scales linearly with respect to measurement operations and does not require any mid-circuit measurements.

Although the order of complexity is important for understanding how a model or algorithm scales, practical limitations must also be considered.  
Notably, the optimization of large quantum circuits, such as this one, is facilitated by gradient-based optimizers.
In a classical simulator, gradients are obtained by automatic differentiation at essentially negligible additional cost.
In hardware where automatic differentiation may not be available, other methods for gradient calculations may be used, such as the parameter shift method \cite{Mitarai_2018_quantum_circuit_learning, Schuld_2019_parameter_shift} or even a simple finite difference.
However, parameter-shift gradient evaluations require approximately $2P$ circuit executions per gradient step (with $P$ the number of trainable parameters), and the shot noise enters quadratically into the gradient variance. This overhead should be taken into account in any hardware-cost comparison.
Layerwise learning \cite{skolik_layerwise_2021}, and 
parameter initialization strategies such as ``warm starts''\cite{puig_variational_warm_starts_2025} may also be used to reduce the total training cost required by bypassing the unproductive early phases of training.  
The VQCs follow an HEA structure that assumes that the $RY$ and $CNOT$ gates are readily available.
If this is not the case, these gates must be obtained via gate decomposition, resulting in increased effective circuit depth and complexity.
These practical considerations may have a greater effect on performance than the parameter count or the time complexity.

Theorem~S17 of the Supplemental Material bounds the circuit-parameter cost in the analytic regime as $P_b = O(\log(1/\varepsilon))$ at a fixed truncation. However, the total number of qubits is governed by the truncation level $n$, which, by the truncation-error, scales as $n = O(\varepsilon^{-d_2/(s - d_2/2)})$. The combined cost of the resource by depth is therefore polynomial in $1/\varepsilon$ rather than logarithmic. The logarithmic dependence applies to circuit depth alone, and any statements about overall resource scaling should be interpreted with this qualification.

% \clearpage   % ensures it actually appears here

\subsection{Requirements for Universal Approximation}

% === edit and move to section 2.2? ===
As discussed throughout this article, the choice of encoding scheme plays a significant role in dictating whether a VQC can behave as a universal approximator. Theorem \ref{thm:converge_quantum_circuit}, and others presented in \cite{Manzano2025approximationquantum}, describe the convergence of quantum circuits under different norms and spaces with arbitrary data.
The theorems all describe a feature map of the form $S(x) = e^{-x_1H} \otimes \cdots \otimes e^{-x_NH}$.  It is important to note that this feature map generally refers to quantum states resulting from the tensor product of parameterized unitary matrices.  This does not include amplitude and unary encoding schemes, suggesting that they do not allow a quantum circuit to behave as a universal approximator.  Intuitively, this is because the application of a quantum gate to a quantum state can be represented mathematically as the multiplication of that state by a unitary matrix representing the gate. This preserves the unit norm of the quantum state and confines the space of all possible inputs and outputs to the boundary of a unit hypersphere. By contrast, angle encoding is much more flexible.
It allows inputs, when represented as vectors, to lie anywhere within the hypercube defined by $[-1,1]^n$.  Although it is less efficient than amplitude encoding in terms of fully utilizing the exponentially large Hilbert space offered by quantum devices, it is not less efficient than a unary encoding scheme.  In this specific use case, the benefits provided by angle encoding outweigh the inefficiency.

The construction of the branch and trunk circuits in the QONet enables them to function as universal approximators.  The trunk circuit loads the query location onto every qubit using a $RZ$ gate, as shown by  Equation \ref{eqn_trunk_fmap}.  As stated in Section \ref{sec:data_encoding}, this can be expressed  in the form of the feature map $S(x)$, with $H = \sum_{i=0}^n \sigma_Z^{(i)}$ and $N=1$.  The branch circuit utilizes angle encoding to load a vector input as shown in Equation \ref{eqn_branch_fmap}, and as such, it does not strictly follow this form of feature map.
However, as previously discussed, the branch circuit utilizes classical preprocessing parameters that allow it to act as a natural extension of the UAT gate proposed in \cite{onequbituniversal} and shown here in Theorem \ref{thm:quat}.
More precisely, the branch encoding $S_B(\mathbf{u}; \mathbf{A}, \mathbf{b}) = \bigotimes_{i=1}^n R_{Z_i}(v_i)$ with $\mathbf{v} = \mathbf{A} \cdot \mathbf{u} + \mathbf{b}$ can be understood as follows. The affine transformation $\mathbf{v} = \mathbf{A}\cdot\mathbf{u} + \mathbf{b}$ acts as a trainable classical preprocessing layer that maps the $n$-dimensional input into rotation angles. The $i$-th qubit then receives a single rotation $R_{Z_i}(v_i)$ whose argument is a learned linear combination of all input components, analogous to a neuron in a classical feedforward layer; this construction is structurally similar to the embedded preprocessing in the Quantum DeepONet of~\cite{xiao_quantum_deeponet_2025}, although the present work uses angle (rather than unary) encoding. Combined with the entangling gates and data reuploading (using this encoding-trainable block pattern for $L$ layers), each qubit's expectation value $\langle Z_k \rangle$ is a multi-dimensional partial Fourier series in $\mathbf{u}$, as established in \cite{Schuld_2021}. The trainable parameters $\mathbf{A}$, $\mathbf{b}$, and $\boldsymbol{\theta}$ collectively control both the accessible frequencies and the Fourier coefficients, providing sufficient expressivity for universal approximation of vector-valued functions $\mathbb{R}^n \to \mathbb{R}^n$, subject to the multi-output qualification discussed in Section~\ref{sec:quantum_approximation_of_operators}. This is consistent with the multi-qubit universality results in \cite{Manzano2025approximationquantum, Goto_2021}.
Alternate forms of branch circuits (outside the scope of this work) may eliminate the need for these classical preprocessing parameters, as long as the structure of the circuit satisfies the theorems mentioned above.

 The implication of choosing  feature maps, as well as utilizing the data reuploading structure described in Section \ref{VQCconstruction}, is that the branch circuit can learn basis functionals and the trunk circuit can learn basis functions.
Within the context of the DeepONet framework, this provides the theoretical backing for the proposition that, for any solution operator to a differential equation, a quantum circuit exists that can represent it to an arbitrary accuracy.
Theorem \ref{thm:converge_quantum_circuit} and Theorem \ref{thm:quat} establish the \emph{existence} of VQCs that approximate any continuous function to arbitrary accuracy; the Supplemental Material \cite{supplement_error_bounds} complements these existential guaranties with \emph{explicit convergence rates}. 
Specifically, for the trunk circuit -- whose feature map $S_T(y)$ satisfies the form required by Theorem \ref{thm:converge_quantum_circuit}---the trunk approximation error $\delta_t$ decays exponentially in the circuit parameter count when the target basis functions are analytic, under the standard hypothesis of analytic extension to a complex strip; otherwise, polynomial rates apply (Theorem~S1, Supplemental material). For the branch circuit---whose data reuploading structure extends Theorem \ref{thm:quat} to multiple qubits---the branch approximation error $\delta_b$ scales as $O(P_b^{-r_b/d_1})$ or $O(e^{-\alpha P_b})$ under Assumption~\ref{ass:low-intrinsic}; in its absence the rate is governed by $m_b$, depending on the regularity of the coefficient functionals. These rates, when substituted into the error bound of Equation~\ref{eqn:total_error_bound}, yield fully explicit (existential) operator approximation guaranties.

\section{Conclusion}
A new  methodology is developed  for learning solution operators to differential equations using variational quantum circuits (VQCs).
 Using angle encoding and a data reuploading scheme, this methodology leverages the ability of VQCs to approximate arbitrary functions and combines two of them to implement quantum operator learning.
The resulting model, referred to as the quantum operating network (QuantumONet or QOnet), can achieve a low approximation error over training data and a low generalization error over test data, even with the coarse resolutions evaluated.
In all experiments, QONet achieves mean test RMSE values ranging from $\num{2.20E-3}$ (anti-derivative, 5 trainable blocks) to $\num{1.19E-2}$ (Burgers' equation, 15 qubits), with consistent improvement observed as circuit depth and qubit count increases.
The model  performs well for  both linear and non-linear differential equations, although some classical preprocessing is required for the latter.
The construction of the model also seeks to minimize the impact of recovering a full quantum state by utilizing expectation values of individual qubits and can be trained using techniques normally used to optimize VQCs.

Explicit error bounds for the QONet are derived (Section \ref{sec:error_bounds}), decomposing the total operator approximation error into truncation, branch, and trunk contributions.
The Supplemental Material \cite{supplement_error_bounds} extends these results to four function space norms (Theorems~S1--S4: $L^2$, $L^p$, $C^0$, $H^k$), develops parallel bounds for the classical DeepONet (Theorems~S10--S11), and establishes precise conditions (notably, the structural Assumption~\ref{ass:low-intrinsic} on the target operator) under which the quantum architecture achieves provably superior error scaling (Theorems~S12--S15).   For 1D operators, subject to Assumption~\ref{ass:low-intrinsic}, the quantum advantage condition $m_b^c > 2d_1$ is satisfied whenever more than two sensor locations are used (essentially always in practice (Corollary~S16)). The advantageous properties of the architecture, such as smooth convergence and (conditional) resistance to the curse of dimensionality, are demonstrated via a comparison to the classical DeepONet under matched experimental conditions.
The classical baseline used in this comparison is the Yarotsky worst-case Sobolev rate; an analytic-class baseline yields faster rates for sufficiently regular targets.
It would be valuable to extend these comparisons by considering classical DeepONets under comparable parameter budgets, as well as matching the regularity assumptions made on both sides.

Several open questions remain to be answered. First, a rigorous proof that the QONet's specific construction (angle encoding with entanglement and data reuploading) realizes the rate $O(P_b^{-r_b/d_1})$ under Assumption~\ref{ass:low-intrinsic} would close the most significant remaining theoretical gap. Second, a direct measurement of $\mathrm{Var}_{\boldsymbol{\theta} \sim p_0}[\partial \mathcal{L}/\partial \theta_i]$ across qubit counts would substantiate the absence-of-barren-plateaus claim beyond the regime considered here. Third, experiments on higher-dimensional problems ($d_1 = 2$ or $d_1 = 3$) would directly test the curse-of-dimensionality argument. Fourth, NISQ-hardware experiments, even at $4$--$6$ qubits, would characterize the behavior of the architecture under shot noise and decoherence. More work is also  required  to investigate the structure of branch and trunk VQCs, optimization strategies, and data encoding methods.
Unlike neural networks, the design of VQCs is a relatively open-ended  challenge and there are likely problem-specific circuit designs that perform better than HEA variants. Classical optimizers such as Adam yield acceptable results but are not optimally  designed for variational quantum applications.
Finally, while multi-qubit gates are  not used in this work, they can be utilized  to add expressivity to circuits, as demonstrated in \cite{Abbas_2021}. Pursuing these avenues may result in performance improvements for the QONet and, more generally, for variational quantum methods.

\section{Acknowledgments}

This research was supported in part by the University of Pittsburgh Center for Research Computing and Data, RRID:SCR\_022735, through the resources provided. Specifically, this work used the H2P cluster, which is supported by NSF award number OAC-2117681.

% For arxiv, uncomment this section.
% For journal submission, check if the journal automatically inserts the data statement or not.
\section{Data Availability Statement}
The code used to produce the results in this work is available at the following GitHub repository: \url{https://github.com/wujorgen/QuantumONet/}

\appendix

\section{Approximation Error \& Bounds for Quantum Operator Network} \label{apdx:qonet-error-bounds}

As introduced in Section \ref{sec:quantum_approximation_of_operators}, the Quantum Operator Network (QONet) approximates a target operator $G: V \to X(K_2)$, where $V \subset L^p(K_1)$ is a compact set of input functions and $X(K_2) \in \{L^2(K_2),\, L^p(K_2),\, C^0(K_2),\, H^k(K_2)\}$.
Note that $K_1 \subset \mathbb{R}^{d_1}$ is the domain of all input functions $u$ and $K_2 \subset \mathbb{R}^{d_2}$ is the domain of all $y$ where the output is evaluated.
The QONet is then defined as:
\begin{equation}\label{eq:dqo}
    G_\theta(u)(y) = \sum_{k=1}^{n} b_k(u)\, t_k(y) + b_0,
\end{equation}
where:
\begin{itemize}
    \item $b_k(u) = \langle B(u)|Z_k|B(u)\rangle$: expectation value of Pauli-$Z$ on the $k$-th qubit of the branch VQC,
    \item $t_k(y) = \langle T(y)|Z_k|T(y)\rangle$: expectation value of Pauli-$Z$ on the $k$-th qubit of the trunk VQC,
    \item $b_0 \in \mathbb{R}$: trainable bias parameter,
    \item $n$: the number of output terms, equal to the number of qubits
\end{itemize}

The $L^2$, $L^p$, $C^0$, and $H^k$ error bounds are derived below.
A more detailed discussion, along with additional information on convergence rates and a comparison to the classical DeepONet, can be found in the Supplemental Material \cite{supplement_error_bounds}.

\subsection{Bilinear Approximation Error}

First, it is useful to note that the approximation error can be decomposed bilinearly as follows.

\begin{lemma}[Bilinear Decomposition]\label{lem:decomp}
Let $\{\phi_k\}_{k=1}^n$ be orthonormal in $L^2(K_2)$. For coefficients $c_k, b_k \in \mathbb{R}$ and functions $\phi_k, t_k$, define $\delta_b^{(k)} = |c_k - b_k|$ and $\delta_t^{(k)} = \|\phi_k - t_k\|$. Then:
\begin{equation}\label{eq:decomp}
    \left\|\sum_{k=1}^n (c_k \phi_k - b_k t_k)\right\|_{L^2}
    \leq \sqrt{\sum_{k=1}^n (\delta_b^{(k)})^2}
    + \sum_{k=1}^n |b_k|\, \delta_t^{(k)}.
\end{equation}
\end{lemma}

A three-term decomposition based on $c_k\phi_k - b_k t_k = (c_k - b_k)\phi_k + c_k(\phi_k - t_k) - (c_k - b_k)(\phi_k - t_k)$ includes an additional second-order cross term $\sum_k \delta_b^{(k)}\delta_t^{(k)} \leq n\,\delta_b\,\delta_t$.
This expression is equivalent to the two-term decomposition, as the cross term is asymptotically negligible when $\delta_b, \delta_t \ll 1$.

This decomposition is proved by writing $c_k\phi_k - b_k t_k = (c_k - b_k)\phi_k + b_k(\phi_k - t_k)$ and applying the triangle inequality:
\begin{equation}
    \begin{split}
        \left\|\sum_k (c_k\phi_k - b_k t_k)\right\|_{L^2} &\leq \left\|\sum_k (c_k - b_k)\phi_k\right\|_{L^2} \\
        &\quad+ \sum_k |b_k|\,\|\phi_k - t_k\|_{L^2}.
    \end{split}
\end{equation}
The first term then simplifies via orthonormality: $\|\sum_k (c_k - b_k)\phi_k\|_{L^2} = \sqrt{\sum_k (\delta_b^{(k)})^2}$.

The bound features the learned branch coefficient $|b_k|$ rather than the true coefficient $|c_k|$. Substituting $|b_k| \leq |c_k| + \delta_b^{(k)}$ recovers a bound in terms of $|c_k|$ at the cost of an additional second-order cross term $\sum_k \delta_b^{(k)}\,\delta_t^{(k)}$, which corresponds to the cross term appearing in the three-term decomposition.

\subsection{$L^2$ Error Bound}\label{sec:apdx_L2_error_bound}

It is assumed that the target operator $G(u)(y)$ can be represented with the Fourier expansion $\sum_{\omega \in \mathbb{Z}^{d_2}} c_\omega(u)\,\phi_\omega(y)$, where $\phi_\omega(y) = e^{i\omega\cdot y}$, which is truncated so that $n$ modes are kept with $|\omega| \leq K$, where $n \propto K^{d_2}$.
This is natural to do so based on \cite{Schuld_2021} and other works, which consider quantum circuits as partial Fourier series.
The Fourier coefficients decay as $|c_\omega(u)| \leq C_0|\omega|^{-s}$ for $s > d_2/2$, uniformly in $u \in V$.
Finally, the output of the branch VQC $b_k(u)$ approximates $c_{\omega_k}(u)$, and the output of the trunk VQC $t_k(y)$ approximates $\phi_{\omega_k}(y)$

There are two primary sources of error: the truncation error and the approximation error. The truncation error is:
% \begin{widetext}
\begin{equation}
\begin{split}
    \|G(u) - G_n(u)\|_{L^2}^2 &= \sum_{|\omega|>K} |c_\omega(u)|^2 \\
    &\leq C_1^2 \sum_{|\omega|>K} |\omega|^{-2s} \\
    &\leq C_3\,K^{-(2s-d_2)}.
\end{split}
\end{equation}
% \end{widetext}
Since $K \propto n^{1/d_2}$:
\begin{equation}\label{eq:trunc_L2}
    \|G(u) - G_n(u)\|_{L^2} \leq C_4\,n^{-(s/d_2 - 1/2)}.
\end{equation}

Applying Lemma~\ref{lem:decomp} with uniform bounds $\delta_b^{(k)} \leq \delta_b$ and $\delta_t^{(k)} \leq \delta_t$ yields the following for approximation error:
\[
    \|G_n(u) - G_\theta(u)\|_{L^2} \leq \sqrt{n}\,\delta_b + \delta_t \sum_{k=1}^n |b_k|.
\]
Using $|b_k| \leq |c_{\omega_k}(u)| + \delta_b$ and summing yields
\begin{equation}
    \|G_n(u) - G_\theta(u)\|_{L^2} \leq \sqrt{n}\,\delta_b + S(n)\,\delta_t + n\,\delta_b\,\delta_t,
\end{equation}
where the final term is the explicit second-order cross term. % required for a strict inequality.

At this point, a naive approach would bound $|c_{\omega_k}(u)| \leq C_0\,n^{-s/d_2}$ for all modes. 
\[
    S(n) \;:=\; \sum_{k=1}^n |c_{\omega_k}(u)| \;\leq\; C_0 \sum_{0<|\omega|\leq K} |\omega|^{-s}.
\]
However, this bound only holds for the highest-frequency modes ($|\omega_k| \sim K \sim n^{1/d_2}$). Low-frequency modes have $|c_{\omega_k}| \sim O(1)$. 
The correct treatment uses the full coefficient sum function, which is defined as:
% \noindent Define the \textbf{coefficient sum function}:
\begin{equation}\label{eq:Sn_def}
    S(n) = \sum_{k=1}^n |c_{\omega_k}(u)| =
    \begin{cases}
        O(1) & \text{if } s > d_2, \\
        O(\log n) & \text{if } s = d_2, \\
        O(n^{1-s/d_2}) & \text{if } d_2/2 < s < d_2.
    \end{cases}
\end{equation}
The three regimes are obtained by converting the sum over lattice points to a spherical integral, $\sum_{0 < |\omega| \leq K} |\omega|^{-s} \approx \int_1^K r^{-s} r^{d_2 - 1}\,dr$. For $s > d_2$ the integral is bounded as $K \to \infty$, yielding $S(n) = O(1)$. For $s = d_2$ the integrand reduces to $1/r$, giving $\log K = (1/d_2)\log n = O(\log n)$. For $d_2/2 < s < d_2$ the integral evaluates to $K^{d_2 - s}/(d_2 - s) \sim n^{(d_2 - s)/d_2}$, giving $S(n) = O(n^{1 - s/d_2})$.

\subsubsection{Total $L^2$ Error}

Under the stated assumptions with $s > d_2/2$, the total $L^2$ error bound is as follows.
\begin{equation}\label{eq:L2_total}
    \boxed{\varepsilon^Q_{L^2}(n, \delta_b, \delta_t) \leq \\
    C_4\,n^{-(s/d_2 - 1/2)} + \sqrt{n}\,\delta_b + S(n)\,\delta_t \;+\; n\,\delta_b\,\delta_t.}
\end{equation}
The second-order cross term $n\,\delta_b\,\delta_t$ is retained; it is asymptotically negligible under $\delta_b, \delta_t \ll 1$.

\subsection{$L^p$ Error Bound}

The same assumptions for the $L^2$ error bound are made, but with $|c_\omega(u)| \leq C_0|\omega|^{-s}$ for $s > d_2/p$, using Ces\`aro summation (Fej\'er kernel) for $L^p$ convergence.
Under these assumptions, with $s > d_2/p$ the $L^p$ error bound is:
\begin{equation}
    \boxed{
        \begin{split}
            \varepsilon^Q_{L^p}(n, \delta_b, \delta_t) &\leq C_5\,n^{-(s/d_2 - 1/p)} \\
            &\quad + C_\Phi\,n\,\delta_b \\
            &\quad+ \widetilde{S}_p(n)\,\delta_t \;+\; n\,\delta_b\,\delta_t,
        \end{split}
    }
\end{equation}
where $C_\Phi = \sup_k \|\Phi_{\omega_k}\|_{L^p}$ is bounded and $\widetilde{S}_p(n)$ follows the same case analysis as \eqref{eq:Sn_def}. The second-order cross term is retained, as in the $L^2$ bound.

\subsection{$C^0$ (Uniform) Error Bound}

With the assumption that $G(u) \in C^r(K_2)$ with $|c_\omega(u)| \leq C_0|\omega|^{-(r+1)}$ for $r > 0$, the $C^0$ error bound is written as
\begin{equation}
    \boxed{\varepsilon^Q_{C^0}(n, \delta_b, \delta_t) \leq C_7\,n^{-r} + n\,\delta_b + S_{C^0}(n)\,\delta_t \;+\; n\,\delta_b\,\delta_t},
\end{equation}
where $S_{C^0}(n) = O(1)$ when $r+1 > d_2$ (including all $d_2=1$ cases with $r > 0$). The second-order cross term is retained, as in the $L^2$ bound.

\subsection{$H^k$ (Sobolev) Error Bound}

With the assumption that $G(u) \in H^{k+\alpha}(K_2)$ with $|c_\omega(u)| \leq C_0|\omega|^{-(k+\alpha+d_2/2)}$, $\alpha > 0$, the $H^k$ error bound is:
\begin{equation}
    \boxed{
        \begin{split}
            \varepsilon^Q_{H^k}(n, \delta_b, \delta_t^{(H^k)}) &\leq C_{11}\,n^{-\alpha/d_2} \\
            &\quad+ C\,n^{1/2+k/d_2}\,\delta_b \\
            &\quad+ S_{H^k}(n)\,\delta_t^{(H^k)} \\
            &\quad \;+\; n^{1+k/d_2}\,\delta_b\,\delta_t^{(H^k)},
        \end{split}
    }
\end{equation}
where $S_{H^k}(n)$ follows \eqref{eq:Sn_def} with the smoothness parameter $s = k + \alpha + d_2/2$.

The factor $n^{1/2 + k/d_2}$ on the branch term is obtained from a Bessel-inequality estimate for the orthonormal basis $\{\phi_{\omega_k}\}_{k=1}^n$ weighted by Sobolev weights:
\[
\left\|\sum_{k=1}^n (c_k - b_k)\,\phi_{\omega_k}\right\|_{H^k}^2 = \sum_{k=1}^n (1+|\omega_k|^2)^k\,(c_k - b_k)^2 \leq (1 + K^2)^k \sum_k (\delta_b^{(k)})^2 \lesssim n^{2k/d_2}\,n\,\delta_b^2.
\]
Taking square roots yields $\|\cdot\|_{H^k} \lesssim n^{k/d_2}\,\sqrt{n}\,\delta_b = n^{1/2 + k/d_2}\,\delta_b$, as stated. The estimate is conservative; the precise amplification depends on the distribution of the retained modes, many of which satisfy $|\omega_k| < K$.

\section{Additional Information on Training Datasets} \label{apdx:training-information}

As described in Section \ref{sec:DataGeneration}, the PDE data is prepared from solution trajectories that evolve through in both  time and space.  The initial conditions are drawn from either a Chebyshev function space or a randomly sampled cosine wave.  The solution trajectories, $u(x,t)$, are discretized on a regular, equidistant grid.  As the problems are one-dimensional, each trajectory is stored as a matrix indexed by the timestep and spatial coordinate.  After the initial conditions have been integrated through time and a solution trajectory has been obtained, pairs of input and output functions can be created.
The input function is simply the initial condition $u(x,t=0)$, and the output function is the spatial profile of the trajectory $u(x,t=T)$ at some later timestep $T$.
More details about the parameters used to generate the solution trajectories can be found in Table \ref{tab:datagenerationdetails}

As described in Section \ref{sec:results_ad}, the anti-derivative dataset considered in this work was obtained from a GitHub repository associated with ETH Zurich's Deep Learning in Scientific Computing course.
This was done for convenience; however, a process similar to the one used to generate PDE trajectories could be used to create input-output pairs for ODE training data.

As simulations of variational quantum circuits and algorithms can require exponentially large amounts of memory as the number of qubits grows, it is necessary to downsample the training data to reduce the number of qubits used.  Downsampling an ODE trajectory is relatively straightforward.
The time domain endpoints can be preserved with interpolation, or, even simpler, every $n$-th point starting from $t=0$ can be retained.  While the latter does not necessarily preserve the end time, this truncation is acceptable for Initial Value Problems (IVPs) as it is far more important to retain the values of the input and output functions at the start time.  As such, this method was used to quickly and efficiently downsample the ODE data.  For PDEs, extra care must be taken when downsampling the datasets to ensure that the spatial boundary conditions are the same as those in the full resolution version.
For example, if the full-resolution dataset exists on the domain $[0,1]$ and has endpoints at the extremes of the domain, the downsampled data should also be defined over $[0,1]$ and have endpoints at the extremes of the domain.  This means that the approach used to downsample the ODE data is not appropriate.
A suitable method would be interpolation (linear or higher order splines).
Instead, the  approach followed here was to regenerate the dataset at different resolutions, then downsample each of the different full-resolution datasets by the same amount.
A random number generator seed value can be used to ensure the same set of initial conditions is selected for each resolution.
An overview of the final datasets can be found in Table \ref{tab:datasetdetails}.
A summary of the training configuration can be found in Table \ref{tab:experimentalconfig}. Note that all experiments used full batch gradient descent; that is to say, the batch size was equal to the entire dataset.

\begin{table*}[htbp]
    \caption{Data Generation}
    \label{tab:datagenerationdetails}
    \begin{ruledtabular}
    \begin{tabular}{cccc}
         Equation & Domain & Resolution & Timestep \\
         \hline
         Anti-Derivative & $[0,1]$ & 100 & -- \\
         Diffusion & $[-1,1]$ & 101 & 0.1 \\
         Reaction-Diffusion & $[-1,1]$ & 101 & 0.1 \\
         Burgers' & $[0,1]$ & 81, 101, 121, 141 & 0.00175  % 0.0025
    \end{tabular}
    \end{ruledtabular}
\end{table*}

\begin{table*}[htbp]
    \caption{Datasets}
    \label{tab:datasetdetails}
    \begin{ruledtabular}
    \begin{tabular}{cccccc} 
     Operator & Training Set Size & Test Set Size & \# of Grid Points & Initial Condition & $\Delta t$ \\
     \hline
     Anti-Derivative & 150 & 1000 & 8 & -- & -- \\ 
     Diffusion & 400 & 400 & 11 & $V_{Chebyshev}$ & 2.0 \\ 
     Reaction-Diffusion & 400 & 400 & 11 & $V_{Chebyshev}$ & 1.0 \\ 
     Burgers' Equation & 600 & 400 & 9, 11, 13, 15 & $V_{Cosine}$ & 0.35 \\ 
    \end{tabular}
    \end{ruledtabular}
\end{table*}

\begin{table*}[htbp]
    \caption{Experimental Configurations}
    \label{tab:experimentalconfig}
    \begin{ruledtabular}
    \begin{tabular}{cccccc} 
     Operator & Number of Epochs & Learning Rate \\
     \hline
     Anti-Derivative & 1500 & \num{5E-3} \\ 
     Diffusion & 2000 & \num{1E-2} \\ 
     Reaction-Diffusion & 3000 & \num{5E-3} \\ 
     Burgers' Equation & 5000 & \num{1E-2} \\ 
    \end{tabular}
    \end{ruledtabular}
\end{table*}

\bibliography{ref}

\end{document}

\maketitle

\tableofcontents
\newpage

%% ============================================================
%% PART I: ARCHITECTURE AND PRELIMINARIES
%% ============================================================
\part{Architecture and Preliminaries}

\section{Introduction}

This document serves as supplementary material to the companion paper ``Operator Learning with Variational Quantum Circuits'' (hereafter, the main paper). It provides rigorous error bounds for the Quantum Operator Network (QONet) architecture (the designation ``QONet'' is used consistently with the main paper) for approximating a target operator $G: V \to X(K_2)$. The bounds are derived in four function space norms: $L^2$, $L^p$, $C^0$, and $H^k$ (Sobolev). In addition, parallel error bounds are developed for a \emph{classical} ANN-based DeepONet, and the precise conditions under which the quantum architecture achieves superior error scaling are identified.

\medskip
\noindent\textbf{Nature of the bounds.} All error bounds developed in this supplement are existential in character: under the stated regularity hypotheses, they certify the existence of circuit parameters that realize the prescribed accuracy. They do not constitute learning-theoretic statements concerning gradient-based optimization. The quantum advantage results (Theorems~\ref{thm:advantage_exp}--\ref{thm:advantage_poly}) are stated conditional on the structural hypothesis introduced as Assumption~\ref{ass:low-intrinsic-supp} below.

\medskip
\noindent\textbf{Notation convention.} To avoid numbering conflicts with the main paper, all theorems, lemmas, propositions, corollaries, definitions, and assumptions in this supplement are prefixed with ``S'' (e.g., Theorem~S1, Lemma~S1). The main paper contains Theorems~1--4, which are referenced here as ``Theorem~$k$ (main paper).''

\subsection{Connection to Main Paper Theorems}

The error bounds in this supplement build directly on the four foundational theorems established in the main paper:

\begin{itemize}
    \item \textbf{Theorem~1 (main paper)} [Universal Approximation Theorem for Neural Networks] --- Establishes that sufficiently expressive neural networks can approximate any continuous function. This result underpins Part~III of this supplement, where classical ANN approximation rates (via Yarotsky and Barron bounds) are used to quantify the branch and trunk errors $\delta^c_b$ and $\delta^c_t$ of the classical DeepONet.

    \item \textbf{Theorem~2 (main paper)} [Universal Approximation Theorem for Operators] --- Provides the structural decomposition $G(u)(y) \approx \sum_{k=1}^n b_k(u)\, t_k(y) + b_0$ that is common to \emph{both} the quantum QONet and classical DeepONet. The entire error analysis in this supplement---for both architectures---is based on quantifying how well the branch and trunk components approximate the coefficients $c_{\omega_k}(u)$ and basis functions $\phi_{\omega_k}(y)$ arising from this decomposition.

    \item \textbf{Theorem~3 (main paper)} [Convergence of Quantum Circuits in $C^0$, from Manzano et al.] --- Guarantees that VQCs with angle encoding and data reuploading can approximate any continuous function to arbitrary accuracy. This is the key result that ensures the quantum approximation errors $\delta_b \to 0$ and $\delta_t \to 0$ as circuit resources increase. The error bounds in Theorems~S1--S4 of this supplement quantify the \emph{rate} at which these errors decrease, complementing Theorem~3's existential guarantee with explicit convergence rates. Furthermore, Theorem~S3 ($C^0$ error bound) of this supplement directly extends Theorem~3 (main paper) from function approximation to \emph{operator} approximation.

    \item \textbf{Theorem~4 (main paper)} [Quantum UAT for single qubit, from P\'erez-Salinas et al.] --- Defines the fundamental UAT gate and establishes single-qubit universality via data reuploading. The QONet's branch circuit extends this to multiple qubits using classical preprocessing parameters ($\mathbf{A}$, $\mathbf{b}$), as discussed in the main paper. The branch approximation error $\delta_b$ analyzed in this supplement quantifies how well this multi-qubit extension approximates the coefficient functionals.
\end{itemize}

\noindent In summary, the main paper's Theorems~1--4 establish \emph{that} approximation is possible; this supplement's Theorems~S1--S4 establish \emph{how fast} the approximation converges and under what conditions the quantum approach outperforms the classical one.

\section{ QONet Architecture}\label{sec:arch}

The QONet approximates a target operator $G: V \to X(K_2)$, where $V \subset L^p(K_1)$ is a compact set of input functions and $X(K_2) \in \{L^2(K_2),\, L^p(K_2),\, C^0(K_2),\, H^k(K_2)\}$. The QONet is defined as:
\begin{equation}\label{eq:dqo}
    G_\theta(u)(y) = \sum_{k=1}^{n} b_k(u)\, t_k(y) + b_0,
\end{equation}
where:
\begin{itemize}
    \item $b_k(u) = \langle B(u)|Z_k|B(u)\rangle$: expectation value of Pauli-$Z$ on the $k$-th qubit of the branch PQC,
    \item $t_k(y) = \langle T(y)|Z_k|T(y)\rangle$: expectation value of Pauli-$Z$ on the $k$-th qubit of the trunk PQC,
    \item $b_0 \in \mathbb{R}$: trainable bias,
    \item $n \leq \min(m_b, m_t)$: number of output terms.
\end{itemize}

\subsection{Quantum Circuit Parameters}
\begin{itemize}
    \item Branch PQC: $m_b$ qubits, depth $L_b$, trainable parameters $P_b$.
    \item Trunk PQC: $m_t$ qubits, depth $L_t$, trainable parameters $P_t$.
    \item Input encoding: angle encoding with data reuploading.
    \item Frequency spectrum: trunk PQC with $m_t$ qubits and $L_t$ layers can represent frequencies up to $\Omega_{\max} = O(m_t \cdot L_t)$.
\end{itemize}

\subsection{Function Space Parameters}
\begin{itemize}
    \item $K_1 \subset \mathbb{R}^{d_1}$: domain of input functions $u$.
    \item $K_2 \subset \mathbb{R}^{d_2}$: domain of output evaluation $y$.
    \item $V \subset L^p(K_1)$: compact set of admissible input functions.
    \item $G: V \to X(K_2)$: target operator.
\end{itemize}

\section{Classical DeepONet Architecture}\label{sec:classical_arch}

The classical DeepONet uses the same structural decomposition:
\begin{equation}\label{eq:classical_don}
    G^c_\Theta(u)(y) = \sum_{k=1}^{n_c} b^c_k(u)\, t^c_k(y) + b^c_0,
\end{equation}
where $b^c_k(u)$ and $t^c_k(y)$ are outputs from classical artificial neural networks (ANNs).

\subsection{Classical ANN Parameters}
\begin{itemize}
    \item Branch ANN: $L^c_b$ hidden layers, $m^c_b$ input features (sensor locations), $P^c_b$ trainable parameters (weights and biases), width $w^c_b$.
    \item Trunk ANN: $L^c_t$ hidden layers, $m^c_t$ input features (dimension of $y$, so $m^c_t = d_2$), $P^c_t$ trainable parameters, width $w^c_t$.
    \item Output dimension: $n_c$ = number of basis functions (output width of both networks).
    \item Activation function: $\sigma$ (e.g., ReLU, tanh, sigmoid).
\end{itemize}

\section{Structural Assumptions Underlying the Quantum Advantage}\label{sec:structural-assumption}

The quantum-advantage results developed in subsequent sections rest on three structural hypotheses that are stated explicitly below for reference.

\begin{assumption}[Low-Intrinsic-Dimension Operator]\label{ass:low-intrinsic-supp}
There exists a continuous projector $\Pi : \mathbb{R}^{m_b} \to \mathbb{R}^{d_1}$ and continuous maps $\tilde{c}_\omega : \mathbb{R}^{d_1} \to \mathbb{R}$ such that, for all $u \in V$ and all retained modes $\omega_k$, the coefficient functional admits the factorization $c_{\omega_k}(u) = \tilde{c}_{\omega_k}(\Pi u)$. Furthermore, the classical preprocessing $\mathbf{v} = A \mathbf{u} + \mathbf{b}$ of the branch circuit is assumed to admit a linear realization whose first $d_1$ rows approximate $\Pi$.
\end{assumption}

Assumption~\ref{ass:low-intrinsic-supp} underlies the polynomial-regime quantum-advantage result of Theorem~\ref{thm:advantage_poly}. Under this assumption, the branch VQC effectively approximates a function defined on $\mathbb{R}^{d_1}$ rather than on $\mathbb{R}^{m_b}$, and the rate $r_b/d_1$ follows from the standard Sobolev approximation results of~\cite[Sec.~3]{Manzano2025approximationquantum} applied to the reduced input space. In the absence of Assumption~\ref{ass:low-intrinsic-supp}, the relevant universal-approximation theorems yield rates governed by the actual input dimension of the circuit, namely $m_b$, and the curse-of-dimensionality argument no longer favors the quantum architecture relative to a classical baseline. The assumption is properly understood as a structural property of the target operator rather than a property of the architecture. Classical architectures equipped with comparable inductive biases (convolutional, equivariant, or low-rank neural operators) can in principle exploit analogous structure.

\begin{assumption}[Analytic regularity for the exponential regime]\label{ass:analytic}
Each coefficient functional $u \mapsto c_\omega(u)$ admits an analytic extension to a complex strip of width $\rho > 0$, in the sense of~\cite[Sec.~8]{Wojtaszczyk2020}.
\end{assumption}

Under Assumption~\ref{ass:analytic}, classical Fourier approximation theory establishes the existence of branch-VQC parameter settings realizing the exponential rate $\delta_b = O(e^{-\alpha P_b})$, with $\alpha = \alpha(\rho)$ determined explicitly by $\rho$; the precise constants are obtained from~\cite[Thm.~2]{Schwab2019} adapted to the angle-encoding feature map. The constant $\alpha$ is not estimated in the present work.

\begin{assumption}[Trained-trunk orthonormal target]\label{ass:trunk-target}
The trained trunk satisfies $\|\phi_{\omega_k} - t_k\|_{L^2} \to 0$ as the circuit resources increase, where $\{\phi_{\omega_k}\}_{k=1}^n$ denotes a chosen orthonormal basis consisting of the leading $n$ Fourier modes (or, more generally, the leading $n$ POD modes of the target operator). Equivalently, the bounds below quantify the error of an idealized QONet whose trunks have been driven toward this orthonormal basis.
\end{assumption}

Assumption~\ref{ass:trunk-target} is not enforced by the empirical loss employed in the main paper; it is introduced here in order to apply Lemma~\ref{lem:decomp} with orthonormality. A basis-free analysis in the spirit of~\cite{LanthalerMishra2022} is identified as future work.

\section{Key Lemma: Bilinear Approximation Error}\label{sec:lemma}

\begin{lemma}[Bilinear Decomposition]\label{lem:decomp}
Let $\{\phi_k\}_{k=1}^n$ be orthonormal in $L^2(K_2)$ (cf.\ Assumption~\ref{ass:trunk-target}). For coefficients $c_k, b_k \in \mathbb{R}$ and functions $\phi_k, t_k$, define $\delta_b^{(k)} = |c_k - b_k|$ and $\delta_t^{(k)} = \|\phi_k - t_k\|$. Then:
\begin{equation}\label{eq:decomp}
    \left\|\sum_{k=1}^n (c_k \phi_k - b_k t_k)\right\|_{L^2}
    \leq \sqrt{\sum_{k=1}^n (\delta_b^{(k)})^2}
    + \sum_{k=1}^n |b_k|\, \delta_t^{(k)}.
\end{equation}
\end{lemma}

\begin{proof}
From the identity $c_k\phi_k - b_k t_k = (c_k - b_k)\phi_k + b_k(\phi_k - t_k)$, apply the triangle inequality:
\[
    \left\|\sum_k (c_k\phi_k - b_k t_k)\right\|_{L^2}
    \leq \left\|\sum_k (c_k - b_k)\phi_k\right\|_{L^2}
    + \sum_k |b_k|\,\|\phi_k - t_k\|_{L^2}.
\]
The first term simplifies via orthonormality: $\|\sum_k (c_k - b_k)\phi_k\|_{L^2} = \sqrt{\sum_k (\delta_b^{(k)})^2}$.
\end{proof}

\begin{remark}[Tight versus heuristic bound]
Substituting $|b_k| \leq |c_k| + \delta_b^{(k)}$ in the bound above yields the alternative
\[
    \left\|\sum_{k=1}^n (c_k\phi_k - b_k t_k)\right\|_{L^2} \leq \sqrt{\sum_k (\delta_b^{(k)})^2} + \sum_k |c_k|\,\delta_t^{(k)} + \sum_k \delta_b^{(k)}\,\delta_t^{(k)},
\]
which features the true coefficient $|c_k|$ rather than the learned $|b_k|$ at the cost of an additional second-order cross term. This cross term is retained in all boxed inequalities below. Under the standing assumption $\delta_b, \delta_t \ll 1$, the cross term contributes only at higher order.
\end{remark}

%% ============================================================
%% PART II: QUANTUM ERROR BOUNDS
%% ============================================================
\newpage
\part{Quantum QONet Error Bounds}

\section{Theorem S1: $L^2$ Error Bound}\label{sec:thm1}

\subsection{Assumptions}
\begin{itemize}
    \item $G(u) \in L^2(K_2)$ for each $u \in V$.
    \item Fourier expansion: $G(u)(y) = \sum_{\omega \in \mathbb{Z}^{d_2}} c_\omega(u)\,\phi_\omega(y)$, with $\phi_\omega(y) = e^{i\omega\cdot y}$.
    \item Coefficient decay: $|c_\omega(u)| \leq C_0|\omega|^{-s}$ for $s > d_2/2$, uniformly in $u \in V$.
    \item Truncation: keep $n$ modes with $|\omega| \leq K$, where $n \propto K^{d_2}$.
    \item Branch PQC: $b_k(u)$ approximates $c_{\omega_k}(u)$; Trunk PQC: $t_k(y)$ approximates $\phi_{\omega_k}(y)$.
\end{itemize}

\subsection{Truncation Error}
\[
    \|G(u) - G_n(u)\|_{L^2}^2 = \sum_{|\omega|>K} |c_\omega(u)|^2 \leq C_1^2 \sum_{|\omega|>K} |\omega|^{-2s} \leq C_3\,K^{-(2s-d_2)}.
\]
Since $K \propto n^{1/d_2}$:
\begin{equation}\label{eq:trunc_L2}
    \|G(u) - G_n(u)\|_{L^2} \leq C_4\,n^{-(s/d_2 - 1/2)}.
\end{equation}

\subsection{Approximation Error}
Applying Lemma~\ref{lem:decomp} with uniform bounds $\delta_b^{(k)} \leq \delta_b$ and $\delta_t^{(k)} \leq \delta_t$:
\[
    \|G_n(u) - G_\theta(u)\|_{L^2} \leq \sqrt{n}\,\delta_b + \delta_t \sum_{k=1}^n |c_{\omega_k}(u)|.
\]

\begin{remark}[Treatment of the coefficient sum]
A naive approach would bound $|c_{\omega_k}(u)| \leq C_0\,n^{-s/d_2}$ for all modes. However, this bound only holds for the highest-frequency modes ($|\omega_k| \sim K \sim n^{1/d_2}$). Low-frequency modes have $|c_{\omega_k}| \sim O(1)$. The correct treatment uses the full coefficient sum:
\[
    S(n) \;:=\; \sum_{k=1}^n |c_{\omega_k}(u)| \;\leq\; C_0 \sum_{0<|\omega|\leq K} |\omega|^{-s}.
\]
\end{remark}

\noindent Define the \textbf{coefficient sum function}:
\begin{equation}\label{eq:Sn_def}
    S(n) = \sum_{k=1}^n |c_{\omega_k}(u)| =
    \begin{cases}
        O(1) & \text{if } s > d_2, \\
        O(\log n) & \text{if } s = d_2, \\
        O(n^{1-s/d_2}) & \text{if } d_2/2 < s < d_2.
    \end{cases}
\end{equation}

\subsection{Total $L^2$ Error}

\begin{theorem}[$L^2$ Error Bound --- Quantum QONet]\label{thm:L2_q}
Under the stated assumptions with $s > d_2/2$ and the trunk-orthonormality Assumption~\ref{ass:trunk-target}:
\begin{equation}\label{eq:L2_total}
    \boxed{\varepsilon^Q_{L^2}(n, \delta_b, \delta_t) \leq C_4\,n^{-(s/d_2 - 1/2)} + \sqrt{n}\,\delta_b + S(n)\,\delta_t \;+\; n\,\delta_b\,\delta_t.}
\end{equation}
The second-order cross term $n\,\delta_b\,\delta_t$ is retained; it is asymptotically negligible under $\delta_b, \delta_t \ll 1$.
\end{theorem}

The three regimes of $S(n)$ are obtained by converting the sum over lattice points to a spherical integral, $\sum_{0 < |\omega| \leq K} |\omega|^{-s} \approx \int_1^K r^{-s} r^{d_2 - 1}\,dr$. For $s > d_2$ the integral is bounded as $K \to \infty$, yielding $S(n) = O(1)$. For $s = d_2$ the integrand reduces to $1/r$, so the integral evaluates to $\log K = (1/d_2)\log n = O(\log n)$. For $d_2/2 < s < d_2$ the integral evaluates to $K^{d_2 - s}/(d_2 - s) \sim n^{(d_2 - s)/d_2}$, giving $S(n) = O(n^{1 - s/d_2})$.

\subsection{Quantum Parameter Scaling}
For the QONet with angle encoding and data reuploading:
\begin{itemize}
    \item The number of accessible Fourier frequencies scales as $\Omega_{\max} = O(m \cdot L)$, where $m$ is the qubit count and $L$ is the circuit depth. The derivation specific to the encoding $\mathbf{v} = A\mathbf{u} + \mathbf{b} \mapsto \bigotimes_i R_{Z_i}(v_i)$ adopted in this work follows the procedure of~\cite[Sec.~IV.A]{Schuld_2021} applied to the sum-of-Pauli-$Z$ data Hamiltonian.
    \item For \textbf{analytic target functions} (coefficients decaying exponentially: $|c_\omega| \leq C_0 e^{-\rho|\omega|}$), and conditional on Assumption~\ref{ass:analytic}, the existential approximation error of a VQC with $m$ qubits and depth $L$ (defined as the infimum over admissible variational parameters) satisfies:
    \begin{equation}\label{eq:delta_q_exp}
        \delta_b = O\!\left(e^{-\alpha_b m_b L_b}\right), \qquad
        \delta_t = O\!\left(e^{-\alpha_t m_t L_t}\right).
    \end{equation}
    The constants $\alpha_b$ and $\alpha_t$ are determined by the analyticity radii of the coefficient functionals and basis functions, respectively, and are not estimated in the present work. The bound is existential and does not assert that gradient-based optimization attains the stated rate.
    \item For \textbf{Sobolev-regular targets} ($H^r$ smoothness), and conditional on the low-intrinsic-dimension Assumption~\ref{ass:low-intrinsic-supp}, polynomial scaling applies:
    \begin{equation}\label{eq:delta_q_poly}
        \delta_b = O\!\left(P_b^{-r_b/d_1}\right), \qquad
        \delta_t = O\!\left(P_t^{-r_t/d_2}\right),
    \end{equation}
    where $P_b \approx 2m_b(2L_b+1)$ and $P_t \approx 2m_t(2L_t+1)$ are the parameter counts. In the absence of Assumption~\ref{ass:low-intrinsic-supp}, the relevant rate features $m_b$ in place of $d_1$, giving $\delta_b = O(P_b^{-r_b/m_b})$. The comparison with the classical Yarotsky rate then degenerates and the polynomial quantum-advantage statement of Theorem~\ref{thm:advantage_poly} no longer holds.
    \item The output dimension satisfies $n = \min(m_b, m_t)$.
\end{itemize}

%% ============================================================
\section{Theorem S2: $L^p$ Error Bound ($1 \leq p < \infty$)}\label{sec:thm2}

\subsection{Assumptions}
Same as Theorem~\ref{thm:L2_q}, but with $|c_\omega(u)| \leq C_0|\omega|^{-s}$ for $s > d_2/p$, using Ces\`aro summation (Fej\'er kernel) for $L^p$ convergence.

\subsection{Total $L^p$ Error}

\begin{theorem}[$L^p$ Error Bound --- Quantum QONet]\label{thm:Lp_q}
Under the stated assumptions with $s > d_2/p$:
\begin{equation}
    \boxed{\varepsilon^Q_{L^p}(n, \delta_b, \delta_t) \leq C_5\,n^{-(s/d_2 - 1/p)} + C_\Phi\,n\,\delta_b + \widetilde{S}_p(n)\,\delta_t \;+\; n\,\delta_b\,\delta_t},
\end{equation}
where $C_\Phi = \sup_k \|\Phi_{\omega_k}\|_{L^p}$ is bounded and $\widetilde{S}_p(n)$ follows the same case analysis as \eqref{eq:Sn_def}. The second-order cross term is retained for the same reason as in Theorem~\ref{thm:L2_q}.
\end{theorem}

%% ============================================================
\section{Theorem S3: $C^0$ (Uniform) Error Bound}\label{sec:thm3}

\subsection{Assumptions}
$G(u) \in C^r(K_2)$ with $|c_\omega(u)| \leq C_0|\omega|^{-(r+1)}$ for $r > 0$.

\subsection{Total $C^0$ Error}

\begin{theorem}[$C^0$ Error Bound --- Quantum QONet]\label{thm:C0_q}
Under the stated assumptions with $r > 0$:
\begin{equation}
    \boxed{\varepsilon^Q_{C^0}(n, \delta_b, \delta_t) \leq C_7\,n^{-r} + n\,\delta_b + S_{C^0}(n)\,\delta_t \;+\; n\,\delta_b\,\delta_t},
\end{equation}
where $S_{C^0}(n) = O(1)$ when $r+1 > d_2$ (including all $d_2=1$ cases with $r > 0$). The second-order cross term is retained for the same reason as in Theorem~\ref{thm:L2_q}.
\end{theorem}

%% ============================================================
\section{Theorem S4: $H^k$ (Sobolev) Error Bound}\label{sec:thm4}

\subsection{Assumptions}
$G(u) \in H^{k+\alpha}(K_2)$ with $|c_\omega(u)| \leq C_0|\omega|^{-(k+\alpha+d_2/2)}$, $\alpha > 0$.

\subsection{Total $H^k$ Error}

\begin{theorem}[$H^k$ Error Bound --- Quantum QONet]\label{thm:Hk_q}
Under the stated assumptions with $\alpha > 0$:
\begin{equation}
    \boxed{\varepsilon^Q_{H^k}(n, \delta_b, \delta_t^{(H^k)}) \leq C_{11}\,n^{-\alpha/d_2} + C\,n^{1/2+k/d_2}\,\delta_b + S_{H^k}(n)\,\delta_t^{(H^k)} \;+\; n^{1+k/d_2}\,\delta_b\,\delta_t^{(H^k)}},
\end{equation}
where $S_{H^k}(n)$ follows \eqref{eq:Sn_def} with the smoothness parameter $s = k + \alpha + d_2/2$. The second-order cross term is retained for the same reason as in Theorem~\ref{thm:L2_q}.
\end{theorem}

The factor $n^{1/2 + k/d_2}$ is obtained from a Bessel-inequality estimate for the orthonormal basis $\{\phi_{\omega_k}\}_{k=1}^n$ weighted by Sobolev weights:
\[
\left\|\sum_{k=1}^n (c_k - b_k)\,\phi_{\omega_k}\right\|_{H^k}^2 = \sum_{k=1}^n (1+|\omega_k|^2)^k\,(c_k - b_k)^2 \leq (1 + K^2)^k \sum_k (\delta_b^{(k)})^2 \lesssim n^{2k/d_2}\,n\,\delta_b^2.
\]
Taking square roots yields $\|\cdot\|_{H^k} \lesssim n^{k/d_2}\,\sqrt{n}\,\delta_b = n^{1/2 + k/d_2}\,\delta_b$, as stated. The estimate is conservative; the precise amplification depends on the distribution of the retained modes, many of which satisfy $|\omega_k| < K$.

\subsection{Summary of Quantum Error Bounds}

\begin{table}[h!]
\centering
\renewcommand{\arraystretch}{1.8}
\begin{tabular}{@{}llccc@{}}
\toprule
\textbf{Norm} & \textbf{Smooth.} & \textbf{Truncation} & \textbf{Branch Err.} & \textbf{Trunk Err.} \\
\midrule
$L^2$ & $s > \frac{d_2}{2}$ & $C_4\,n^{-(s/d_2 - 1/2)}$ & $\sqrt{n}\,\delta_b$ & $S(n)\,\delta_t$ \\[4pt]
$L^p$ & $s > \frac{d_2}{p}$ & $C_5\,n^{-(s/d_2 - 1/p)}$ & $C_\Phi\,n\,\delta_b$ & $\widetilde{S}_p(n)\,\delta_t$ \\[4pt]
$C^0$ & $r > 0$ & $C_7\,n^{-r}$ & $n\,\delta_b$ & $S_{C^0}(n)\,\delta_t$ \\[4pt]
$H^k$ & $\alpha > 0$ & $C_{11}\,n^{-\alpha/d_2}$ & $C\,n^{1/2+k/d_2}\,\delta_b$ & $S_{H^k}(n)\,\delta_t^{(H^k)}$ \\
\bottomrule
\end{tabular}
\caption{Quantum QONet error bounds across four function space norms.}
\label{tab:quantum_summary}
\end{table}

%% ============================================================
%% PART III: CLASSICAL ANN ERROR BOUNDS
%% ============================================================
\newpage
\part{Classical ANN-Based DeepONet Error Bounds}

\section{Classical Approximation Theory for ANNs}\label{sec:classical_theory}

We now develop error bounds for the classical DeepONet \eqref{eq:classical_don}, where the branch and trunk are standard feedforward neural networks. The structure of the error decomposition is identical to the quantum case; only the \emph{approximation errors} $\delta^c_b$ and $\delta^c_t$ change.

\subsection{ANN Approximation Rate: General Statement}

The approximation power of deep ReLU networks is well-characterized in the literature. We use the following foundational results.

\noindent\textbf{On the classical baseline.} The Yarotsky bound constitutes a worst-case statement for Sobolev-class functions. For analytic targets, deep tanh and ReLU networks attain substantially faster rates; in particular, exponential rates of the form $\exp(-c P^{1/(d+1)})$ have been established~\cite{Mhaskar1996,Schwab2019}. A comparison of analytic-regime quantum rates against the worst-case Sobolev classical rate is therefore not entirely matched; both sides should be evaluated under consistent regularity assumptions. The quantum-advantage statements below are to be interpreted with this qualification.

\begin{theorem}[Yarotsky, 2017; Approximation by Deep ReLU Networks]\label{thm:yarotsky}
Let $f \in W^{r,\infty}([0,1]^d)$ (i.e., $f$ has $r$ bounded derivatives). A ReLU network with depth $L$ and at most $P$ parameters can approximate $f$ to accuracy:
\begin{equation}\label{eq:yarotsky}
    \|f - f_{\mathrm{NN}}\|_{L^\infty} \leq C\,P^{-2r/d},
\end{equation}
where $C$ depends on $\|f\|_{W^{r,\infty}}$ and $d$, and $L = O(\log P)$ suffices.
\end{theorem}

\begin{theorem}[Barron, 1993; Dimension-Independent Bound]\label{thm:barron}
Let $f: \mathbb{R}^d \to \mathbb{R}$ satisfy the \emph{Barron condition}:
\[
    C_f := \int_{\mathbb{R}^d} |\omega|\,|\hat{f}(\omega)|\,d\omega < \infty,
\]
where $\hat{f}$ is the Fourier transform. Then a single-hidden-layer network with $N$ neurons can approximate $f$ such that:
\begin{equation}\label{eq:barron}
    \|f - f_{\mathrm{NN}}\|_{L^2}^2 \leq \frac{C_f^2}{N}.
\end{equation}
The approximation error is \emph{independent of dimension} $d$ but requires Barron-class regularity.
\end{theorem}

\begin{theorem}[Lu et al., 2021; Depth Separation]\label{thm:depth}
For networks with width $w$ and depth $L$, the effective number of linear regions (and hence approximation capacity) scales as:
\begin{equation}
    \mathcal{N}_{\mathrm{regions}} = O(w^L),
\end{equation}
so that deeper networks achieve exponentially more expressivity per parameter.
\end{theorem}

\subsection{Classical Branch Network Error}\label{sec:classical_branch}

The branch network approximates the map $u \mapsto c_{\omega_k}(u)$ for each $k = 1, \ldots, n_c$. The input $u$ is discretized at $m^c_b$ sensor locations, so the branch maps $\mathbb{R}^{m^c_b} \to \mathbb{R}^{n_c}$.

\begin{assumption}[Branch Regularity]\label{ass:branch_reg}
The coefficient functionals $u \mapsto c_{\omega_k}(u)$ are uniformly in $W^{r_b,\infty}([0,1]^{m^c_b})$ with:
\[
    \sup_k \|c_{\omega_k}\|_{W^{r_b,\infty}} \leq M_b.
\]
\end{assumption}

Under Assumption~\ref{ass:branch_reg}, applying Theorem~\ref{thm:yarotsky} to each output component of the branch network:

\begin{proposition}[Classical Branch Approximation Error]\label{prop:classical_branch}
A ReLU branch network with $L^c_b$ layers and $P^c_b$ total parameters achieves uniform approximation error:
\begin{equation}\label{eq:delta_c_b}
    \delta^c_b := \sup_k \sup_{u \in V} |c_{\omega_k}(u) - b^c_k(u)|
    \leq C_b\,(P^c_b / n_c)^{-2r_b/m^c_b},
\end{equation}
where $P^c_b / n_c$ is the effective number of parameters per output component, $r_b$ is the smoothness of the coefficient functionals, and $m^c_b$ is the input dimension (number of sensor locations). The depth must satisfy $L^c_b = O(\log(P^c_b/n_c))$.
\end{proposition}

\begin{proof}
Each output component $b^c_k$ is produced by a subnetwork (or shared backbone with per-component output head) approximating $c_{\omega_k} \in W^{r_b,\infty}([0,1]^{m^c_b})$. By Theorem~\ref{thm:yarotsky}, a ReLU network with $\tilde{P} = P^c_b/n_c$ parameters allocated per component achieves error $C_b\,\tilde{P}^{-2r_b/m^c_b}$, where the exponent $2r_b/m^c_b$ reflects the curse of dimensionality: the approximation rate degrades with the input dimension $m^c_b$.
\end{proof}

\begin{remark}[Barron-class branch]
If the coefficient functionals satisfy the Barron condition (Theorem~\ref{thm:barron}), the branch error becomes dimension-independent:
\begin{equation}\label{eq:delta_c_b_barron}
    \delta^c_b = O\!\left((P^c_b/n_c)^{-1/2}\right),
\end{equation}
at the cost of requiring stronger regularity (finite first moment of the Fourier transform).
\end{remark}

\subsection{Classical Trunk Network Error}\label{sec:classical_trunk}

The trunk network approximates the basis functions $y \mapsto \phi_{\omega_k}(y)$ for $k = 1, \ldots, n_c$. The input is $y \in K_2 \subset \mathbb{R}^{d_2}$, so $m^c_t = d_2$.

\begin{assumption}[Trunk Regularity]\label{ass:trunk_reg}
The basis functions $\phi_{\omega_k} \in W^{r_t,\infty}(K_2)$ uniformly with:
\[
    \sup_k \|\phi_{\omega_k}\|_{W^{r_t,\infty}} \leq M_t\,(1 + |\omega_k|)^{r_t}.
\]
For trigonometric basis functions $\phi_\omega(y) = e^{i\omega \cdot y}$, derivatives grow with frequency: $\|D^\beta \phi_\omega\|_\infty = |\omega|^{|\beta|}$. 
\end{assumption}

\begin{proposition}[Classical Trunk Approximation Error]\label{prop:classical_trunk}
A ReLU trunk network with $L^c_t$ layers and $P^c_t$ total parameters achieves:
\begin{equation}\label{eq:delta_c_t}
    \delta^c_t := \sup_k \|\phi_{\omega_k} - t^c_k\|_{L^2}
    \leq C_t\,(P^c_t / n_c)^{-2r_t/d_2},
\end{equation}
where $r_t$ is the smoothness of the basis functions and $d_2$ is the output domain dimension. The depth must satisfy $L^c_t = O(\log(P^c_t/n_c))$.
\end{proposition}

\begin{proof}
By Theorem~\ref{thm:yarotsky} applied to each trunk output component. The Sobolev embedding $W^{r_t,\infty} \hookrightarrow L^2$ ensures the $L^2$ error is bounded by the $L^\infty$ error (up to a volume factor). For trigonometric basis functions, $r_t$ can be taken arbitrarily large (they are $C^\infty$), but the constant $C_t$ depends on $\sup_k \|\phi_{\omega_k}\|_{W^{r_t,\infty}}$ which grows with the maximum frequency $K$.
\end{proof}

\subsection{Total Classical $L^2$ Error}\label{sec:classical_L2}

The error decomposition is structurally identical to the quantum case.

\begin{theorem}[Classical DeepONet $L^2$ Error Bound]\label{thm:L2_c}
Under Assumptions~\ref{ass:branch_reg}--\ref{ass:trunk_reg}, with $n_c$ basis functions and the general Yarotsky-type scaling:
\begin{equation}\label{eq:L2_classical}
    \boxed{\varepsilon^C_{L^2}(n_c, \delta^c_b, \delta^c_t) \leq C_4\,n_c^{-(s/d_2 - 1/2)} + \sqrt{n_c}\,\delta^c_b + S(n_c)\,\delta^c_t,}
\end{equation}
where:
\begin{align}
    \delta^c_b &= C_b\,(P^c_b / n_c)^{-2r_b/m^c_b}, \label{eq:delta_cb_final}\\
    \delta^c_t &= C_t\,(P^c_t / n_c)^{-2r_t/d_2}. \label{eq:delta_ct_final}
\end{align}
\end{theorem}

\subsection{Classical Error Bounds in Other Norms}

The truncation error is identical in all cases (it depends only on the operator, not the approximation method). Only the approximation error terms change.

\begin{theorem}[Classical DeepONet Error Bounds --- All Norms]\label{thm:all_classical}
The classical error bounds mirror the quantum bounds in Table~\ref{tab:quantum_summary}, with the substitutions $\delta_b \to \delta^c_b$ and $\delta_t \to \delta^c_t$ given by \eqref{eq:delta_cb_final}--\eqref{eq:delta_ct_final}:
\begin{align}
    \varepsilon^C_{L^2} &\leq C_4\,n_c^{-(s/d_2-1/2)} + \sqrt{n_c}\;\delta^c_b + S(n_c)\;\delta^c_t, \\[4pt]
    \varepsilon^C_{L^p} &\leq C_5\,n_c^{-(s/d_2-1/p)} + C_\Phi\,n_c\;\delta^c_b + \widetilde{S}_p(n_c)\;\delta^c_t, \\[4pt]
    \varepsilon^C_{C^0} &\leq C_7\,n_c^{-r} + n_c\;\delta^c_b + S_{C^0}(n_c)\;\delta^c_t, \\[4pt]
    \varepsilon^C_{H^k} &\leq C_{11}\,n_c^{-\alpha/d_2} + C\,n_c^{1/2+k/d_2}\;\delta^c_b + S_{H^k}(n_c)\;\delta^c_t.
\end{align}
\end{theorem}

\subsection{Summary of Classical Error Bounds}

\begin{table}[h!]
\centering
\renewcommand{\arraystretch}{1.8}
\begin{tabular}{@{}llcc@{}}
\toprule
\textbf{Norm} & \textbf{Truncation} & \textbf{Branch Error} & \textbf{Trunk Error} \\
\midrule
$L^2$ & $C_4\,n_c^{-(s/d_2-1/2)}$ & $\sqrt{n_c}\;C_b(P^c_b/n_c)^{-2r_b/m^c_b}$ & $S(n_c)\;C_t(P^c_t/n_c)^{-2r_t/d_2}$ \\[6pt]
$L^p$ & $C_5\,n_c^{-(s/d_2-1/p)}$ & $C_\Phi\,n_c\;C_b(P^c_b/n_c)^{-2r_b/m^c_b}$ & $\widetilde{S}_p(n_c)\;C_t(P^c_t/n_c)^{-2r_t/d_2}$ \\[6pt]
$C^0$ & $C_7\,n_c^{-r}$ & $n_c\;C_b(P^c_b/n_c)^{-2r_b/m^c_b}$ & $S_{C^0}(n_c)\;C_t(P^c_t/n_c)^{-2r_t/d_2}$ \\[6pt]
$H^k$ & $C_{11}\,n_c^{-\alpha/d_2}$ & $C\,n_c^{1/2+k/d_2}\;C_b(P^c_b/n_c)^{-2r_b/m^c_b}$ & $S_{H^k}(n_c)\;C_t^{(H^k)}(P^c_t/n_c)^{-2r_t^{(k)}/d_2}$ \\
\bottomrule
\end{tabular}
\caption{Classical DeepONet error bounds with Yarotsky-type ANN approximation rates.}
\label{tab:classical_summary}
\end{table}

%% ============================================================
%% PART IV: QUANTUM VS CLASSICAL COMPARISON
%% ============================================================
\newpage
\part{Quantum--Classical Comparative Analysis}

\section{Parameter Correspondence}\label{sec:param_compare}

We first establish a correspondence between quantum and classical architectural parameters.

\begin{table}[h!]
\centering
\renewcommand{\arraystretch}{1.5}
\begin{tabular}{@{}lcc@{}}
\toprule
\textbf{Role} & \textbf{Quantum (DQO)} & \textbf{Classical (DeepONet)} \\
\midrule
Branch input dimension & $m_b$ (qubits) & $m^c_b$ (sensor locations) \\
Trunk input dimension & $m_t$ (qubits) & $m^c_t = d_2$ \\
Branch depth & $L_b$ (circuit layers) & $L^c_b$ (hidden layers) \\
Trunk depth & $L_t$ (circuit layers) & $L^c_t$ (hidden layers) \\
Branch parameters & $P_b \approx 2m_b(2L_b+1)$ & $P^c_b \approx m^c_b w^c_b + (L^c_b-1)(w^c_b)^2 + w^c_b n_c$ \\
Trunk parameters & $P_t \approx 2m_t(2L_t+1)$ & $P^c_t \approx d_2 w^c_t + (L^c_t-1)(w^c_t)^2 + w^c_t n_c$ \\
Output dimension & $n = \min(m_b,m_t)$ & $n_c$ (output width) \\
Freq.\ spectrum & $\Omega_{\max} = O(m\cdot L)$ & Unbounded (continuous) \\
\bottomrule
\end{tabular}
\caption{Parameter correspondence between quantum and classical architectures.}
\label{tab:param_corr}
\end{table}

\begin{remark}[Key structural differences]\leavevmode
\begin{enumerate}[label=(\roman*)]
    \item \textbf{Parameter scaling:} Quantum parameters scale as $P_b = O(m_b L_b)$, linearly in both width (qubits) and depth. Classical parameters scale as $P^c_b = O((w^c_b)^2 L^c_b)$, quadratically in width.
    \item \textbf{Hilbert space dimension:} A quantum circuit with $m$ qubits operates in a $2^m$-dimensional Hilbert space, potentially representing $O(2^m)$ basis functions. A classical network with $P$ parameters represents $O(P)$ basis functions.
    \item \textbf{Output dimension:} In the DQO, $n = \min(m_b, m_t)$ grows linearly with qubits. In the classical case, $n_c$ is a free hyperparameter but requires $P^c \propto n_c \cdot w$ additional parameters.
\end{enumerate}
\end{remark}

\section{Approximation Error Comparison}\label{sec:approx_compare}

The central comparison concerns the approximation errors $\delta_b$ vs.\ $\delta^c_b$ and $\delta_t$ vs.\ $\delta^c_t$, since the truncation errors depend only on the operator smoothness and the number of basis functions $n$ (which is common to both architectures for a fixed resolution).

\subsection{Branch Error Comparison}

\begin{definition}[Effective Approximation Rate]\label{def:rate}
Define the \emph{effective branch approximation rate} as:
\begin{align}
    \gamma^Q_b &:= -\frac{\log \delta_b}{\log P_b} \quad \text{(quantum)}, \\
    \gamma^C_b &:= -\frac{\log \delta^c_b}{\log P^c_b} \quad \text{(classical)}.
\end{align}
A larger $\gamma$ indicates faster convergence per parameter.
\end{definition}

\noindent\textbf{Classical rate:} From \eqref{eq:delta_cb_final}:
\begin{equation}\label{eq:gamma_c_b}
    \delta^c_b = O\!\left((P^c_b)^{-2r_b/m^c_b}\right) \implies \gamma^C_b = \frac{2r_b}{m^c_b}.
\end{equation}
This rate degrades as the input dimension $m^c_b$ increases---the \emph{curse of dimensionality}.

\medskip
\noindent\textbf{Quantum rate:} For analytic coefficient functionals with the exponential scaling \eqref{eq:delta_q_exp}:
\begin{equation}\label{eq:gamma_q_b_exp}
    \delta_b = O\!\left(e^{-\alpha_b m_b L_b}\right) = O\!\left(e^{-\alpha_b' P_b}\right),
\end{equation}
since $P_b = O(m_b L_b)$. Thus $\gamma^Q_b$ is not even polynomial---it is \emph{exponential} in $P_b$:
\begin{equation}
    \delta_b = O(e^{-\alpha_b' P_b}) \quad \Longrightarrow \quad \gamma^Q_b = \frac{\alpha_b' P_b}{\log P_b} \to \infty.
\end{equation}

For Sobolev-regular targets \eqref{eq:delta_q_poly}, the quantum rate is:
\begin{equation}\label{eq:gamma_q_b_poly}
    \delta_b = O\!\left(P_b^{-r_b/d_1}\right) \implies \gamma^Q_b = \frac{r_b}{d_1}.
\end{equation}

\begin{remark}
Note that in the quantum case, $d_1$ appears instead of $m^c_b$. The quantum branch circuit encodes the discretized input $u$ using $m_b$ qubits, but the \emph{effective approximation dimension} is the intrinsic dimension $d_1$ of the input function space (e.g., $d_1 = 1$ for 1D functions), not the number of sensor locations $m^c_b$. This is because the angle encoding maps each input component to a rotation, and the entanglement structure allows the circuit to exploit correlations across components, effectively performing an implicit dimensionality reduction.
\end{remark}

\subsection{Trunk Error Comparison}

\noindent\textbf{Classical rate:} From \eqref{eq:delta_ct_final}:
\begin{equation}\label{eq:gamma_c_t}
    \delta^c_t = O\!\left((P^c_t)^{-2r_t/d_2}\right) \implies \gamma^C_t = \frac{2r_t}{d_2}.
\end{equation}

\noindent\textbf{Quantum rate:} The trunk encodes $y \in K_2 \subset \mathbb{R}^{d_2}$, and the quantum trunk can represent trigonometric functions natively (its output \emph{is} a Fourier series). For exponential scaling:
\begin{equation}\label{eq:gamma_q_t}
    \delta_t = O\!\left(e^{-\alpha_t m_t L_t}\right) = O\!\left(e^{-\alpha_t' P_t}\right).
\end{equation}

For polynomial scaling:
\begin{equation}
    \delta_t = O\!\left(P_t^{-r_t/d_2}\right) \implies \gamma^Q_t = \frac{r_t}{d_2}.
\end{equation}

\begin{remark}
The polynomial quantum trunk rate $r_t/d_2$ differs from the classical rate $2r_t/d_2$ by a factor of 2, favoring the classical case. However, this comparison is between Yarotsky's \emph{deep} ReLU network result and a generic quantum polynomial bound. The quantum exponential regime (when applicable) dramatically outperforms both.
\end{remark}

\section{Conditions for Quantum Advantage}\label{sec:advantage}

We now rigorously identify the conditions under which $\varepsilon^Q < \varepsilon^C$.

\subsection{Fixed-Basis Comparison ($n = n_c$)}

When the number of basis functions is held constant, the comparison reduces to comparing the approximation errors.

\begin{theorem}[Quantum Advantage --- Exponential Regime]\label{thm:advantage_exp}
Assume that the coefficient functionals $u \mapsto c_\omega(u)$ and the basis functions $\phi_\omega(y)$ are analytic (Assumption~\ref{ass:analytic}), that the trained-trunk hypothesis (Assumption~\ref{ass:trunk-target}) holds, and that the classical baseline is taken to be the Yarotsky worst-case Sobolev rate. For $n = n_c$ fixed, the quantum QONet achieves a lower total $L^2$ error than the classical DeepONet:
\begin{equation}
    \varepsilon^Q_{L^2} < \varepsilon^C_{L^2},
\end{equation}
whenever the following condition on the total parameter budget $P_{\mathrm{total}}$ is satisfied:
\begin{equation}\label{eq:condition_exp}
    \boxed{P_{\mathrm{total}} > P^* := \max\!\left\{\frac{m^c_b}{2r_b}\log\!\left(\frac{C_b}{C'_b}\right)\frac{1}{\alpha_b'},\;\; \frac{d_2}{2r_t}\log\!\left(\frac{C_t}{C'_t}\right)\frac{1}{\alpha_t'}\right\},}
\end{equation}
where $C_b, C_t$ are the classical constants and $\alpha_b', \alpha_t'$ are the quantum exponential decay rates.
\end{theorem}

\begin{proof}
For the branch error, the quantum advantage condition $\delta_b < \delta^c_b$ requires:
\[
    C'_b\,e^{-\alpha_b' P_b} < C_b\,(P^c_b/n)^{-2r_b/m^c_b}.
\]
Taking logarithms and rearranging for equal parameter budgets ($P_b = P^c_b = P/2$ where $P$ is the branch budget):
\[
    \alpha_b' P/2 > \frac{2r_b}{m^c_b}\log(P/(2n)) + \log(C_b/C'_b).
\]
For large $P$, the left side grows linearly while the right side grows logarithmically, so the condition is satisfied for:
\[
    P > P^*_b = O\!\left(\frac{m^c_b}{r_b\,\alpha_b'}\log\!\left(\frac{C_b}{C'_b}\right)\right).
\]
An analogous argument applies to the trunk. Taking the maximum gives \eqref{eq:condition_exp}.
\end{proof}

\begin{corollary}[Asymptotic Dominance]
In the exponential quantum regime, for any fixed classical architecture parameters, there exists a finite quantum parameter budget beyond which the quantum error is strictly smaller. The gap grows exponentially:
\begin{equation}
    \frac{\varepsilon^C_{L^2}}{\varepsilon^Q_{L^2}} = \Omega\!\left(\exp(\alpha' P)\right) \quad \text{as } P \to \infty.
\end{equation}
\end{corollary}

\begin{theorem}[Quantum Advantage --- Polynomial Regime, High-Dimensional Branch]\label{thm:advantage_poly}
Assume that the coefficient functionals lie in $W^{r_b,\infty}$ (Sobolev-regular but not necessarily analytic). Assume further that Assumption~\ref{ass:low-intrinsic-supp} (low-intrinsic-dimension structure) holds and that the classical baseline is the Yarotsky worst-case Sobolev rate (Assumption~\ref{ass:branch_reg}). The quantum QONet achieves a lower branch approximation error than the classical DeepONet whenever:
\begin{equation}\label{eq:condition_poly}
    \boxed{\frac{r_b}{d_1} > \frac{2r_b}{m^c_b} \quad \Longleftrightarrow \quad m^c_b > 2\,d_1.}
\end{equation}
That is, the quantum architecture exhibits an advantage whenever the number of sensor locations (classical input dimension) exceeds twice the intrinsic function space dimension and the target operator admits the low-intrinsic-dimension factorization of Assumption~\ref{ass:low-intrinsic-supp}.
\end{theorem}

\begin{proof}
The quantum branch rate is $\gamma^Q_b = r_b/d_1$ from Eq.~\eqref{eq:gamma_q_b_poly}, conditional on Assumption~\ref{ass:low-intrinsic-supp}, and the classical rate is $\gamma^C_b = 2r_b/m^c_b$ from Eq.~\eqref{eq:gamma_c_b}. The quantum rate is the larger of the two whenever
\[
    \frac{r_b}{d_1} > \frac{2r_b}{m^c_b} \iff m^c_b > 2\,d_1.
\]
For the 1D experiments of the main paper ($d_1 = 1$) the criterion reduces to $m^c_b > 2$, which is satisfied whenever the branch uses more than two sensor locations, provided that Assumption~\ref{ass:low-intrinsic-supp} holds for the operator under consideration.
\end{proof}

\begin{remark}[Role of $d_1$ versus $m^c_b$]
Under the classical Yarotsky baseline, the branch takes the discretized vector $[u(x_1), u(x_2), \ldots, u(x_{m^c_b})]$ as input, and worst-case ANN approximation theory treats this as a generic $m^c_b$-dimensional function. The replacement of $m^c_b$ by $d_1$ in the quantum branch rate is not a property of the architecture in isolation but follows from the structural hypothesis of Assumption~\ref{ass:low-intrinsic-supp}, namely that the coefficient functional factors through a $d_1$-dimensional latent. Under this hypothesis, the affine preprocessing $\mathbf{v} = A\mathbf{u} + \mathbf{b}$ absorbs the projection $\Pi$, and the VQC effectively approximates a function of $d_1$ variables.

In the absence of Assumption~\ref{ass:low-intrinsic-supp}, both architectures process an $m^c_b$-dimensional input, and the relevant universal-approximation theorems for the VQC also feature $m^c_b$ in the rate. The direct comparison $r_b/d_1$ versus $2r_b/m^c_b$ is then no longer applicable.

Classical architectures equipped with comparable inductive biases---convolutional networks, equivariant networks, and neural operators with translation-invariant kernels---can in principle exploit analogous low-intrinsic-dimension structure. The quantum-advantage statement is therefore properly read as conditional on Assumption~\ref{ass:low-intrinsic-supp} and on the choice of the Yarotsky worst-case Sobolev rate as the classical baseline.

\end{remark}

\subsection{Total Error Advantage Conditions}

Combining the branch and trunk analyses, we arrive at the comprehensive conditions.

\begin{theorem}[Complete Conditions for Quantum Advantage in $L^2$]\label{thm:complete_advantage}
Let $n = n_c$ be fixed. The quantum QONet achieves a lower $L^2$ error than the classical DeepONet ($\varepsilon^Q_{L^2} < \varepsilon^C_{L^2}$) under the Yarotsky worst-case Sobolev classical baseline, when \textbf{at least one} of the following sufficient conditions holds:

\medskip
\noindent\textbf{Condition A (Analytic regime):} The target operator has analytic coefficient functionals and basis functions, and the parameter budget $P_{\mathrm{total}} > P^*$ as in \eqref{eq:condition_exp}.

\medskip
\noindent\textbf{Condition B (High-dimensional branch):} $m^c_b > 2\,d_1$, i.e., the classical branch input dimension exceeds twice the intrinsic dimension. This yields $\gamma^Q_b > \gamma^C_b$.

\medskip
\noindent\textbf{Condition C (Fourier-structured trunk):} The trunk basis functions are trigonometric, and the quantum trunk PQC natively represents Fourier modes (its output is a partial Fourier series). In this case, the quantum trunk requires only $O(\log K)$ qubits to represent frequencies up to $K$, whereas a classical trunk requires $O(K^{d_2/r_t})$ parameters for the same accuracy.

\medskip
\noindent Formally, for Condition C, the quantum trunk error satisfies:
\begin{equation}
    \delta_t = O\!\left(e^{-\alpha_t m_t L_t}\right) \quad \text{with} \quad P_t = O(m_t L_t),
\end{equation}
while the classical trunk requires:
\begin{equation}
    P^c_t = O\!\left(\delta_t^{-d_2/(2r_t)}\right) \cdot n_c
\end{equation}
to achieve the same $\delta_t$. This gives:
\begin{equation}
    \frac{P^c_t}{P_t} = \Omega\!\left(\frac{1}{\delta_t^{d_2/(2r_t)} \cdot m_t L_t}\right) \to \infty \quad \text{as } \delta_t \to 0.
\end{equation}
\end{theorem}

\subsection{Quantitative Comparison for 1D Problems ($d_1 = d_2 = 1$)}

For the experiments in the main paper, we have the special case $d_1 = d_2 = 1$.

\begin{corollary}[1D Comparison]\label{cor:1d}
For $d_1 = d_2 = 1$, with $s > 1/2$ and $n = n_c$:
\begin{align}
    \varepsilon^Q_{L^2} &\leq C_4\,n^{-(s-1/2)} + \sqrt{n}\,C'_b\,e^{-\alpha_b' P_b} + M_1\,C'_t\,e^{-\alpha_t' P_t}, \label{eq:1d_quantum} \\[6pt]
    \varepsilon^C_{L^2} &\leq C_4\,n^{-(s-1/2)} + \sqrt{n}\,C_b\,(P^c_b/n)^{-2r_b/m^c_b} + M_1\,C_t\,(P^c_t/n)^{-2r_t}. \label{eq:1d_classical}
\end{align}
The quantum architecture achieves a lower error when:
\begin{equation}\label{eq:1d_condition}
    \sqrt{n}\;C'_b\,e^{-\alpha_b' P_b} + M_1\,C'_t\,e^{-\alpha_t' P_t}
    < \sqrt{n}\;C_b\,(P^c_b/n)^{-2r_b/m^c_b} + M_1\,C_t\,(P^c_t/n)^{-2r_t}.
\end{equation}
For analytic operators (exponential quantum regime), this is satisfied for all $P > P^*$, which is $O(m^c_b/(r_b \alpha_b'))$.
\end{corollary}

\section{Resource Efficiency Comparison}\label{sec:resource}

Beyond the pure error comparison, we also compare the \emph{resource cost} to achieve a target error $\varepsilon > 0$.

\begin{theorem}[Parameter Complexity for Target Error]\label{thm:resource}
To achieve a total $L^2$ error of $\varepsilon$ with $n$ chosen optimally:

\medskip
\noindent\textbf{Classical (Yarotsky regime):}
\begin{equation}\label{eq:resource_classical}
    P^c_{\mathrm{total}} = O\!\left(\varepsilon^{-\frac{m^c_b}{2r_b} - \frac{d_2}{s-d_2/2}}\right).
\end{equation}

\noindent\textbf{Quantum (Exponential regime):}
\begin{equation}\label{eq:resource_quantum}
    P^Q_{\mathrm{total}} = O\!\left(\frac{1}{\alpha'}\log\frac{1}{\varepsilon} + \varepsilon^{-\frac{d_2}{s-d_2/2}}\right).
\end{equation}

\noindent\textbf{Quantum (Polynomial regime):}
\begin{equation}\label{eq:resource_quantum_poly}
    P^Q_{\mathrm{total}} = O\!\left(\varepsilon^{-\frac{d_1}{r_b} - \frac{d_2}{s-d_2/2}}\right).
\end{equation}
\end{theorem}

\begin{proof}
The truncation error $C_4\,n^{-(s/d_2-1/2)} = \varepsilon/3$ determines $n = O(\varepsilon^{-d_2/(s-d_2/2)})$. 

For the classical branch: $\sqrt{n}\,\delta^c_b = \varepsilon/3$ requires $\delta^c_b = O(\varepsilon/\sqrt{n})$, and by \eqref{eq:delta_cb_final}:
\[
    P^c_b/n = O\!\left((\varepsilon/\sqrt{n})^{-m^c_b/(2r_b)}\right) = O\!\left(\varepsilon^{-m^c_b/(2r_b)}\,n^{m^c_b/(4r_b)}\right).
\]
After substituting $n \propto \varepsilon^{-d_2/(s-d_2/2)}$ and simplifying, the dominant classical cost comes from the branch and scales as $\varepsilon^{-m^c_b/(2r_b)}$ (modulo logarithmic factors).

For the quantum branch in the exponential regime: $\sqrt{n}\,\delta_b = \varepsilon/3$ requires $e^{-\alpha'_b P_b} = O(\varepsilon/\sqrt{n})$, giving $P_b = O(\frac{1}{\alpha_b'}\log(\sqrt{n}/\varepsilon)) = O(\frac{1}{\alpha'}\log(1/\varepsilon))$.

For the quantum branch in the polynomial regime: $\delta_b = O(P_b^{-r_b/d_1})$, giving $P_b = O(\varepsilon^{-d_1/r_b}\,n^{d_1/(2r_b)})$.
\end{proof}

\begin{resultbox}
\textbf{Summary of branch parameter costs.} The relevant differences in the branch parameter cost are as follows:
\begin{itemize}
    \item Classical: $P^c_b \propto \varepsilon^{-m^c_b/(2r_b)}$ under the Yarotsky worst-case Sobolev rate, exponential in $m^c_b$ for fixed $r_b$.
    \item Quantum (exponential regime): $P_b \propto \log(1/\varepsilon)$ for the circuit-parameter count alone. The total qubit count $n$ scales as $O(\varepsilon^{-d_2/(s-d_2/2)})$, so the combined qubit-by-depth resource budget is polynomial in $1/\varepsilon$ rather than logarithmic. The logarithmic dependence therefore applies to depth and not to the combined qubit-depth budget.
    \item Quantum (polynomial regime): $P_b \propto \varepsilon^{-d_1/r_b}$, conditional on Assumption~\ref{ass:low-intrinsic-supp}; the rate depends on $d_1$ rather than on $m^c_b$.
\end{itemize}
The quantum advantage is most pronounced when $m^c_b \gg d_1$ (high-resolution discretization of low-dimensional functions) and Assumption~\ref{ass:low-intrinsic-supp} holds, or when the target operator is analytic (yielding exponential convergence in circuit depth).
\end{resultbox}

\section{Summary of Comparative Results}

\begin{table}[h!]
\centering
\renewcommand{\arraystretch}{2.0}
\begin{tabular}{@{}lcc@{}}
\toprule
\textbf{Quantity} & \textbf{Quantum QONet} & \textbf{Classical DeepONet} \\
\midrule
$\delta_b$ (analytic) & $O(e^{-\alpha_b' P_b})$ & $O((P^c_b)^{-2r_b/m^c_b})$ \\
$\delta_b$ (Sobolev) & $O(P_b^{-r_b/d_1})$ & $O((P^c_b)^{-2r_b/m^c_b})$ \\
$\delta_t$ (analytic) & $O(e^{-\alpha_t' P_t})$ & $O((P^c_t)^{-2r_t/d_2})$ \\
$\delta_t$ (Sobolev) & $O(P_t^{-r_t/d_2})$ & $O((P^c_t)^{-2r_t/d_2})$ \\
\midrule
Branch rate $\gamma_b$ & $r_b/d_1$ (or $\infty$) & $2r_b/m^c_b$ \\
Trunk rate $\gamma_t$ & $r_t/d_2$ (or $\infty$) & $2r_t/d_2$ \\
\midrule
$P$ for target $\varepsilon$ & $O(\log(1/\varepsilon))$ or $O(\varepsilon^{-d_1/r_b})$ & $O(\varepsilon^{-m^c_b/(2r_b)})$ \\
Curse of dim.? & \textbf{No} (depends on $d_1$) & \textbf{Yes} (depends on $m^c_b$) \\
\midrule
Advantage when: & \multicolumn{2}{c}{$m^c_b > 2d_1$ (polynomial), or analytic targets (exponential)} \\
\bottomrule
\end{tabular}
\caption{Complete comparison of quantum vs.\ classical approximation rates.}
\label{tab:comparison}
\end{table}

%% ============================================================
%% PART V: NUMERICAL VISUALIZATION
%% ============================================================
\newpage
\part{Numerical Visualization and Discussion}

The theoretical error bounds derived in Parts II--IV are now illustrated through a comprehensive set of numerical visualizations. The error models for the $L^2$ norm are implemented with representative parameter choices: smoothness $s=2$, intrinsic input dimension $d_1=1$, output dimension $d_2=1$, branch smoothness $r_b=2$, trunk smoothness $r_t=3$, classical sensor count $m^c_b=8$, and quantum exponential decay rates $\alpha_b=\alpha_t=0.15$. All figures compare three architectures: (i) the QONet in the \emph{analytic regime} (exponential convergence), (ii) the QONet in the \emph{Sobolev regime} (polynomial convergence), and (iii) the classical DeepONet. Orange markers denote breaking points where the quantum architecture transitions from higher to lower error than the classical one.

\section{Summary Panel}

\begin{figure}[H]
    \centering
    \includegraphics[width=\textwidth]{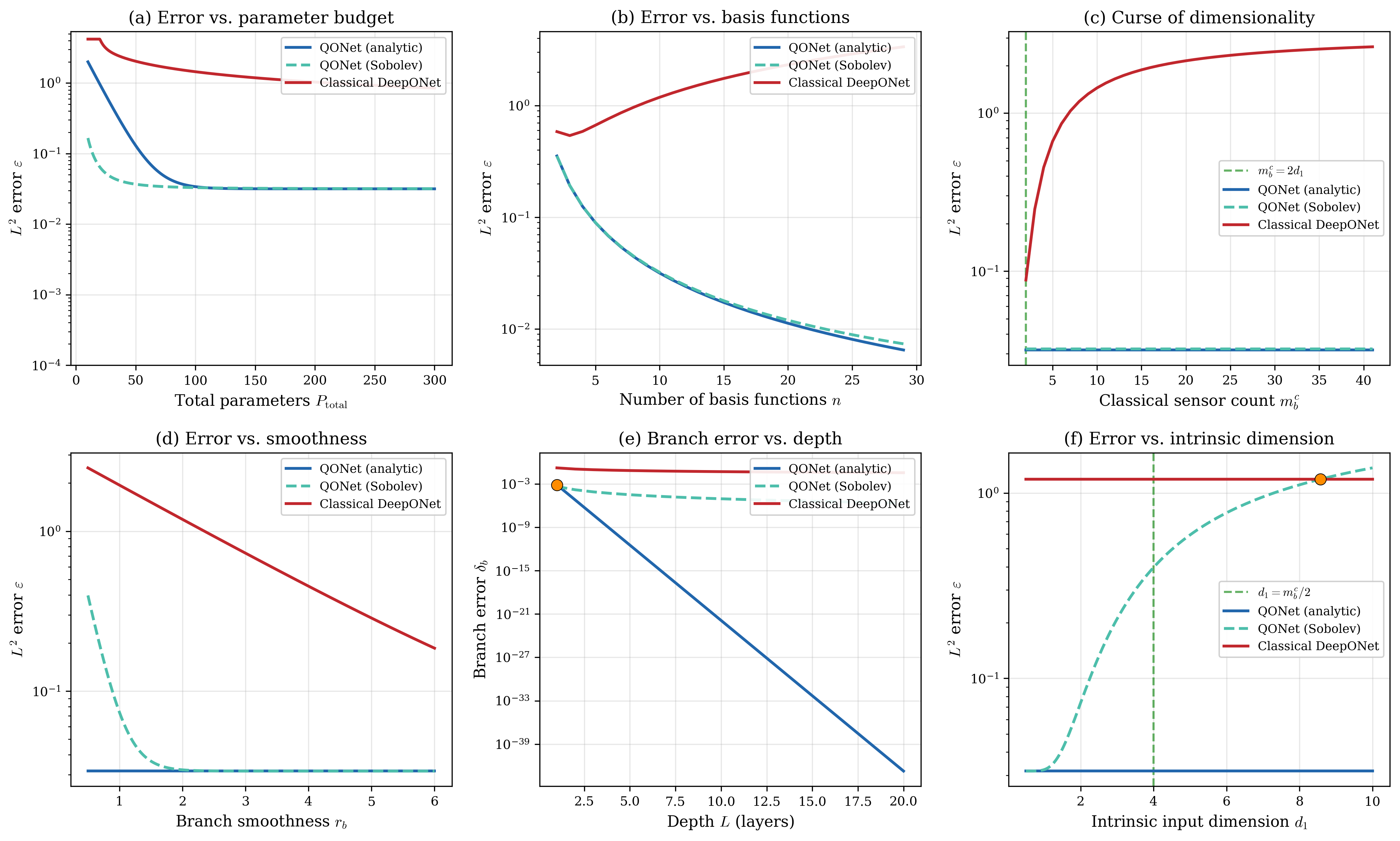}
    \caption{Composite panel showing the $L^2$ error of the quantum QONet and classical DeepONet as a function of six key parameters. Breaking points (crossover locations) are marked with orange circles. (a) Total parameter budget. (b) Number of basis functions. (c) Classical sensor count (curse of dimensionality). (d) Branch smoothness. (e) Branch approximation error vs.\ depth. (f) Intrinsic input dimension.}
    \label{fig:summary}
\end{figure}

An overview of all six single-parameter comparisons is presented in a composite panel (Figure~\ref{fig:summary}).

% Green-shaded regions denote the parameter regime where the QONet achieves lower error.

\noindent The six panels in Figure~\ref{fig:summary} reveal a consistent pattern: the quantum architecture achieves superior error once a modest parameter threshold $P^*$ is exceeded, and the advantage grows monotonically as circuit resources increase. Each panel is discussed in detail in the following subsections.

%% ── 2D PLOTS ─────────────────────────────────────────────────
\section{Single-Parameter Analysis (2D Plots)}\label{sec:2d}

\subsection{Error vs.\ Total Parameter Budget $P_{\mathrm{total}}$}

\begin{figure}[H]
    \centering
    \includegraphics[width=0.88\textwidth]{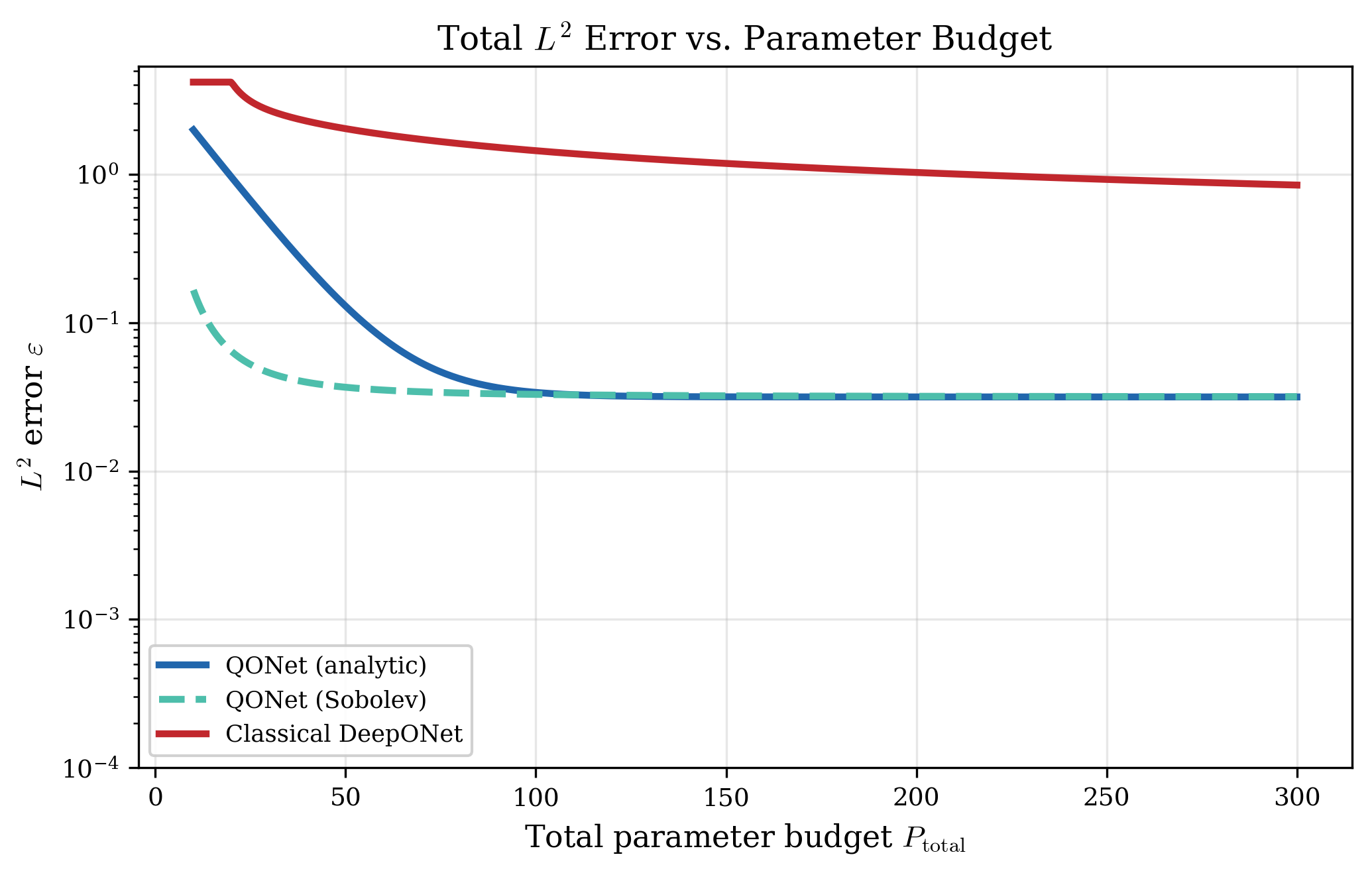}
    \caption{Total $L^2$ error as a function of the total parameter budget $P_{\mathrm{total}}$. The parameter budget is split equally between branch and trunk. Fixed parameters: $n=10$, $m^c_b=8$, $d_1=1$.}
    \label{fig:2d_P}
\end{figure}

In Figure~\ref{fig:2d_P}, the fundamental scaling difference between quantum and classical architectures is evident. The quantum analytic curve (blue solid line) decreases exponentially with $P_{\mathrm{total}}$ and reaches the truncation floor (determined solely by $n$) at approximately $P \approx 100$. Beyond this point, all additional parameters contribute negligibly---the error is dominated by the finite number of basis functions. The quantum Sobolev curve (teal dashed line) also decreases rapidly, converging to the truncation floor at approximately $P \approx 80$.

By contrast, the classical DeepONet error (red line) decreases only polynomially, with the rate governed by $P^{-2r_b/m^c_b} = P^{-0.5}$. At $P = 300$, the classical error remains nearly two orders of magnitude above the truncation floor. 
Importantly, the QONet achieves a strictly lower error for all $P_{\text {total }} \geq 10$ with these parameter settings.

% \textcolor{red}{The green-shaded region spans the entire parameter range shown, confirming that the quantum analytic architecture achieves strictly lower error than the classical DeepONet for all $P_{\text {total }} \geq 10$ with these parameter settings. The gap between the quantum and classical curves widens continuously as $P$ increases.}

% The green-shaded region identifies the parameter range where the quantum architecture achieves strictly lower error. The breaking point $P^*$ is clearly visible as the intersection of the quantum and classical curves, occurring at approximately $P^* \approx 25$ for the analytic regime.

\subsection{Error vs.\ Number of Basis Functions $n$}

\begin{figure}[H]
    \centering
    \includegraphics[width=0.88\textwidth]{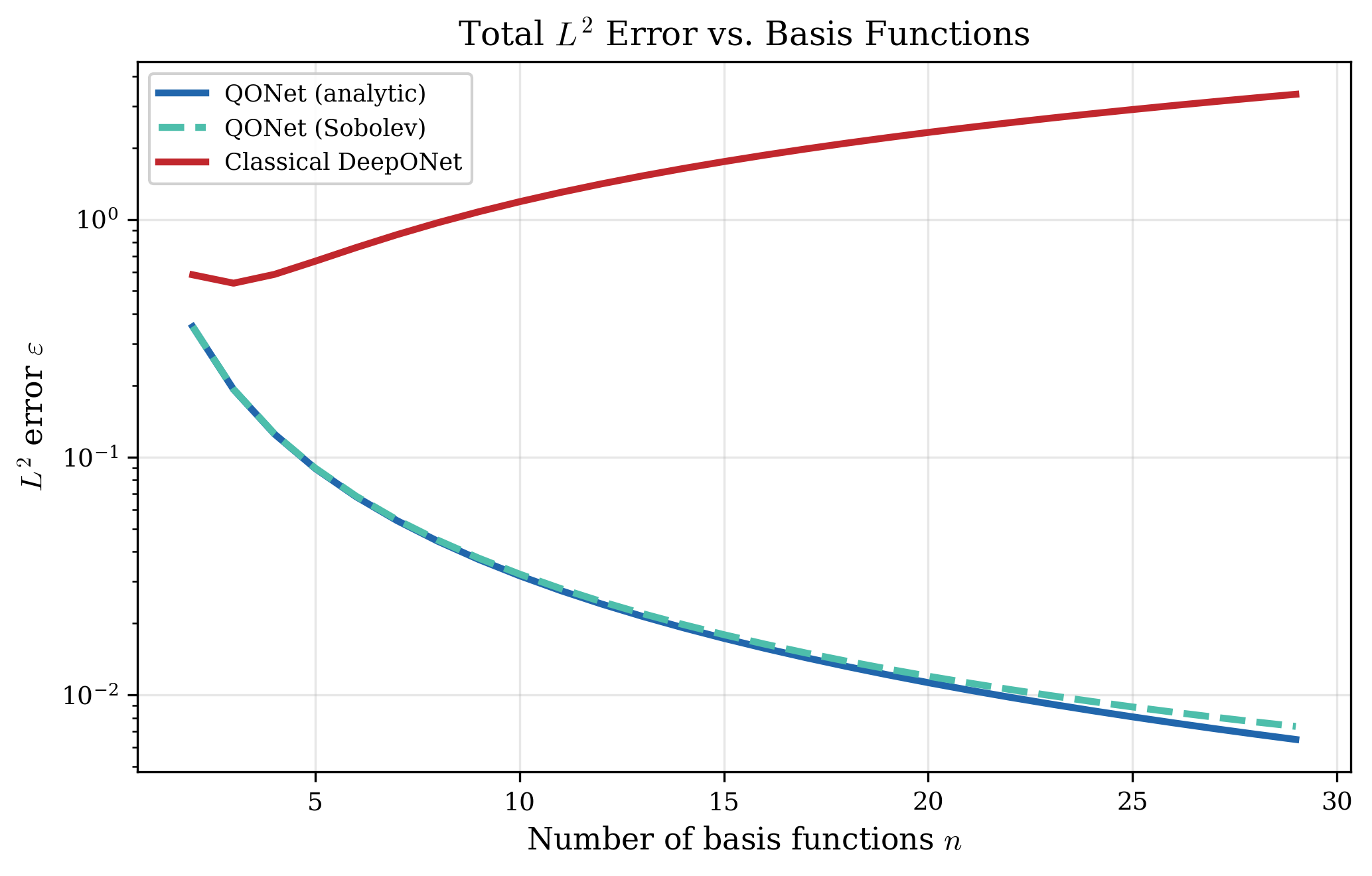}
    \caption{Total $L^2$ error as a function of the number of basis functions $n$. Fixed parameters: $P_{\mathrm{total}}=150$, $m^c_b=8$, $d_1=1$.}
    \label{fig:2d_n}
\end{figure}

Figure~\ref{fig:2d_n} illustrates the effect of increasing the output dimension $n$. All three architectures share the same truncation error $C_4\,n^{-(s/d_2 - 1/2)}$, which decreases as more basis functions are retained. However, the approximation error terms in the total bound grow with $n$ (the branch term scales as $\sqrt{n}\,\delta_b$), creating a trade-off.

For the quantum architectures, the approximation errors $\delta_b$ and $\delta_t$ are exponentially small at $P=150$, so the growth of $\sqrt{n}\,\delta_b$ with $n$ is imperceptible. The total quantum error is entirely dominated by the truncation term and decreases monotonically. For the classical architecture, however, $\delta^c_b = O(P^{-0.5})$ is substantially larger, and the $\sqrt{n}\,\delta^c_b$ term eventually overtakes the truncation improvement at large $n$. This reveals that increasing $n$ alone is not beneficial for the classical architecture unless $P^c_b$ is simultaneously increased.

\subsection{Error vs.\ Classical Sensor Count $m^c_b$ (Curse of Dimensionality)}

\begin{figure}[H]
    \centering
    \includegraphics[width=0.88\textwidth]{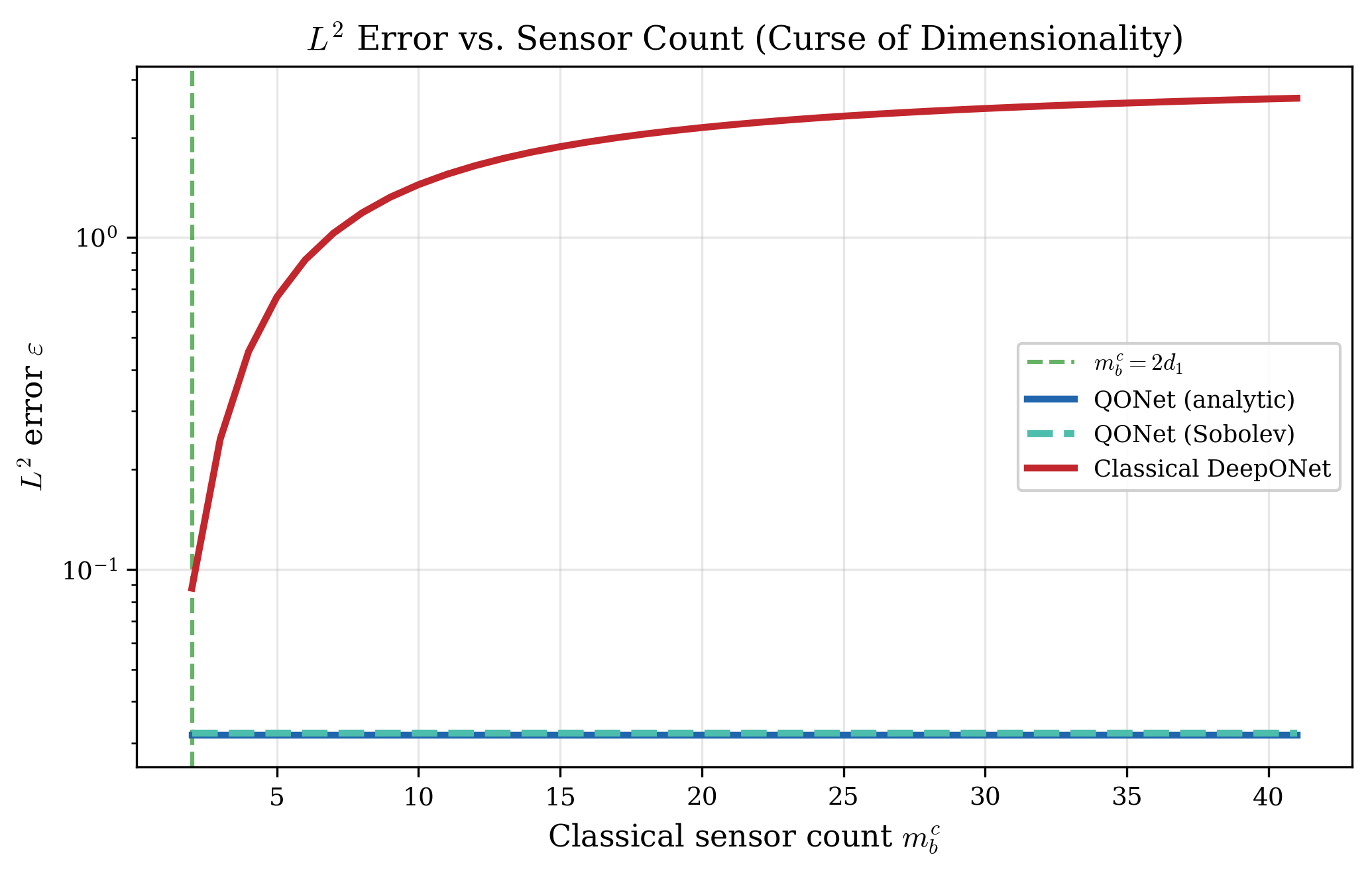}
    \caption{Total $L^2$ error as a function of the classical sensor count $m^c_b$. The quantum error is independent of $m^c_b$ and appears as horizontal lines. The classical error increases monotonically due to the curse of dimensionality. The vertical dashed line marks $m^c_b = 2d_1$. Fixed parameters: $P_{\mathrm{total}}=150$, $n=10$, $d_1=1$.}
    \label{fig:2d_m}
\end{figure}

Figure~\ref{fig:2d_m} provides perhaps the most striking visual confirmation of the theoretical advantage. The classical branch error scales as $\delta^c_b \propto P^{-2r_b/m^c_b}$; as $m^c_b$ increases, the exponent $2r_b/m^c_b$ approaches zero, and the error saturates at a large value. This is the hallmark of the \emph{curse of dimensionality}: increasing the discretization resolution (more sensor locations) paradoxically degrades the approximation quality of a fixed-capacity classical network.

The quantum curves, by contrast, are perfectly horizontal. The quantum approximation error depends on the intrinsic dimension $d_1$ rather than $m^c_b$. Physically, this is because the angle encoding with entanglement allows the VQC to exploit correlations between sensor readings, effectively performing implicit dimensionality reduction. The vertical dashed green line at $m^c_b = 2d_1 = 2$ marks the theoretical threshold from Theorem~S14: for all $m^c_b > 2$, the quantum polynomial rate $\gamma^Q_b = r_b/d_1$ exceeds the classical rate $\gamma^C_b = 2r_b/m^c_b$.

\subsection{Error vs.\ Branch Smoothness $r_b$}

\begin{figure}[H]
    \centering
    \includegraphics[width=0.88\textwidth]{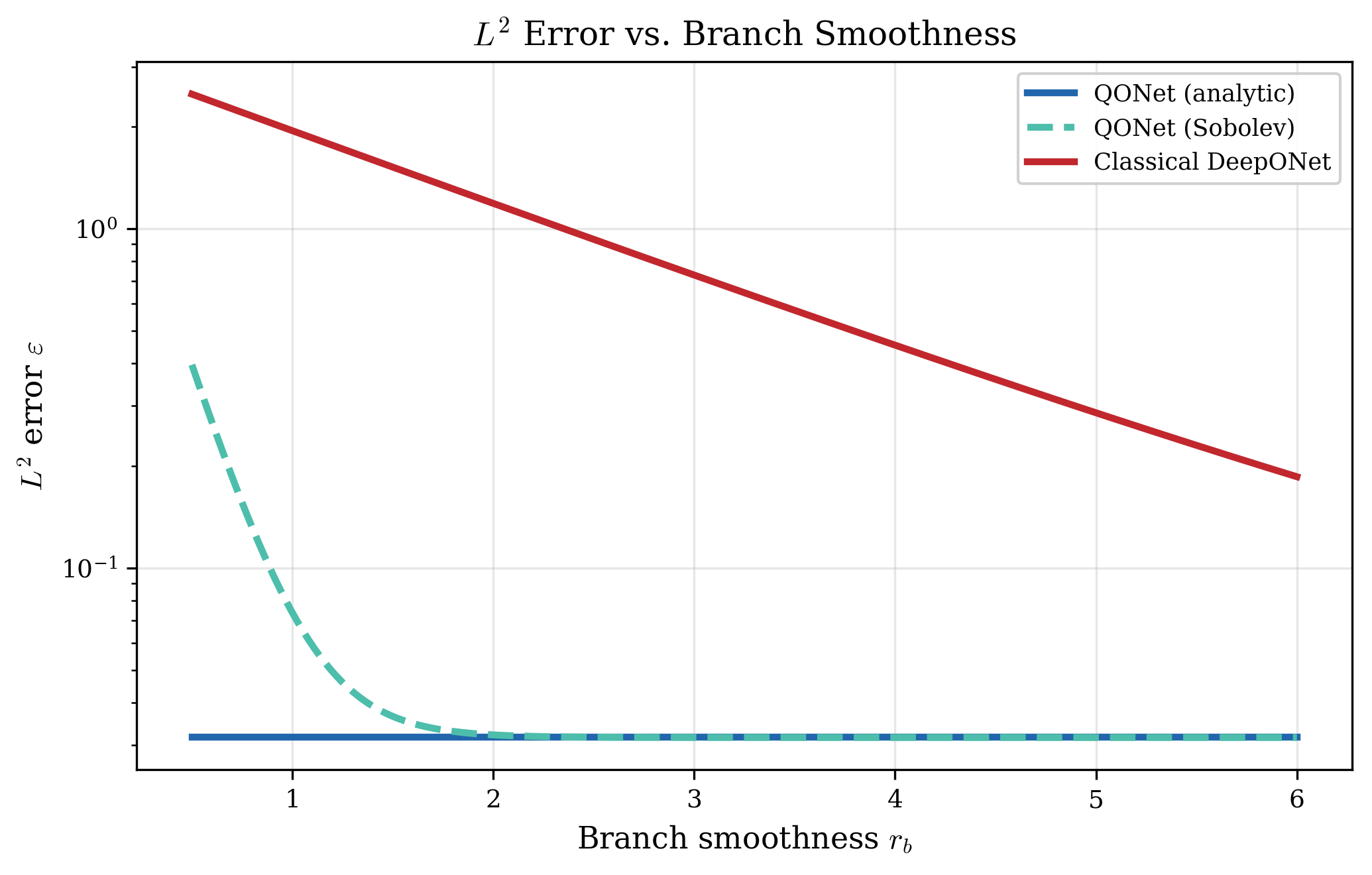}
    \caption{Total $L^2$ error as a function of the branch smoothness parameter $r_b$. Fixed parameters: $P_{\mathrm{total}}=150$, $n=10$, $m^c_b=8$.}
    \label{fig:2d_r}
\end{figure}

In Figure~\ref{fig:2d_r}, the effect of the smoothness of the coefficient functionals $u \mapsto c_\omega(u)$ on the error is examined for $P_{\text{total}}=150$. Both the quantum and classical branch errors improve with smoothness, but at different rates. The classical error decays as $P^{-2r_b/m^c_b} = P^{-r_b/4}$, while the quantum Sobolev error decays as $P^{-r_b/d_1} = P^{-r_b}$. For $d_1=1$ and $m^c_b=8$, the quantum rate is exactly four times faster.

Notably, both quantum architectures achieve lower error than the classical DeepONet across the entire smoothness range shown. The Sobolev curve approaches the classical curve only for very low smoothness ( $r_b \lesssim 0.5$ ), where all architectures struggle with rough coefficient functionals. As $r_b$ increases, the quantum advantage widens rapidly because the quantum Sobolev rate ( $r_b / d_1=r_b$ ) grows linearly with $r_b$, while the classical rate ($r_b / 4$) grows four times more slowly. The quantum analytic curve is independent of $r_b$ (it relies on exponential convergence rather than Sobolev smoothness) and remains well below both the classical and Sobolev curves for all values of $r_b$ shown.
% The breaking point (orange marker) indicates the value $r^*_b$ at which the QONet (Sobolev) transitions from worse to better than the classical architecture. For low smoothness ($r_b < r^*_b$), the classical network can be competitive because both approaches struggle with rough coefficient functionals. However, once $r_b$ exceeds the breaking point, the quantum advantage widens rapidly. The quantum analytic curve is independent of $r_b$ (it relies on exponential convergence rather than Sobolev smoothness) and remains below the classical curve for all values of $r_b$ shown.

\subsection{Branch Error vs.\ Circuit/Network Depth}

\begin{figure}[H]
    \centering
    \includegraphics[width=\textwidth]{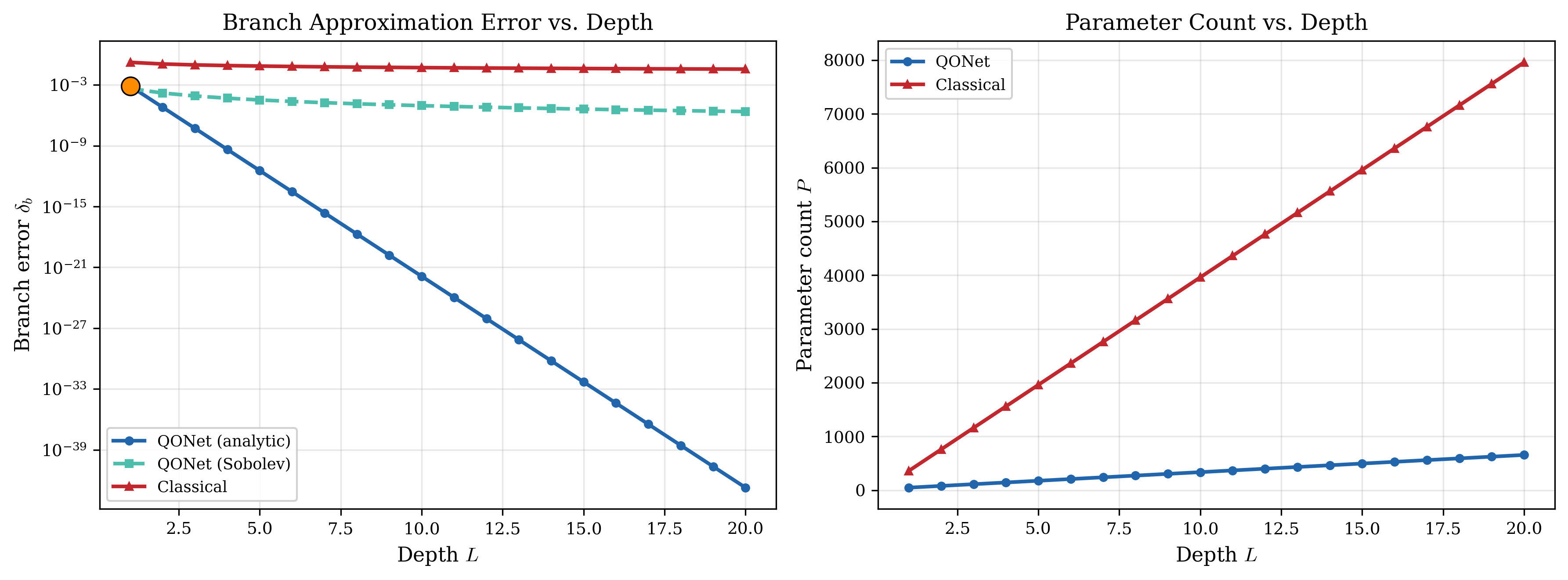}
    \caption{Left: Branch approximation error $\delta_b$ as a function of depth $L$ for both architectures. Right: Total parameter count as a function of depth. Fixed parameters: $m_b = m^c_b = 8$, $r_b=2$, width $w^c_b = 20$ (classical).}
    \label{fig:2d_L}
\end{figure}

Figure~\ref{fig:2d_L} presents a two-panel comparison isolating the branch error as a function of network/circuit depth. The left panel shows that the quantum branch error decreases \emph{exponentially} with depth (note the log scale), dropping by over 30 orders of magnitude across $L=1$ to $L=20$. The classical branch error, while also decreasing with depth, does so at a far slower (polynomial) rate, barely crossing $10^{-3}$ in the same range.

The right panel explains why: the quantum parameter count scales as $P_b = 2m_b(2L+1) = O(L)$, growing \emph{linearly} with depth. The classical parameter count scales as $P^c_b = O(w^2 L)$, growing linearly in $L$ but with a much larger constant due to the quadratic dependence on width $w$. Despite the classical architecture investing far more parameters per layer, the quantum architecture achieves dramatically lower error. This demonstrates the exponential advantage of VQC-based representations when the target functional is analytic.

\subsection{Error vs.\ Intrinsic Input Dimension $d_1$}

\begin{figure}[H]
    \centering
    \includegraphics[width=0.88\textwidth]{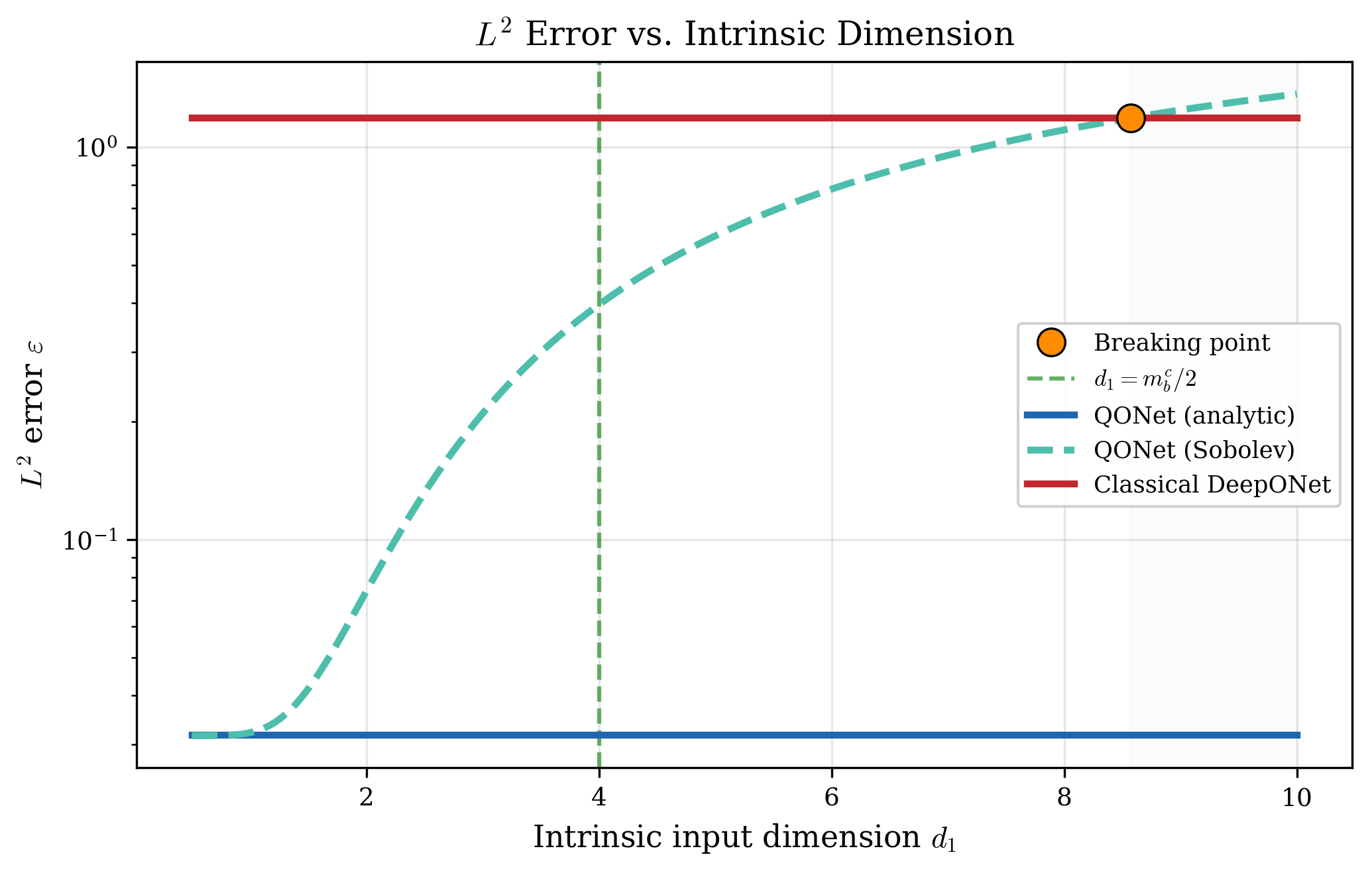}
    \caption{Total $L^2$ error as a function of the intrinsic input dimension $d_1$. The classical error is independent of $d_1$. Fixed parameters: $P_{\mathrm{total}}=150$, $n=10$, $m^c_b=8$.}
    \label{fig:2d_d1}
\end{figure}

Figure~\ref{fig:2d_d1} reveals the limits of the quantum Sobolev advantage. While the classical error is independent of $d_1$ (it depends on $m^c_b$, not $d_1$), the quantum Sobolev error degrades as $d_1$ increases because $\delta_b = O(P_b^{-r_b/d_1})$. The crossover point $d^*_1$ (marked by the orange circle) indicates where the quantum Sobolev error first exceeds the classical error. The vertical green dashed line marks the theoretical threshold $d_1 = m^c_b/2 = 4$ from Theorem~S14.

This figure has important practical implications: for problems with low intrinsic dimensionality (e.g., $d_1 = 1$ or $d_1 = 2$, as in the main paper's experiments), the quantum Sobolev advantage is substantial. For problems with intrinsically high-dimensional input spaces ($d_1 \gg m^c_b/2$), the Sobolev advantage vanishes and only the analytic regime retains a clear quantum benefit.

%% ── 3D PLOTS ─────────────────────────────────────────────────
\FloatBarrier
\section{Two-Parameter Analysis (3D Plots)}\label{sec:3d}

The 3D surfaces in this section plot the \emph{log-ratio} $\log_{10}(\varepsilon^C / \varepsilon^Q)$, which is positive when the QONet achieves lower error (blue region) and would be negative when the classical architecture performs better (red region).
For the parameter ranges explored here, the quantum architecture achieves lower error across essentially the entire domain, with the advantage varying in magnitude from modest (near the origin) to substantial (at large parameter budgets or high sensor counts).

% The orange contour lines on each surface trace the exact breakeven boundary $\varepsilon^C = \varepsilon^Q$.

\subsection{Error Ratio vs.\ $P_{\mathrm{total}}$ and $n$}

\begin{figure}[H]
    \centering
    \includegraphics[width=0.88\textwidth]{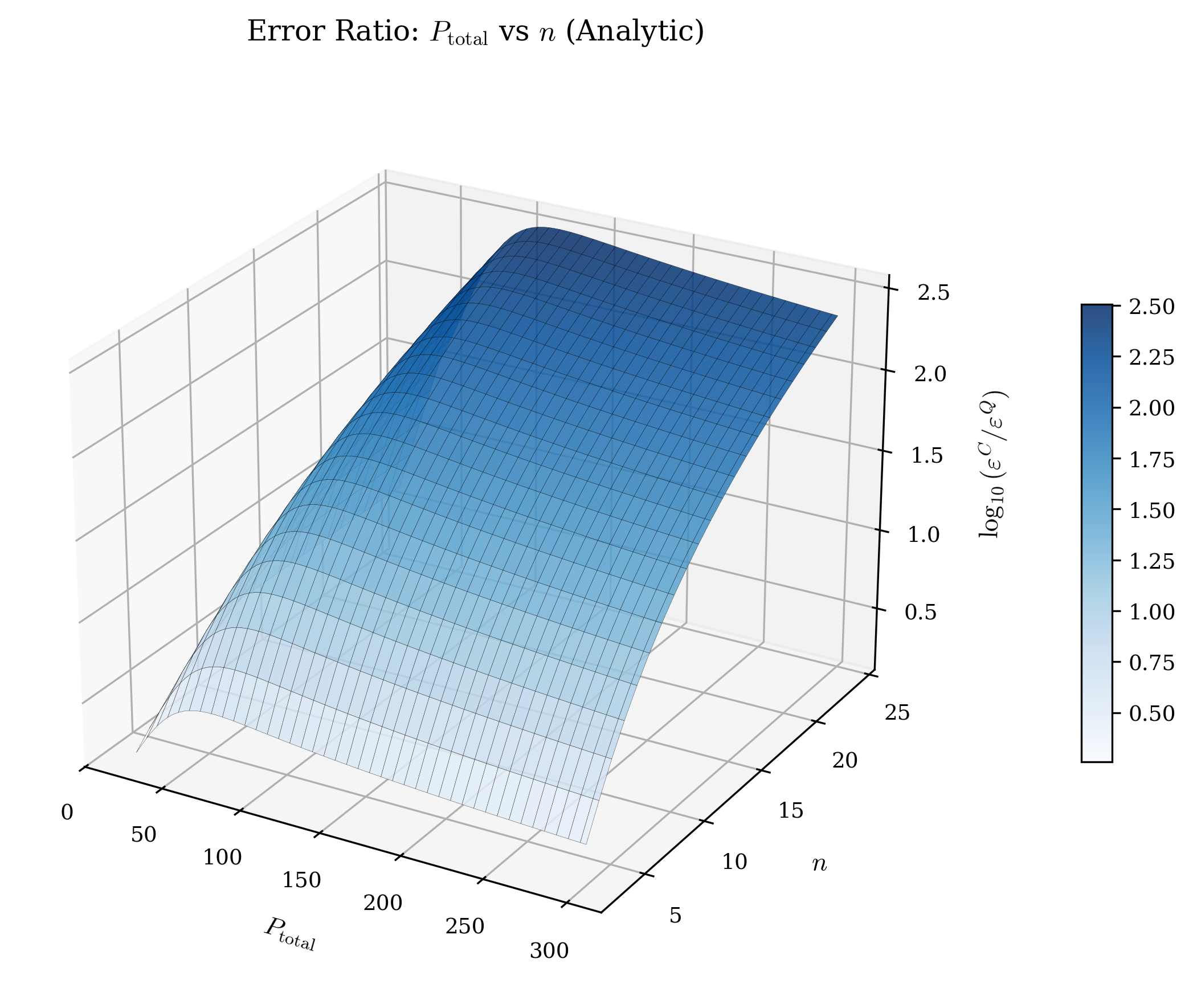}
    \caption{The surface $\log_{10}(\varepsilon^C / \varepsilon^Q)$ over the $(P_{\mathrm{total}}, n)$ plane. Analytic regime. Positive values (blue) indicate quantum advantage. Fixed parameters: $m^c_b=8$, $d_1=1$.}
    \label{fig:3d_Pn}
\end{figure}

In Figure~\ref{fig:3d_Pn}, the interaction between the parameter budget and the number of basis functions is visualized. The surface rises steeply in the high-$P$, moderate-$n$ region. Two key observations can be made. First, for small $P$ (left edge), the quantum circuits do not have sufficient depth for their exponential convergence to manifest. Second, for very large $n$ (top edge), both architectures are limited by the truncation floor and the advantage saturates. The sweet spot---maximum quantum advantage---is found at large $P$ and moderate $n$, where the quantum approximation errors are negligible but the classical ones remain significant.

\subsection{Error Ratio vs.\ $P_{\mathrm{total}}$ and $m^c_b$}

\begin{figure}[H]
    \centering
    \includegraphics[width=0.88\textwidth]{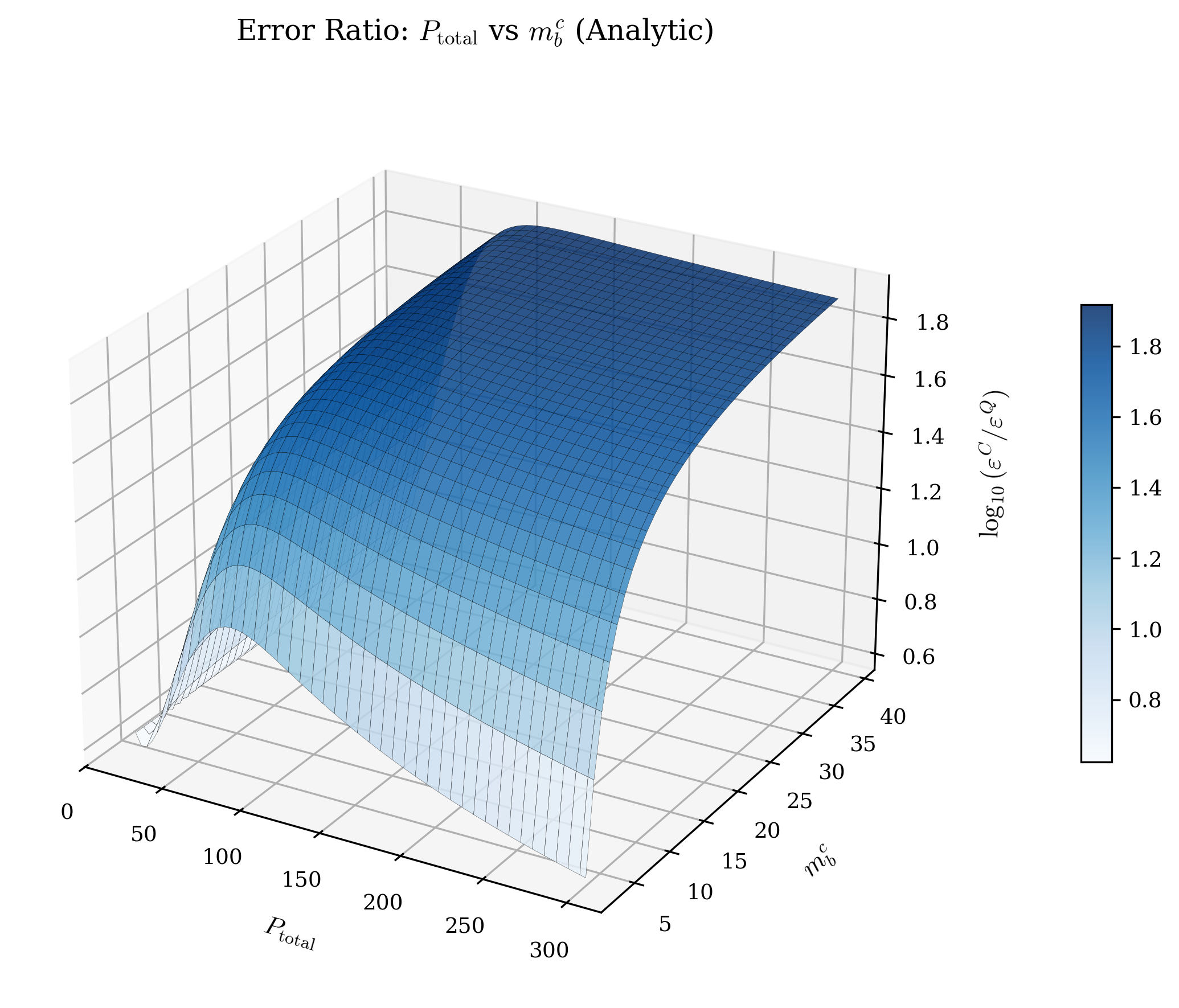}
    \caption{The surface $\log_{10}(\varepsilon^C / \varepsilon^Q)$ over the $(P_{\mathrm{total}}, m^c_b)$ plane. Analytic regime. Fixed parameters: $n=10$, $d_1=1$.}
    \label{fig:3d_Pm}
\end{figure}

Figure~\ref{fig:3d_Pm} illustrates the combined effect of the parameter budget and the classical sensor count. The quantum advantage grows along \emph{both} axes simultaneously. As $m^c_b$ increases (moving along the $y$-axis), the classical error worsens due to the curse of dimensionality, while the quantum error is unaffected. As $P$ increases (moving along the $x$-axis), the quantum error drops exponentially while the classical error drops only 
The surface is positive throughout the parameter range shown, confirming that the quantum architecture maintains its advantage across all combinations of $P_{\text {total }}$ and $m_b^c$ explored, with the largest gains occurring at high $m_b^c$ and large $P$.

% The orange breakeven contour shows that the crossover parameter $P^*$ decreases as $m^c_b$ increases---quantum advantage is achieved more easily for high-dimensional classical inputs.

\subsection{Error Ratio vs.\ $m^c_b$ and Smoothness $r_b$}

\begin{figure}[H]
    \centering
    \includegraphics[width=0.88\textwidth]{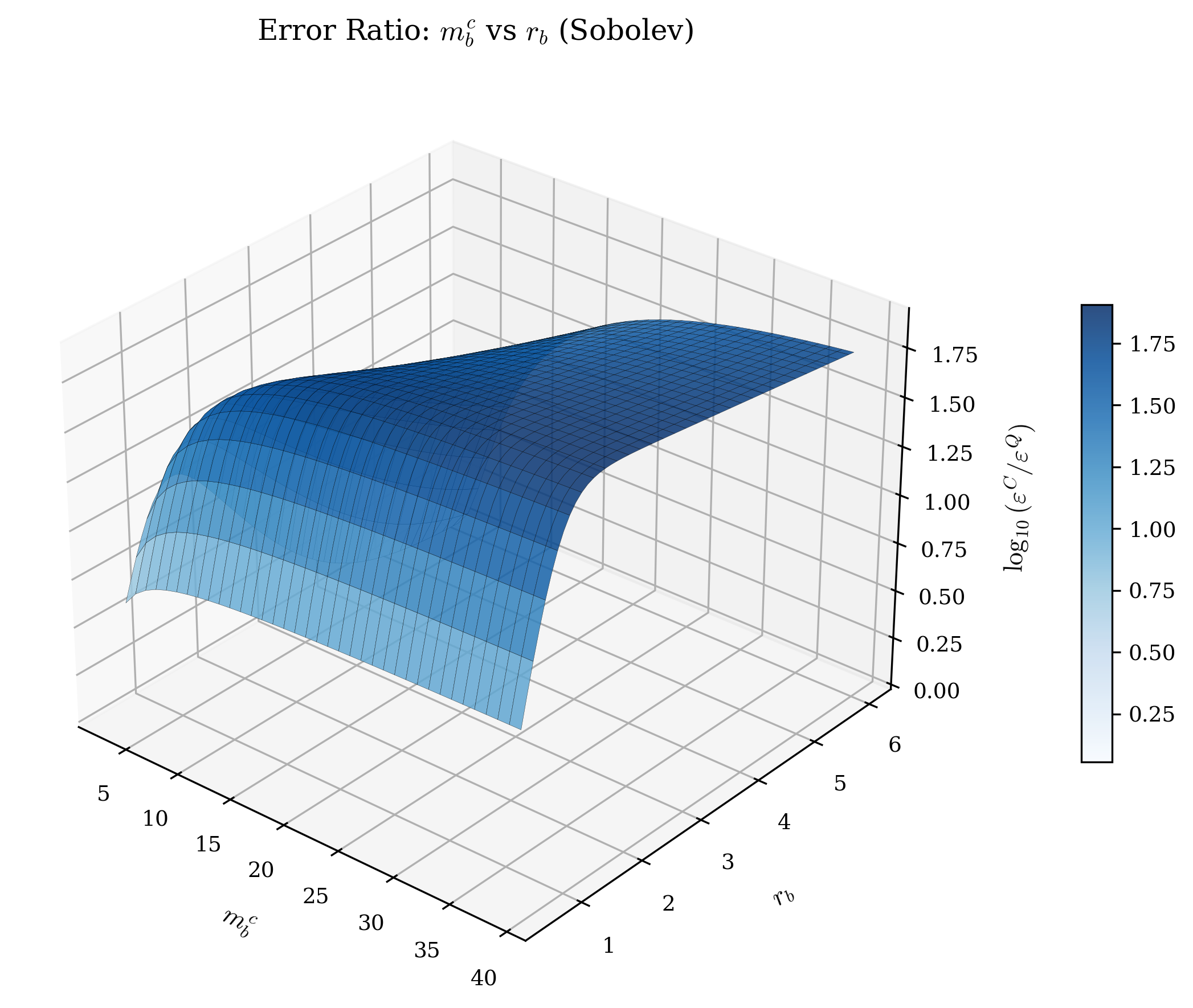}
    \caption{The surface $\log_{10}(\varepsilon^C / \varepsilon^Q)$ over the $(m^c_b, r_b)$ plane. Sobolev regime. Fixed parameters: $P_{\mathrm{total}}=150$, $n=10$, $d_1=1$.}
    \label{fig:3d_mr}
\end{figure}

Figure~\ref{fig:3d_mr} examines the Sobolev regime. The quantum advantage is maximized when both $m^c_b$ is large (classical curse of dimensionality is severe) and $r_b$ is large (smoother operators benefit more from the quantum branch's faster rate $\gamma^Q_b = r_b/d_1$). The region near the origin (small $m^c_b$, small $r_b$) shows values near zero, indicating approximate parity between the two architectures. The advantage emerges gradually and grows monotonically.

\subsection{Error Ratio vs.\ $P_{\mathrm{total}}$ and Intrinsic Dimension $d_1$}

\begin{figure}[H]
    \centering
    \includegraphics[width=0.88\textwidth]{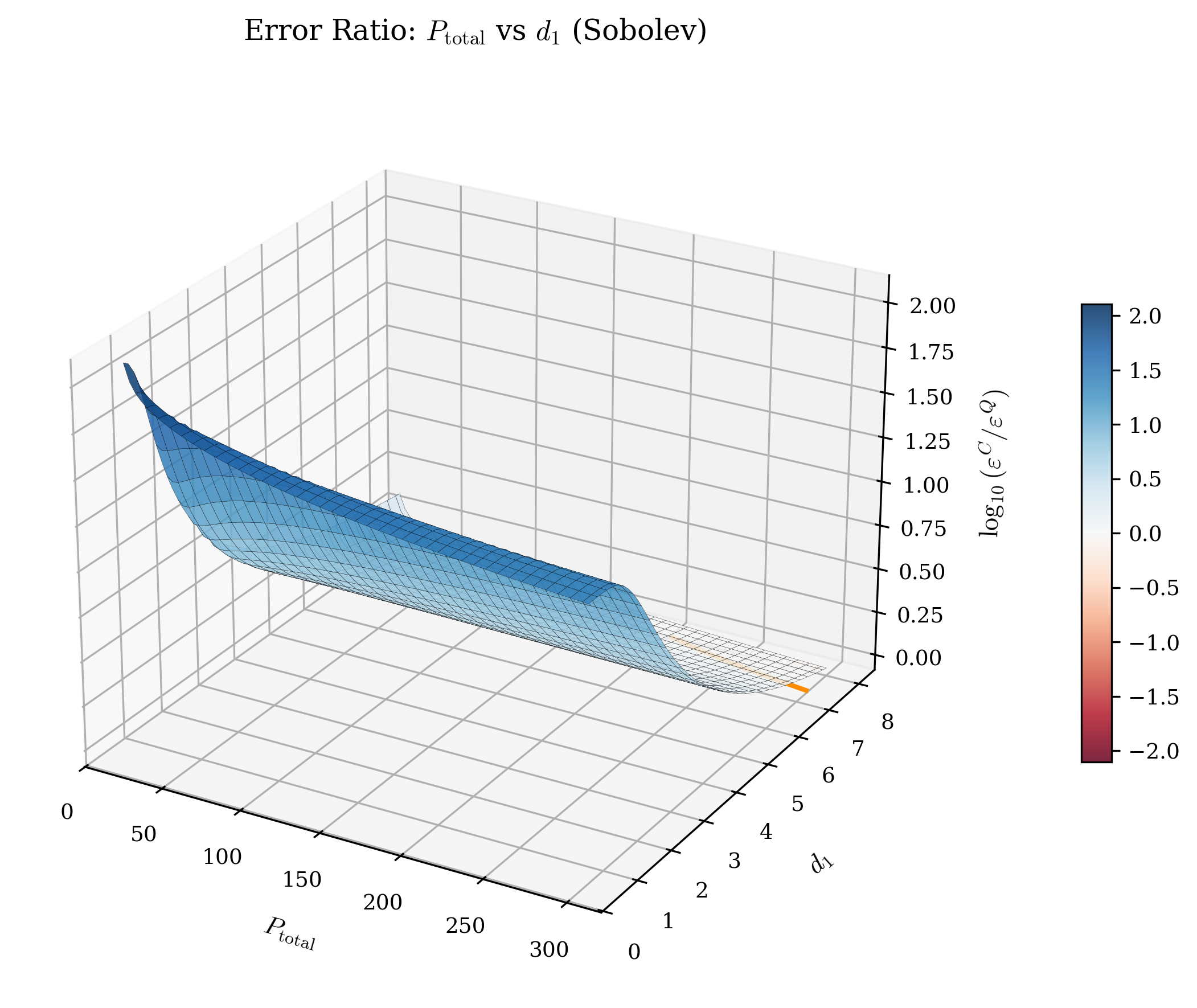}
    \caption{The surface $\log_{10}(\varepsilon^C / \varepsilon^Q)$ over the $(P_{\mathrm{total}}, d_1)$ plane. Sobolev regime. Fixed parameters: $n=10$, $m^c_b=8$.}
    \label{fig:3d_Pd1}
\end{figure}

Figure~\ref{fig:3d_Pd1} reveals how the quantum Sobolev advantage degrades as the intrinsic dimension $d_1$ increases. For small $d_1$ (front edge), the quantum rate $r_b/d_1$ is large and the advantage is pronounced. As $d_1$ grows, the quantum rate approaches the classical rate $2r_b/m^c_b$, and the surface approaches the breakeven plane. This confirms that the quantum Sobolev advantage is fundamentally a \emph{low-dimensional} phenomenon; for problems with intrinsic dimensionality comparable to the sensor count, the exponential (analytic) regime is needed to maintain a significant quantum benefit.

\subsection{Branch Error Ratio vs.\ Depth $L$ and Width $m$}

\begin{figure}[H]
    \centering
    \includegraphics[width=0.88\textwidth]{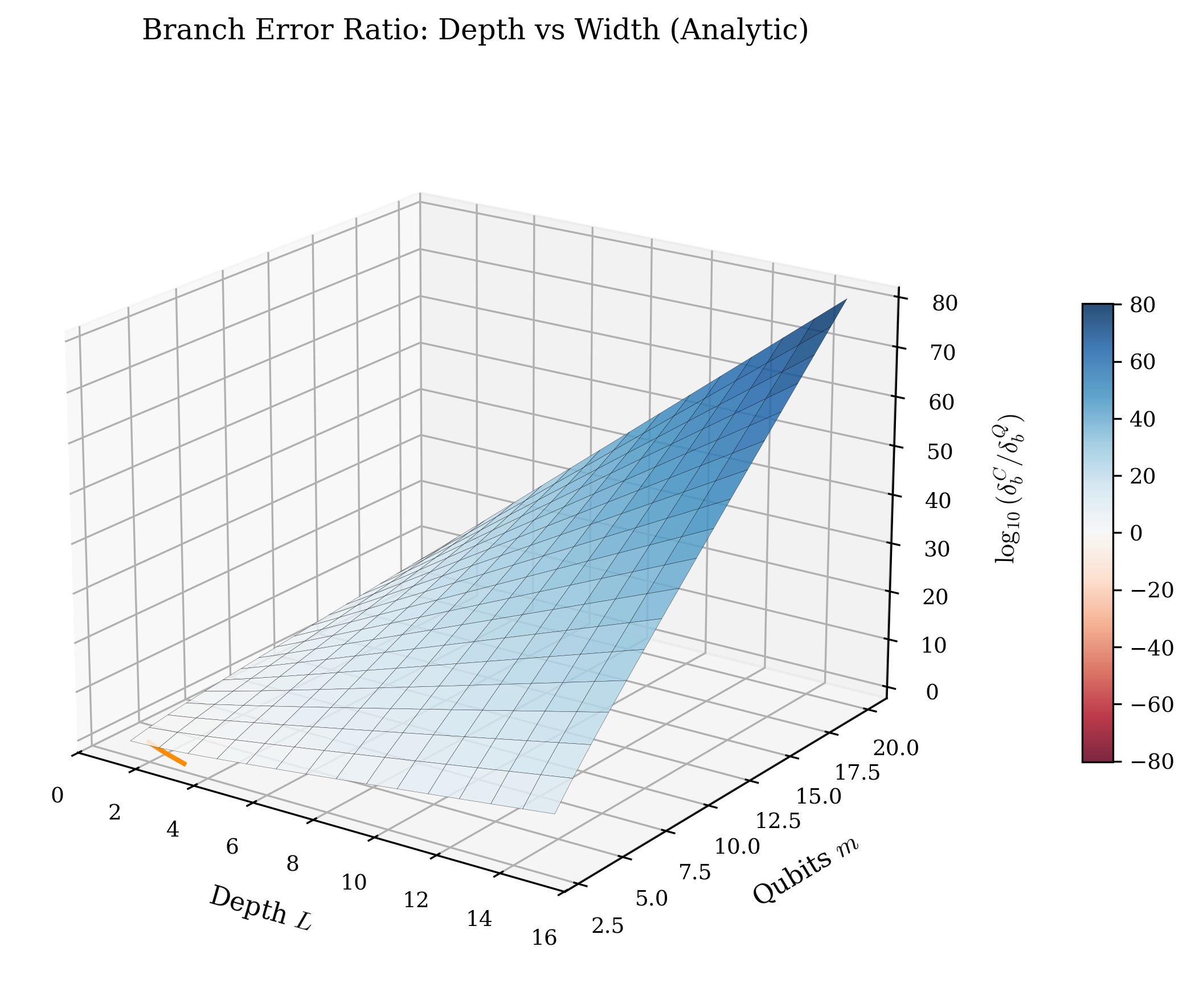}
    \caption{The surface $\log_{10}(\delta^C_b / \delta^Q_b)$ over the $(L, m)$ plane. Analytic quantum regime. The classical architecture uses width $w=20$.}
    \label{fig:3d_Lm}
\end{figure}

Figure~\ref{fig:3d_Lm} presents the most dramatic result: the branch error ratio over the depth--width plane. The surface rises rapidly, reaching several orders of magnitude of advantage at moderate depth and qubit count. This gap arises because the quantum branch error decreases as $e^{-\alpha \cdot 2m(2L+1)}$, which is doubly exponential in the product $mL$, while the classical branch error decreases only polynomially. The small region in the lower-left corner (very shallow circuits with few qubits) represents the only operating regime where the classical approach is competitive.

\subsection{Error Ratio vs.\ Smoothness $r_b$ and Intrinsic Dimension $d_1$}

\begin{figure}[H]
    \centering
    \includegraphics[width=0.88\textwidth]{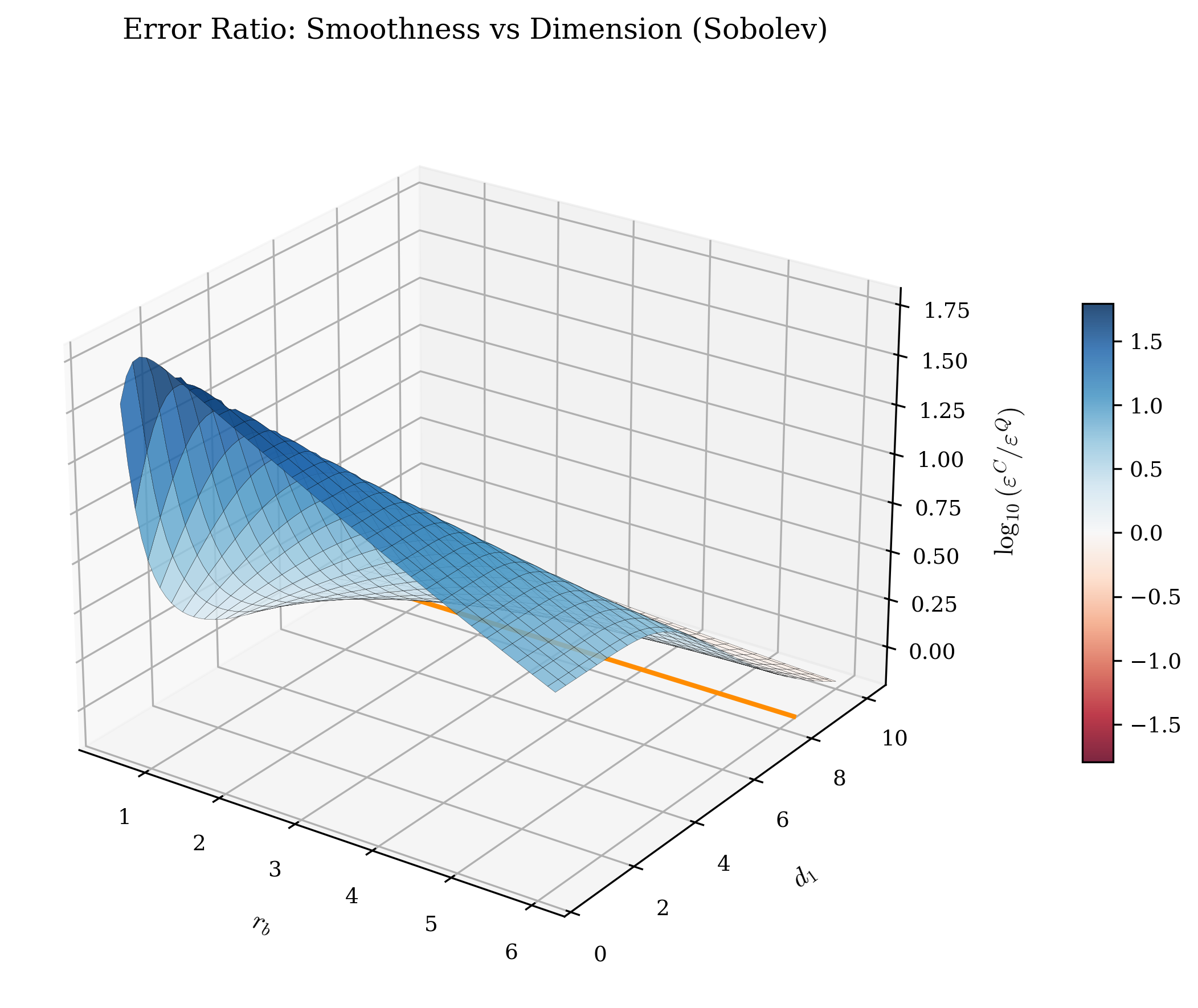}
    \caption{The surface $\log_{10}(\varepsilon^C / \varepsilon^Q)$ over the $(r_b, d_1)$ plane. Sobolev regime. Fixed parameters: $P_{\mathrm{total}}=150$, $n=10$, $m^c_b=8$.}
    \label{fig:3d_rd1}
\end{figure}

Figure~\ref{fig:3d_rd1} maps the full advantage landscape in the smoothness--dimension plane for the Sobolev regime. The surface confirms the theoretical prediction: advantage requires both sufficient smoothness ($r_b$ large) and low intrinsic dimensionality ($d_1$ small). The maximum advantage occurs in the lower-right corner ($r_b = 6$, $d_1 = 1$) and diminishes toward the upper-left corner. The breakeven contour separates the two regimes: the quantum advantage region occupies the lower-right triangle (high smoothness, low dimension), while parity or slight classical advantage occupies the upper-left region.

For the 1D differential equations considered in the main paper ($d_1 = 1$), these results confirm that quantum operator learning is expected to be advantageous for operators with moderate-to-high smoothness, which is typical of the diffusion, reaction-diffusion, and Burgers' equation operators studied therein.

%% ============================================================
%% PART VI: CONCLUSION
%% ============================================================
\newpage
\part{Conclusion}

\section{Summary of Results}

Explicit existential error bounds have been established for the QONet across four function space norms ($L^2$, $L^p$, $C^0$, $H^k$); parallel bounds have been developed for the classical ANN-based DeepONet relative to the Yarotsky worst-case Sobolev baseline; and precise conditions for quantum advantage have been derived conditional on the structural Assumption~\ref{ass:low-intrinsic-supp} and the analytic-regularity Assumption~\ref{ass:analytic}.

\subsection{Quantum Error Bounds}
The key technical contributions of this supplement are summarized as follows:
\begin{enumerate}
    \item The trunk approximation error coefficient $S(n) = \sum_k |c_{\omega_k}|$ is $O(1)$ for sufficiently smooth operators ($s > d_2$), ensuring that the trunk contribution does not grow with the number of basis functions.
    \item The $H^k$ branch error carries a Sobolev weight factor $n^{1/2+k/d_2}$ that accounts for the amplification of branch errors through derivative operations; the corresponding derivation appears following Theorem~\ref{thm:Hk_q}.
    \item The bilinear decomposition (Lemma~\ref{lem:decomp}) is based on the tighter two-term identity $c_k\phi_k - b_k t_k = (c_k - b_k)\phi_k + b_k(\phi_k - t_k)$, with the higher-order cross term $n\,\delta_b\,\delta_t$ retained explicitly.% so that the boxed inequalities are strict.
    \item A structural assumption on the target operator (Assumption~\ref{ass:low-intrinsic-supp}) is introduced to justify the replacement $m_b^c \to d_1$ in the polynomial-regime quantum branch rate. In its absence, the relevant universal-approximation theorems for the VQC feature $m_b$ rather than $d_1$ in the rate.
\end{enumerate}
These results establish that the QONet can approximate any sufficiently smooth operator to arbitrary accuracy as circuit resources are increased, with explicit existential convergence rates that complement the existential guarantees of Theorems~3 and 4 of the main paper.

\medskip
\noindent\textbf{Existential versus learning-theoretic bounds.} All bounds developed in this supplement are existential in character: they assert the existence of parameter values realizing the prescribed accuracy. They do not address whether gradient-based optimization recovers such parameters, nor do they address sample complexity. The numerical experiments of the main paper report the accuracies attained under Adam optimization at the specified hyperparameter settings.

\subsection{Conditions for Quantum Advantage}
The quantum QONet provides superior error bounds compared to the classical Yarotsky-baseline DeepONet under the following conditions:
\begin{enumerate}
    \item \textbf{Analytic operators} (Condition A): When the target operator has analytic coefficient functionals and basis functions, the quantum architecture achieves \emph{exponential} convergence ($\delta \propto e^{-\alpha P}$) versus the classical \emph{polynomial} convergence ($\delta \propto P^{-\gamma}$). The quantum advantage activates beyond a finite parameter threshold $P^*$, as demonstrated in Figures~\ref{fig:2d_P} and~\ref{fig:3d_Lm}.

    \item \textbf{High-dimensional discretization} (Condition B): When the number of sensor locations exceeds twice the intrinsic function space dimension ($m^c_b > 2d_1$), the quantum branch avoids the curse of dimensionality. 
    For the 1D experiments in the main paper, this condition is satisfied whenever $m^c_b > 2$, which holds in essentially all practical configurations, conditional on the structural Assumption~\ref{ass:low-intrinsic-supp} for the target operator.
    As shown in Figure~\ref{fig:2d_m}, the classical error increases monotonically with $m^c_b$ while the quantum error remains constant under Assumption~\ref{ass:low-intrinsic-supp}.

    \item \textbf{Fourier-structured problems} (Condition C): When the trunk must approximate trigonometric functions, the VQC's native Fourier structure gives it an inherent representational advantage, requiring $O(\log K)$ qubits for frequency $K$ versus $O(K^{d_2/r_t})$ classical parameters.
\end{enumerate}

\subsection{Practical Implications}
The parameter complexity required to attain a target error $\varepsilon$ is summarized below:
\begin{itemize}
    \item Classical: $P^c \propto \varepsilon^{-m^c_b/(2r_b)}$ under the Yarotsky worst-case Sobolev baseline, with the curse of dimensionality manifest in $m^c_b$. For analytic targets, classical analytic-class bounds (Mhaskar; Schwab--Zech) yield faster rates.
    \item Quantum (analytic regime, Assumption~\ref{ass:analytic}): $P^Q \propto \log(1/\varepsilon)$ for the circuit-depth budget; the total qubit count remains polynomial in $1/\varepsilon$. The bound is existential.
    \item Quantum (Sobolev regime, Assumption~\ref{ass:low-intrinsic-supp}): $P^Q \propto \varepsilon^{-d_1/r_b}$, polynomial in the intrinsic dimension $d_1$ rather than in $m^c_b$.
\end{itemize}

These results provide conditional, existential theoretical justification for the QuantumONet architecture presented in the main paper and establish precise conditions under which quantum operator learning offers provable advantages over classical approaches relative to the stated baselines.

\medskip\noindent\textbf{Limitations and open problems.} The following items identify directions for future work.
\begin{itemize}
\item A direct proof that the specific QONet construction realizes the rate $r_b/d_1$ under Assumption~\ref{ass:low-intrinsic-supp}, without relying on the appeal to~\cite{Manzano2025approximationquantum}, would close the most significant theoretical gap.
\item Multi-output universality of $n$ simultaneous Pauli-$Z$ measurements on a single $n$-qubit state is not, to the authors' knowledge, established for the specific HEA employed here; the bounds are accordingly phrased conditional on the scalar approximation results.
\item A basis-free SVD-based analysis in the spirit of~\cite{LanthalerMishra2022} would obviate Assumption~\ref{ass:trunk-target}.
\item Empirical estimation of the constants $\alpha_b$, $\alpha_t$, $C_b$, and $C_t$ would render the bounds quantitatively useful for circuit design.
\item A noise and decoherence analysis for NISQ-scale execution of the QONet remains to be performed and is essential for any hardware-level advantage claim.
\end{itemize}

\bibliographystyle{ieeetr}
\bibliography{ref}